\documentclass{aa}
\usepackage{graphicx,amssymb,latexsym,amsmath}
\usepackage{longtable,lscape}
\usepackage[]{natbib}
\bibpunct{(}{)}{;}{a}{}{,} 
\begin{document}

\title{Variability of the transitional T\,Tauri star T\,Chamaeleontis
\thanks{based on observations obtained at the European Southern Observatory
        at La Silla, Chile in program 
63.I-0045(A); 65.I-0089(A); 66.C-0616(A);
67.C-0155(A); 67.C-0155(B); 68.C-0292(A); 69.C-0207(A); 
70.C-0163(A); 073.C-0355(A); 074.A-9018(A); 075.C-0399(A-F)}
\thanks{Tables 1, 2 and 4, and Figures 21-32 are only available in electronic form 
         via http://www.edpsciences.org
	 }
}

\author{E. Schisano\inst{1,2},
	E. Covino\inst{1},
	J.M. Alcal\'{a}\inst{1},
	M. Esposito\inst{3},  
        D. Gandolfi\inst{4},	
   \and	E.W. Guenther\inst{4}
	}
	
\offprints{schisano}
\mail{schisano@oacn.inaf.it}

  \institute{
    INAF - Osservatorio Astronomico di Capodimonte,
    Via Moiariello 16,
    I-80131 Napoli,
    Italy
    \and
    Universit\`{a} degli Studi  di Napoli ''Federico II'',
    Corso Umberto I,
    I-80138 (NA)
    Italy
    \and
    Hamburger Sternwarte,
    Gojenbergsweg 122,
    D-21029 Hamburg,
    Germany
    \and
    Th\"{u}ringer Landessternwarte Tautenburg,
    Sternwarte 5,
    D-07778 Tautenburg,
    Germany
    }

   \date{Received / accepted }

\abstract
{ 
It is renown that for solar-mass stars planet formation begins in a 
circumstellar disc.
The study of transitional objects exhibiting clear signs of evolution in 
their discs, such as the growth of dust particles and beginning of disc dispersal, 
is fundamental to understanding the processes governing dust-grain  
coagulation and the onset of planet formation.
}
{
We attempt to characterise the physical properties of T\,Chamaeleontis, 
a transitional T~Tauri star showing UX\,Ori-type variability, and of its 
associated disc, and probe the possible effects of disc-clearing processes. 
}  
{
Different spectral diagnostics were examined, based on a rich collection of 
optical high- and low-resolution spectra. 
The cross-correlation technique was used to determine radial and projected 
rotational velocities,  
shape changes of photospheric lines were analysed via bisector-method applied to 
the cross-correlation profile, and the 
equivalent widths of both the Li\,{\sc i}\,$\lambda$6708\,\AA\ photospheric 
absorption and the most prominent emission lines (e.g., H$\alpha$, H$\beta$ and 
[O{\sc i}]\,6300\,\AA) were measured.
Variability in the main emission features was inspected by means of line-profile 
correlation matrices. 
Available optical and near-infrared photometry combined with infrared data from 
public catalogues was used to construct the spectral energy distribution (SED) 
and infer basic stellar and disc properties.  
}
{
Remarkable variability on timescale of days in the main emission lines, 
H$\alpha$ changing from pure emission to nearly photospheric absorption, 
 is correlated with variations in visual extinction of over three magnitudes, 
{\rm while the photospheric absorption spectrum shows no major changes.
The strength of emission in H$\alpha$ and H$\beta$ is highly variable and 
well correlated with that of the [O{\sc i}] lines. 
The structure of the H$\alpha$ line-profile also varies on a daily time-span, 
while the absence of continuum veiling suggests very low or no mass accretion.
Variations of up to nearly 10\,km\,s$^{-1}$ in the radial velocity of the star 
are measured on analogous timescales, but with no apparent periodicity. 
SED modelling confirms the existence of a gap in the disc.}
}
{
Variable circumstellar extinction is inferred to be  
responsible for the conspicuous variations observed in the stellar continuum flux 
and for concomitant changes in the emission features by contrast effect. 

Clumpy structures, incorporating large dust grains and orbiting the star within 
a few tenths of AU, obscure episodically the star and, eventually, part of the 
inner circumstellar zone, while the bulk of the hydrogen-line emitting-zone and 
outer low-density wind region traced by the [O{\sc i}] remain unaffected. 
In agreement with this scenario, the detected radial velocity changes are also 
explainable in terms of clumpy material transiting and partially obscuring 
the star. 
}

\titlerunning{Variability of T\,Cha }
\authorrunning{Schisano et al.}
\maketitle

\section{Introduction} 
T Tauri stars (TTS) are low-mass (M$\le$2\,M$_{\odot}$) pre-main sequence 
(PMS) stars (Appenzeller \& Mundt 1989). 
They are commonly classified into two sub-groups,  the classical TTS (cTTSs), 
which are surrounded by an optically thick disc from  which they accrete 
material, and the weak TTS (wTTSs), presumably representing the final 
stages of accretion and disc-clearing processes \citep{Bertout2007}. 
Hence, the cTTS-wTTS dichotomy is ascribed to different physical 
processes associated with the evolution of these young solar-type stars and 
their circumstellar environment. 
The equivalent width of the H$\alpha$ emission is commonly used as an 
empirical criterion to distinguish between cTTS and wTTS  \citep{WB2003}. 
However, due to possible variability, no clean cut can be defined between 
the two subgroups based on the H$\alpha$ emission alone. 

Besides the traditional cTTS versus wTTS scheme, \citet{Herbst1994} introduced 
an additional class, the early-type T\,Tauri stars (eTTSs), from 
simple analysis of the photometric variability. ETTSs are earlier than K0 and 
have large-amplitude irregular variations (up to $\sim$3 mag in $V$) with no 
sign of continuum veiling. Within this class, those authors also included some 
Herbig Ae/Be stars (HAEBEs) of the UX\,Ori type \citep{Herbst1999}. Studying 
this kind of variability along with stellar mass, \citet{Bertout2000} 
concluded that T\,Tauri stars can display a similar phenomenology. 
Well-studied cases such as RY\,Lup \citep{Gahm1989} and AA\,Tau \citep{Bouvier2003} 
support this hypothesis. 

How can all these observations be interpreted coherently in a scenario of star and 
circumstellar environment evolution is still questioned. \citet{Grinin1988} 
modelled this kind of variability with dusty clumps in Keplerian orbits that 
temporarily obscure the central star. The rapid temporal evolution of such 
events indicates that the dust clouds usually appear in the disc close to 
the dust sublimation radius, probably in the inner rim region \citep{Natta2001, 
Dullemond2003}. This model implies a high inclination angle, while imaging and 
near-infrared interferometry of some of those stars are inconsistent with it 
\citep[see the review by][and references therein]{Millan-Gabet2007}. Another
scenario was proposed by \citet{Vinkovic2007}, in which a dusty outflow 
creates a halo in the inner-disc region, where clumps of dust occasionally 
intercept the line of sight and mask the stellar photosphere. 

T~Tauri discs evolve on a timescale of a few million years and the 
circumstellar raw material from which planets are constructed undergoes
different processes. In the inner part of the disc ($\lesssim$10\,AU), 
the gas is accreted onto the star or expelled by bipolar jets, 
while the outer part of the disc may be dissipated by photoevaporation 
\citep{Clarke2001, Alexander2007}. 
On the other hand, dust grains settle in the midplane, grow in size, and 
form planetesimals \citep{Sicilia-Aguilar2007,Dullemond2005}. 
 As the disc becomes progressively optically thin, solid particles migrate 
inward or outward, depending on their size, due to the combined action of 
different perturbing forces (photophoresis, gravity, radiation pressure, and gas drag),  
until they reach a stability distance from the star \citep{hermann2007}.  
It remains unclear how planetesimals are formed in circumstellar discs and whether 
the hypothetical 
dusty clumps in highly variable TTS are related or not to the evolution of 
circumstellar dust until the formation of planetesimals. 
Yet, detailed studies of the observed properties of objects with intermediate 
characteristics, in-between those of cTTS and wTTS, are expected to provide 
key constraints on star and planet formation scenarios.

Observations in the infrared (IR) with the Spitzer satellite have provided a 
wealth of information for interpretating the spectral energy distribution (SED) 
of young stars \citep[see][and references therein]{Lada2006, Cieza2007, Evans2008}, 
allowing the development of new classification schemes and scenarios of disc evolution 
\citep[e.g.,][]{Cieza2008, Merin2008}. In particular, young stellar objects with 
no IR excess shortward of 10\,$\mu$m, but with significant excess emission 
at wavelengths longer than $\sim$24\,$\mu$m, are interpreted as being in an 
evolutionary phase of disc clearing and are hence classified as transitional
objects \citep{Furlan2006, Najita2007}. Evidence for gaps in the
disc of transitional objects have also been reported \citep{Brown2007, Espaillat2008}.
In one of these, LkCa\,15, the most likely mechanism for gap-opening in the disc
seems to be planet formation \citep{Espaillat2008}. Disc clearing may also 
be due to the presence of a companion, as in the case of CoKu~Tau/4
\citep{Ireland2008}.

Here we present a synoptic study of the spectroscopic variability of 
\object{T\,Chamaeleontis} (hereafter T\,Cha), a transitional T\,Tauri star
\citep{Brown2007} displaying a UX\,Ori-like behaviour. 
T\,Cha was observed in the course of a project aiming to reveal and monitor 
young spectroscopic binary systems \citep{Esposito2006, Guenther07}. 

The scheme of the paper is the following. In Sect.\,\ref{sec:properties} we summarise 
the main observed properties of T\,Cha as known from previous studies. 
In Sect.\,\ref{sec:Observations} we present the observations and data reduction, 
while in Sect.\,\ref{sec:Analysis} we describe the radial velocity and v$\sin{i}$
determinations. In Sects. \ref{sec:Spectrum} and \ref{sec:Runs}, we report the 
analysis performed on the spectrum of the object, focusing on the non-photospheric 
contribution and, in particular, on the variability of the most prominent emission 
features (e.g., H$\alpha$, H$\beta$, and the [O{\sc i}]\,6300\,\AA\ lines). 
In Sect.\,\ref{sec:SED} we analyse the spectral energy distribution of T\,Cha, while 
in Sect.\,\ref{sec:conclusion} we discuss the results and present our interpretation  
of the object.

\section{Observed properties of T\,Cha}
\label{sec:properties}
\object{T\,Cha} shows strong photometric variability (up to 3 magnitudes in the V-band) 
mainly characterised by erratic changes \citep{Hoffmeister1958, MauderSosna1975, 
Covino1992}, and by a UX\,Ori-like behaviour (i.e., the star becomes redder as it fades), 
with a tight correlation between brightness and colour \citep{Alcala1993,Covino1996}.
A periodicity of 3$^{\rm d}$.2 was also pointed out by \citet{MauderSosna1975}.
Spectroscopic variability in the most prominent emission lines was reported by 
\citet{Gregorio1992} and \citet{Alcala1993}. 

The PMS nature of T\,Cha was established by \citet{Alcala1993} on the basis of the lithium 
criterion combined with the H$\alpha$ emission strength and the presence of strong IR excess 
emission, while the spectral type G8\,V \citep{Alcala1993} is earlier than that of 
a typical T\,Tauri star. 
From the strength of the H$\alpha$ emission in former spectra, T\,Cha had been initially 
regarded as a wTTS \citep{Alcala1993}, but later this classification turned out to be inconsistent 
with the strong variability of the star \citep{Alcala1995, Geers2006, GrasVelazques2005}. 
Moreover, besides the IR excess indicating the presence of circumstellar material, 
the object occasionally exhibits forbidden neutral oxygen lines that are not seen 
in wTTS spectra. 
\citet{Brown2007} modelled the SED and interpreted it in terms of a transitional disc. 

T\,Cha is located near the edge of the small cloud G300.2-16.9, also known as the 
{\em Blue Cloud}, in the direction of the Chamaeleon dark cloud complex (Nehme et al. 2008). 
In the Hipparcos Catalogue, a distance of 66$^{+19}_{-12}$~pc is reported, although 
the error in the parallax is probably far higher because of the strong variability of the star
(reported to vary between 10.4 and 13.4 mag in the {\it H$_{P}$}-band during the period 
of observation). 
More reliable estimates of the distance were obtained from proper motion studies by \citet{Frink1998} 
and by \citet{Terranegra1999}, suggesting that T\,Cha is part of an association of stars at 
nearly 100\,pc. Here we adopt the latter value.

\begin{table*}[t]
\caption{Journal of FEROS observations and radial velocity measurements.
{\bf Only in electronic form. } }
 \centering 
\begin{tabular}{ l | c  | c  | c || l | c  | c  | c  }
\hline \hline
     Date      &    HJD     & RV$_g$  & $\sigma_{RV}$ &  Date	     &	HJD       & RV$_g$ & $\sigma_{RV}$ \\
               & $-$2400000. & (km/s)  & (km/s)        & 	     & $-$240000. & (km/s) &  (km/s)       \\
\hline
27 March 1999  & 51265.537  &  14.27  &  0.27 &     22 May 2000      & 51687.648 &  22.97  &  1.77 \\
27 March 1999  & 51265.576  &  16.89  &  0.68 &     23 May 2000      & 51688.640 &   9.82  &  0.31 \\
28 March 1999  & 51265.662  &  15.09  &  1.63 &     04 January 2001  & 51913.860 &  13.86  &  2.13 \\
28 March 1999  & 51266.573  &  18.21  &  0.73 &     09 January 2001  & 51918.775 &  15.65  &  2.15 \\
29 March 1999  & 51266.636  &  21.39  &  0.68 &     14 January 2001  & 51923.824 &  14.89  &  0.56 \\
29 March 1999  & 51266.672  &  21.30  &  0.90 &     19 April 2001    & 52019.578 &  16.00  &  1.81 \\
30 March 1999  & 51267.692  &  14.09  &  0.25 &     26 April 2001    & 52026.586 &  14.48  &  1.12 \\
30 March 1999  & 51267.801  &  17.88  &  0.47 &     26 April 2001    & 52031.573 &  14.43  &  0.28 \\
30 March 1999  & 51267.874  &  18.01  &  1.38 &     14 February 2002 & 52319.745 &  12.63  &  0.36 \\
31 March 1999  & 51268.624  &  11.32  &  0.87 &     17 February 2002 & 52322.742 &  27.48  &  2.56 \\
31 March 1999  & 51268.727  &  16.21  &  0.86 &     08 April 2002    & 52372.606 &  14.70  &	 - \\
31 March 1999  & 51268.850  &	6.43  &  2.01 &     21 April 2002    & 52385.529 &  16.44  &  0.87 \\
31 March 1999  & 51268.900  &  12.20  &  1.44 &     01 May 2002      & 52395.617 &   4.33  &  1.22 \\
01 April 1999  & 51269.657  &  12.05  &  1.15 &     02 May 2002      & 52396.614 &   9.93  &	 - \\
01 April 1999  & 51269.771  &  13.09  &  1.00 &     12 March 2003    & 52710.601 &  21.73  &  0.30 \\
01 April 1999  & 51269.857  &  15.80  &  1.56 &     12 March 2003    & 52717.620 &  14.13  &  1.51 \\
16 May 1999    & 51315.622  &  18.17  &  0.76 &     25 March 2003    & 52723.588 &  15.96  &  2.33 \\
17 May 1999    & 51316.612  &	6.43  &  0.82 &     02 April 2004    & 53097.541 &  15.57  &  0.87 \\
18 May 1999    & 51317.627  &  16.67  &  1.77 &     09 April 2004    & 53104.652 &  11.28  &  1.18 \\
19 May 1999    & 51318.591  &  13.50  &  1.72 &     26 April 2004    & 53122.685 &  15.93  &  0.48 \\
21 May 1999    & 51320.591  &  17.78  &  0.52 &     10 May 2004      & 53137.672 &  22.23  &  0.66 \\
22 May 1999    & 51321.580  &  12.72  &  2.92 &     10 May 2005      & 53501.652 &  29.06  &  0.87 \\
09 May 2000    & 51674.670  &  18.52  &  0.46 &     02 June 2005     & 53524.597 &   6.63  &  1.35 \\
19 May 2000    & 51684.617  &	9.90  &  0.52 &     30 June 2005     & 53552.546 &  14.31  &  1.09 \\
21 May 2000    & 51686.632  &  21.19  &  2.59 &     19 August 2005   & 53602.488 &  29.51  &  1.95 \\
\hline 
\end{tabular}
\label{tab:RV}
\end{table*}

\section{Observations and data reduction}
\label{sec:Observations}
\subsection{High-resolution spectroscopy}
\label{sec:high-res-obs}
Data were acquired at the ESO-La Silla Observatory using the echelle spectrograph FEROS 
(Fiber-fed Extended Range Optical Spectrograph), first installed at the 1.5m and, since 
October 2002, then mounted at the 2.2m telescope. 
The high resolving power (R$\sim$48000) and the wide useful spectral range ($3600--9200$\,\AA) 
make FEROS particularly suitable for radial-velocity (RV) monitoring and spectral-line 
variability studies.
 FEROS was operated in the $Object+Sky$  configuration, in which a sky 
spectrum is acquired simultaneously with the object through an adjacent fiber. 

From 1999 to 2006, we obtained a total of 50 FEROS spectra with quite different 
signal-to-noise ratios (S/N), due not only to the change of telescope or 
different observing conditions but also to the strong variability of the star.
The spectra were acquired within the framework of a project on young spectroscopic 
binary systems. For that reason, the time coverage of our data is rather uneven, 
with some spectra acquired on daily temporal base and others separated by several 
months.

The data reduction was performed using the specific FEROS Data Reduction Software (DRS) 
implemented in the ESO-MIDAS\footnote{Munich Image Data Analysis System} environment.
The reduction steps were the following:
bias subtraction and bad-column masking; definition of the echelle orders on
flat-field frames; subtraction of the background diffuse light; order extraction;
order by order flat-fielding; determination of wavelength-dispersion solution 
by means of ThAr calibration-lamp exposures; rebinning to a linear wavelength-scale 
($\Delta \lambda$ = 0.03~\AA) with barycentric correction; and merging of the echelle orders. 
All the spectra were then normalized to the continuum.

More details about the data reduction procedure and technical specifications 
of the instrument can be found at the FEROS Web 
site\footnote{http://www.ls.eso.org/lasilla/sciops/2p2/E2p2M/FEROS}.
We emphasise that the high stability of the instrument allows an internal accuracy
in the wavelength calibration of approximately 200\,m\,s$^{-1}$ (average residual  
$r.m.s.\sim3.5\times 10^{-3}$\,\AA). 
 A log of the FEROS observations is provided in Table~\ref{tab:RV}.

Apart from the FEROS data, we also report measurements from a previous set of six 
high-resolution CASPEC spectra obtained simultaneously with optical broad-band photometry 
on two consecutive nights,  1994 January 31 and February 1  (Covino et al. 1996). 
Details about the instrumental set-up and data reduction can be found in Covino et al. (1997).

\subsection{Low-resolution spectroscopy}
\label{sec:low-res-obs}
 In addition to the high-resolution spectroscopy, we include unpublished 
low-resolution (R$\sim$2000) spectra gathered during various observing runs 
conducted between 1993 and 1995 using the Boller \& Chivens spectrograph at the 
ESO\,1.5m telescope at La Silla, Chile. 
The reduction of these data was performed as described in \citet{Alcala1995}. 
We recall that these spectra were calibrated in relative fluxes using a standard star. 


\section {Cross-correlation function analysis}
\label{sec:Analysis}

\begin{figure*} 
\begin{center}
\begin{minipage}{.45\linewidth}
\centering
\hspace{-5mm}
\resizebox{\hsize}{!}{\includegraphics[width=8cm, trim= 15mm 0mm 30mm 0mm, clip]{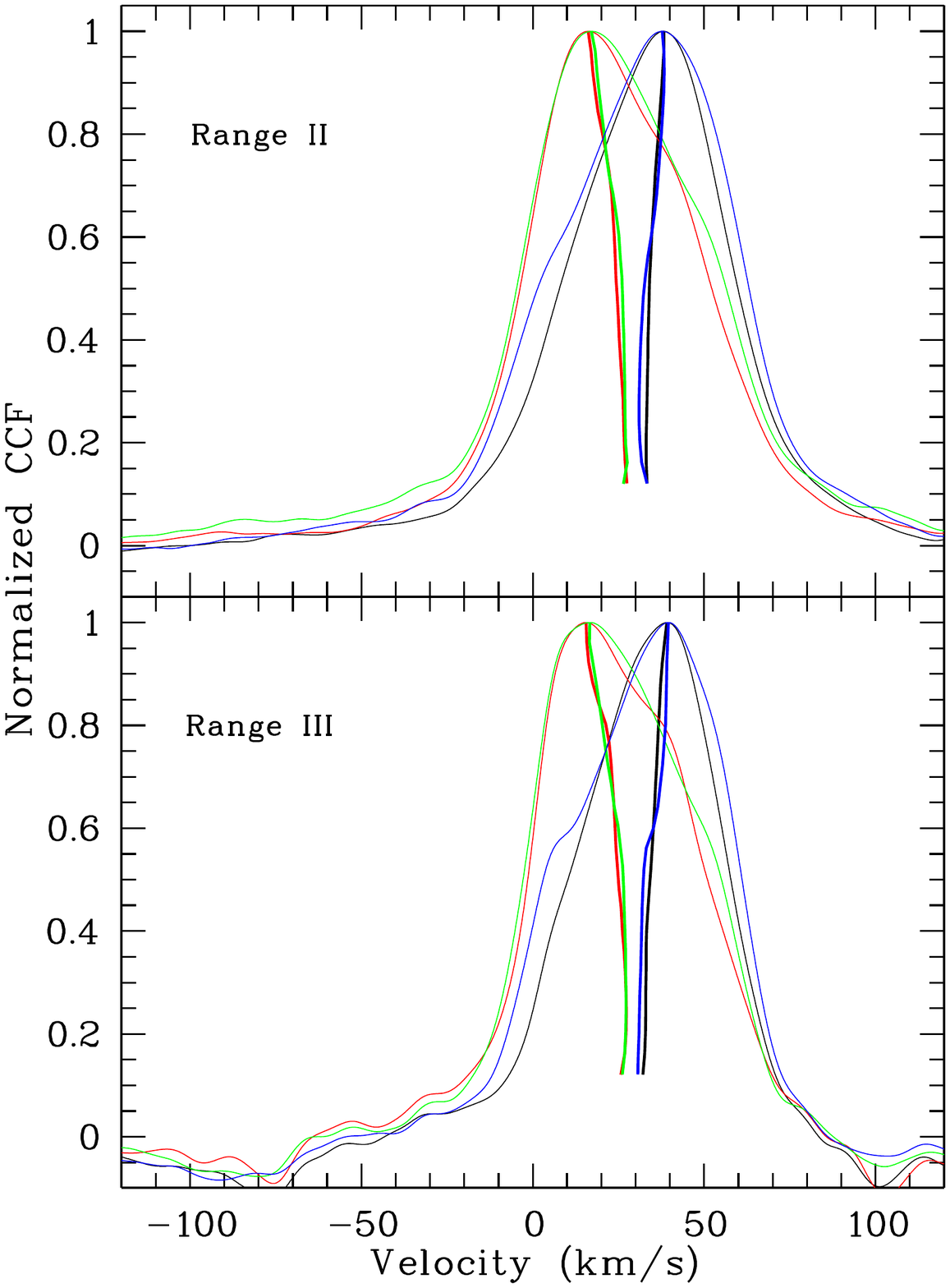}}
\end{minipage}
\begin{minipage}{.45\linewidth}
\centering
\vspace{-1.0cm}
\resizebox{\hsize}{!}{\includegraphics[width=8.3cm, trim= 10mm 0mm 30mm 20mm, clip]{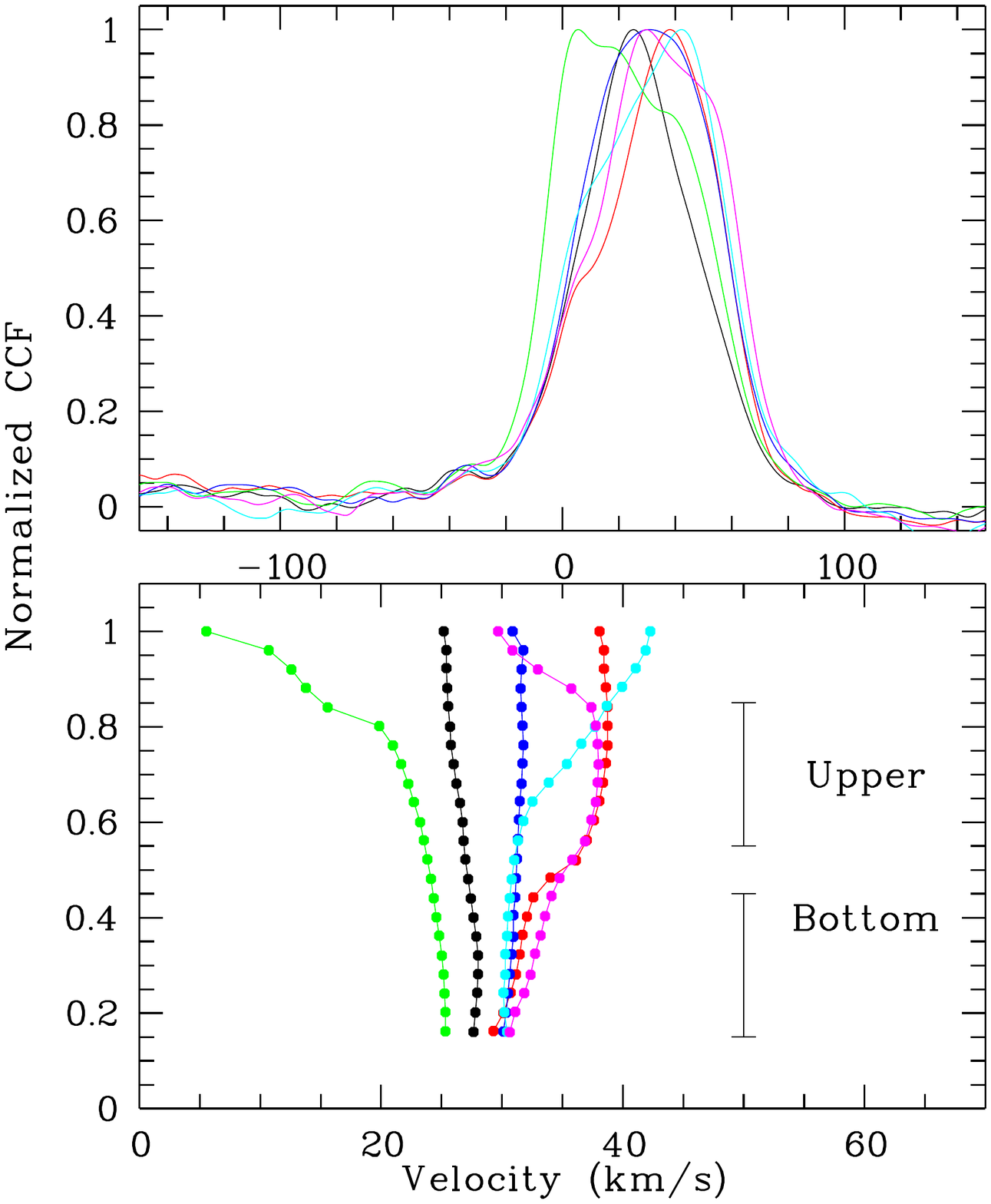}}
\end{minipage}\\
\vspace{-1.5cm}
\begin{minipage}[t]{.47\linewidth}
\caption{Changes in the CCF in the period 19-23 May 2000, in two different 
wavelength ranges. The different colours refer to the following dates: 
19 May ($black$), 21 May ($red$) 22 May, ($green$), 23 May ($blue$). 
The corresponding bisectors are also traced. 
Velocity is relative to the template star HD\,152391. }
\label{Fig1}
\end{minipage}
\hspace{5mm}
\begin{minipage}[t]{.47\linewidth}
\caption{CCF peaks between 16 and 22 May 1999 (upper panel). 
Velocity is relative to the template star. 
The different colours refer to the following dates: 16 May ($black$), 17 May ($red$), 
18 May ($green$), 19 May ($blue$), 21 May ($yellow$), 22 May ($pink$). 
The lower panel shows, on expanded scale, the corresponding CCF bisectors. 
}
\label{Fig2}
\end{minipage}
\end{center}
\end{figure*}

\begin{figure*} 
\centering

  \begin{minipage}[t]{0.5\linewidth}
  \centering
 \resizebox{\hsize}{!}{\rotatebox[]{-90}{\includegraphics[width=8.cm,height=7.5cm,trim = 0mm 0mm 20mm 0mm, clip]{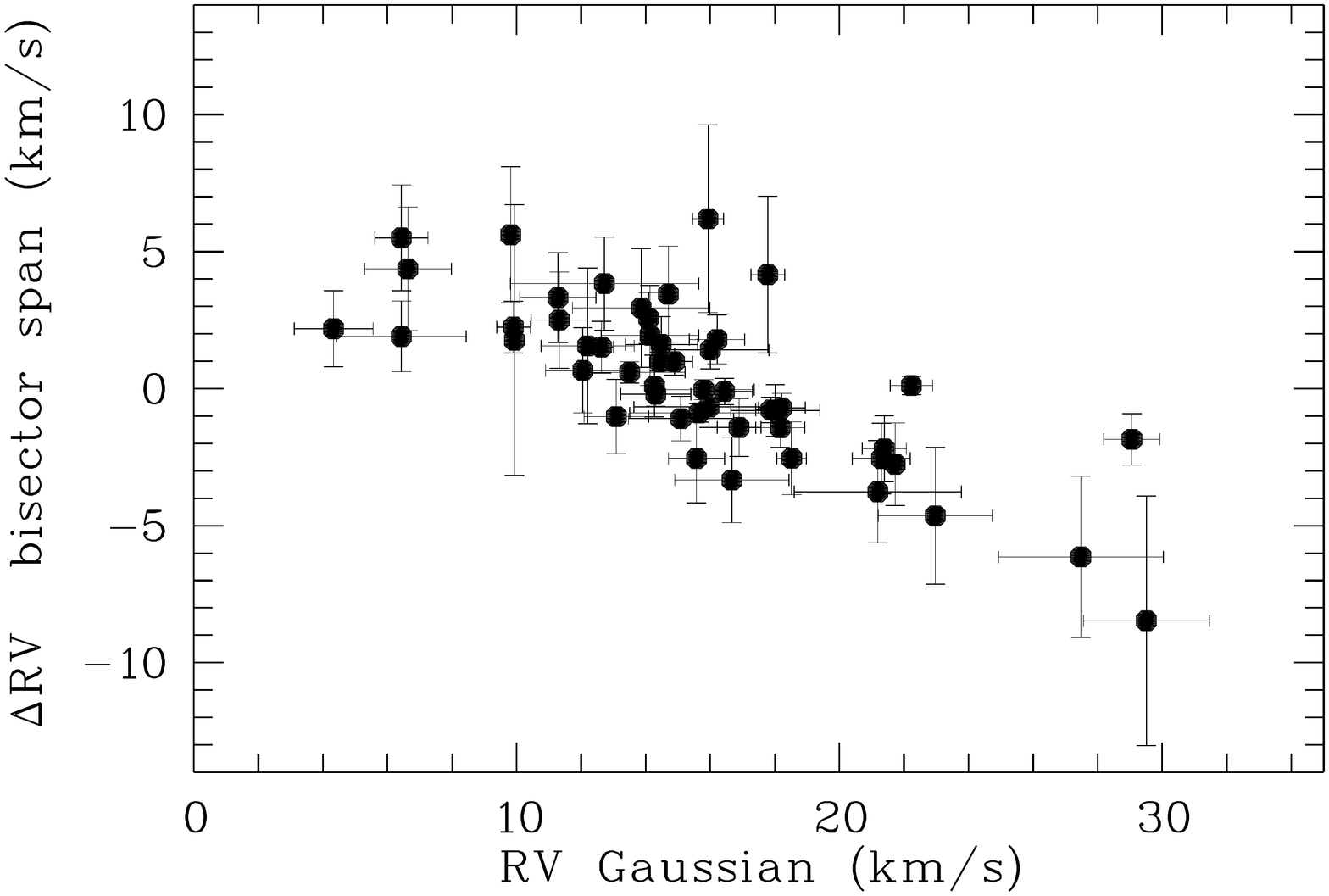}}}
  \label{Fig3a}
  \end{minipage}%
  \begin{minipage}[t]{0.5\linewidth}
  \centering
 
 \resizebox{\hsize}{!}{\rotatebox[]{-90}{\includegraphics[width=8.cm,height=7.5cm,trim = 0mm 0mm 20mm 0mm, clip]{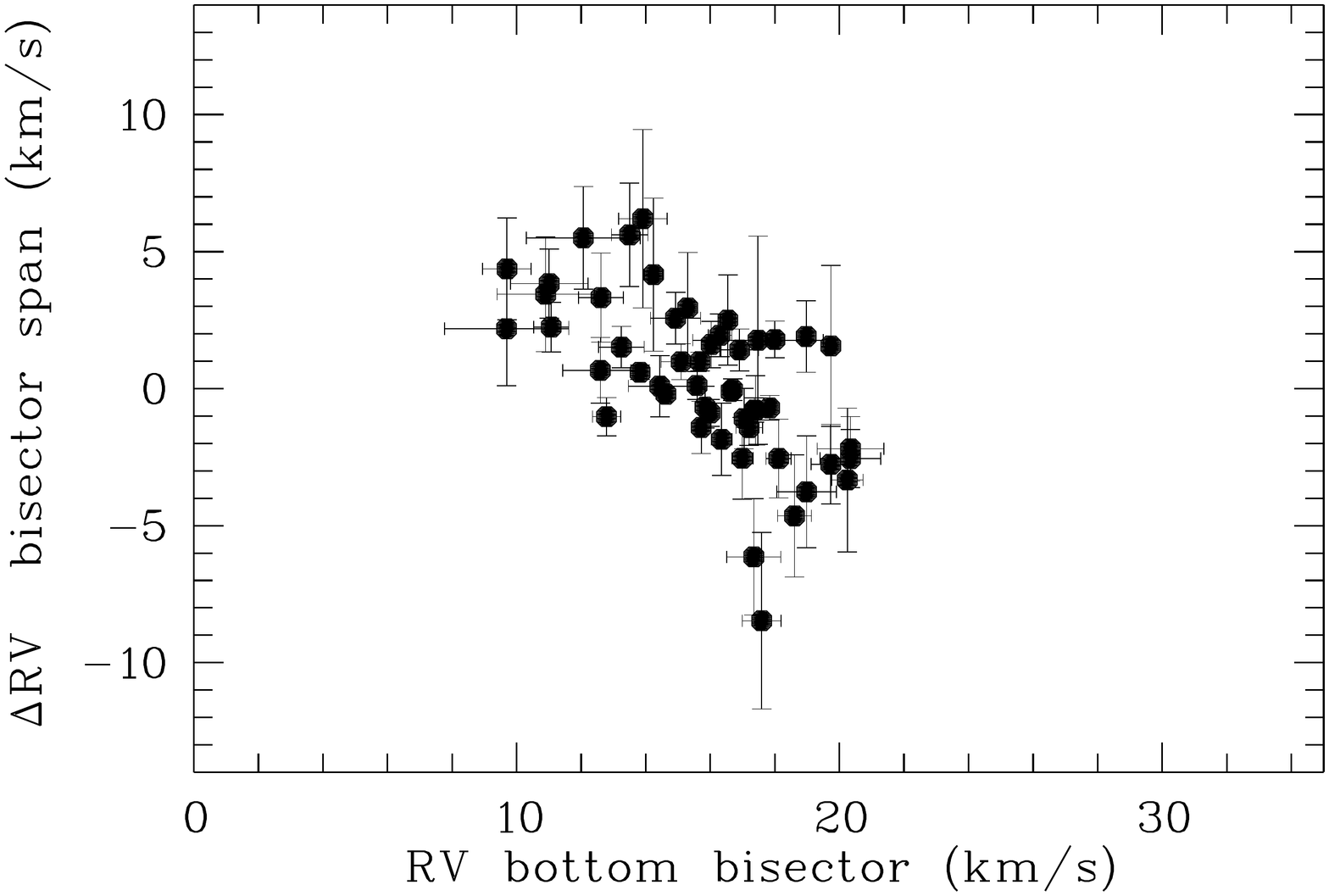}}}
  \label{Fig3b}
  \end{minipage}\\
\vspace{-.5cm}
\caption{The radial velocity measurements from Gaussian-fit method versus the ``bisector velocity span'' 
(i.e. the difference between the upper and bottom bisector mean velocities) are plotted in the $left$ panel. 
The $right$ panel shows, instead, the mean radial velocity of the CCF bisector bottom versus the ``bisector 
velocity span''. 
} 	   
   \label{Fig3}
\end{figure*}

\subsection{Radial velocity determinations}
\label{sec:rv}
We used the cross-correlation technique to determine the radial velocity of T\,Cha from 
the FEROS spectra.
As a template, we chose a spectrum of the G8.5\,V star HD152391 
(RV=44.8\,km\,s$^{-1}$ and $v\sin{i}=3$\,km\,s$^{-1}$) \citep{Nordstrom2004}, 
observed with the same instrumental set-up. 

We followed the prescription by \citet{Esposito2006}, i.e., a Gaussian was fitted to the peak 
of the cross-correlation function (CCF), computed in six distinct spectral ranges. 
In some cases, the Gaussian fit failed, and in such cases we calculated the moments of the CCF 
profile and took the first-order moment as an RV estimator. In the end, we adopted as RV value 
and associated error, the average of the RV determined in the different ranges and 
their r.m.s. dispersion, respectively.
All RV measurements are reported in Table~\ref{tab:RV}. The RV values vary with 
time ranging between 6 and 30\,km\,s$^{-1}$. 
A period search was performed on these data, but no clear indication of periodicity was found.

\subsection{CCF bisector analysis}
\label{sec:bisector}
The Gaussian fit may sometimes appear inadequate due to asymmetries in the CCF peak. 
Asymmetries in stellar absorption lines can arise for several reasons and may be conveniently 
represented by line bisectors (Gray 2005). 
In some cases, variations in the bisector shape 
may arise due to photospheric spots crossing the stellar disc as the star rotates, 
whereas uncorrelated variations in both bisector position and shape may indicate 
an unresolved spectroscopic system \citep{Santos2002,Dall2006}.

From the correlation between RV changes and bisector orientation, \citet{Queloz2001}, 
in the case of HD\,166435, stated that the periodicity in their RV data was not due to 
an unseen secondary object orbiting the star, but to stellar activity. 
We adopted the same tool to verify whether RV shifts measured by a Gaussian 
fit are purely due to changes in the CCF peak shape, or reflect true changes in the 
radial velocity of the star. 
An example of the variability of the CCF peak (and corresponding bisector) of T\,Cha 
on daily timescales is shown in Figs.~\ref{Fig1} and \ref{Fig2}.
For each of the six spectral ranges considered in Sect.~\ref{sec:rv}, 
we divided the corresponding CCF bisector into two intervals as in \citet{Queloz2001}: 
an upper part, where the strongest changes occur, and a lower one, where the bisector 
position appears more stable (see in Fig.~\ref{Fig2}). 
The RV difference between the top and bottom bisector's mean velocities yields 
information about the bisector orientation.
This quantity, called ``bisector velocity span'' \citep{TonerGray1988,Queloz2001,Dall2006}, 
is a good measure of the changes in bisector orientation, and hence of the asymmetry of 
the peak. 
 The variations in the central part of the CCF, which reflect the higher stability 
of the wings compared to the peak can be easily discerned.  \\
The comparison between the RV values obtained from a Gaussian-fit and by a bisector analysis 
is shown in Fig.~\ref{Fig3}.
The left panel shows how the RV values from a Gaussian-fit are well correlated with the 
bisector velocity span\footnote{all RV measurements are reported 
in Table~\ref{tab:high-res-spec}, available electronically}.
The Gaussian function reproduces well only one of the two sides of an asymmetric CCF peak, 
leading to a systematic shift in the measured RV.
On the other hand, the bisector values, at the bottom part of the profile, versus the 
bisector span show a weaker correlation than the Gaussian-fit values 
(Fig.~\ref{Fig3} right panel). 

The RV determinations derived from the bisector method, although of lower amplitude than 
the values derived from the Gaussian-fit method, confirm that RV is indeed variable, considering 
the average uncertainty of 1.7\,km\,s$^{-1}$.
We used Fourier analysis, in the formulation of the periodogram given by \citet{Scargle1982}, 
to search for possible periodicities in the radial velocity data, although no clear values were
identified.  This might in part be due to the extremely uneven temporal sampling. 
We had only one relatively long run of six consecutive nights, but when analysing this run alone, 
no hint of periodicity was found on a timescale of a few days.

Although the presence of blended spectroscopic components cannot be excluded, 
the lack of a clear periodicity prevents us from drawing firm conclusions about the 
possible binarity or multiple-nature of T\,Cha. 
 As in \citet{Queloz2001}, the correlations between the RV values from the Gaussian-fit 
method and the ``velocity span'' indicate as most plausible explanation the presence of 
inhomogeneities 
of variable extent moving across the photospheric disc, either intrinsic to the star 
(e.g., cool spots), or external to it (e.g., clustered orbiting material). 
Hence, in the following, T\,Cha will be treated as a single star.
For the purposes of photospheric spectrum subtraction (see Sect.\,\ref{sec:spec-sub}), 
we used the bottom bisector's  mean RV value.

\begin{table*}
\caption{Radial velocity determinations, both from gaussian fit and bisector method, and equivalent widths of H$\alpha$ 
       and [O{\sc i}]\,6300\AA\ measured on FEROS spectra. {\bf Only in electronic form. }
       }\label{tab:high-res-spec}
\begin{tabular}{ l   r   c   r   c   r   c   r   c   r   c  }
\hline \hline
    HJD     &  RV$_g$ & $\Delta$RV$_g$ & RV$_b$ & $\Delta$RV$_b$ & W$_{\rm H\alpha}$ & $\sigma_{\rm H\alpha}$ & W$_{\rm H\beta}$  & $\sigma_{\rm H\beta}$ & W$_{6300}$ & $\sigma_{6300}$ \\
$-$2450000  & (km/s)  & (km/s) & (km/s) & (km/s) &  (\AA)   &  (\AA) &  (\AA)    &   (\AA) &    (\AA)  &  (\AA)   \\
\hline
 51265.537  &  14.27  &  0.27 & 14.64  & 1.07  &   $-$6.26  &  1.06  &  $-$1.37  &    0.78 &   $-$0.81 &   0.05   \\
 51265.576  &  16.89  &  0.68 & 16.54  & 1.14  &   $-$6.56  &  0.50  &  $-$0.34  &    0.29 &   $-$1.00 &   0.14   \\
 51265.662  &  15.09  &  1.63 & 17.21  & 0.61  &   $-$7.50  &  0.30  &  $-$0.56  &    0.24 &   $-$1.01 &   0.11   \\
 51266.573  &  18.21  &  0.73 & 17.81  & 0.54  &  $-$10.64  &  0.42  &  $-$2.00  &    0.29 &   $-$1.34 &   0.01   \\ 
 51266.636  &  21.39  &  0.68 & 20.57  & 1.81  &  $-$12.46  &  0.37  &  $-$2.02  &    0.38 &   $-$1.48 &   0.15   \\
 51266.672  &  21.30  &  0.90 & 20.70  & 1.98  &  $-$12.36  &  0.80  &  $-$2.89  &    0.32 &   $-$1.56 &   0.11   \\
 51267.692  &  14.09  &  0.25 & 14.35  & 2.16  &   $-$2.89  &  0.39  &  $-$0.01  &    0.21 &   $-$0.58 &   0.12   \\
 51267.801  &  17.88  &  0.47 & 17.58  & 0.48  &   $-$2.72  &  0.26  &   0.03	 &    0.23 &   $-$0.43 &   0.06   \\
 51267.874  &  18.01  &  1.38 & 17.81  & 0.82  &   $-$2.55  &  0.29  &   0.03	 &    0.22 &   $-$0.34 &   0.03   \\
 51268.624  &  11.32  &  0.87 & 15.40  & 1.88  &   $-$2.83  &  0.36  &   0.32	 &    0.23 &   $-$0.40 &   0.03   \\ 
 51268.727  &  16.21  &  0.86 & 17.10  & 1.16  &   $-$5.32  &  0.35  &  $-$0.41  &    0.20 &   $-$0.39 &   0.05   \\ 
 51268.850  &	6.43  &  2.01 & 18.39  & 1.29  &   $-$5.05  &  0.41  &   0.07	 &    0.25 &   $-$0.35 &   0.04   \\ 
 51268.900  &  12.20  &  1.44 & 18.75  & 2.07  &   $-$3.96  &  0.35  &   0.10	 &    0.36 &   $-$0.39 &   0.03   \\  
 51269.657  &  12.05  &  1.15 & 12.76  & 1.46  &   $-$6.18  &  0.47  &  $-$0.31  &    0.28 &   $-$0.78 &   0.05   \\ 
 51269.771  &  13.09  &  1.00 & 13.42  & 0.91  &   $-$5.19  &  0.63  &  $-$0.76  &    0.40 &   $-$0.77 &   0.11   \\ 
 51269.857  &  15.80  &  1.56 & 16.56  & 0.29  &   $-$4.23  &  0.55  &  $-$0.95  &    0.44 &   $-$0.80 &   0.17   \\ 
 51315.622  &  18.17  &  0.76 & 17.65  & 0.83  &  $-$22.68  &  0.39  &  $-$2.88  &    0.51 &   $-$1.88 &   0.25   \\
 51316.612  &	6.43  &  0.82 & 10.54  & 3.80  &  $-$23.93  &  0.95  &  $-$5.89  &    0.70 &   $-$2.06 &   0.22   \\ 
 51317.627  &  16.67  &  1.77 & 20.90  & 1.78  &  $-$20.20  &  1.03  &  $-$4.30  &    0.39 &   $-$0.88 &   0.06   \\
51318.591$^\dag$ & 13.50 & 1.72 & 13.42 & 0.49 &  $-$20.65  &  0.28  &  $-$3.03  &    0.32 &   $-$1.22 &   0.07   \\
 51320.591  &  17.78  &  0.52 & 12.61  & 2.59  &   $-$1.42  &  0.30  &   1.08	 &    0.17 &   $-$0.28 &   0.03   \\
 51321.580  &  12.72  &  2.92 &  9.64  & 2.72  &   $-$7.56  &  0.55  &  $-$0.07  &    0.29 &   $-$0.64 &   0.04   \\ 
 51674.670  &  18.52  &  0.46 & 17.39  & 1.38  &   $-$0.75  &  0.38  &   0.87	 &    0.67 &   $-$0.35 &   0.03   \\
 51684.617  &	9.90  &  0.52 & 10.45  & 1.69  &   $-$2.39  &  0.34  &   0.43	 &    0.25 &   $-$0.69 &   0.03   \\
 51686.632  &  21.19  &  2.59 & 19.51  & 2.50  &   $-$1.13  &  0.22  &   0.59	 &    0.39 &   $-$0.52 &   0.04   \\
 51687.648  &  22.97  &  1.77 & 19.71  & 2.94  &   $-$0.88  &  0.21  &   0.86	 &    0.29 &   $-$0.39 &   0.03   \\
 51688.640  &	9.82  &  0.31 & 11.37  & 3.17  &   $-$2.34  &  0.35  &  $-$0.06  &    0.43 &   $-$0.48 &   0.07   \\
 51913.860  &  13.86  &  2.13 & 13.69  & 1.88  &   $-$1.46  &  0.41  &   0.55	 &    0.19 &   $-$0.23 &   0.04   \\
 51918.775  &  15.65  &  2.15 & 15.90  & 0.43  &   $-$1.86  &  0.29  &   0.15	 &    0.23 &   $-$0.22 &   0.04   \\
 51923.824  &  14.89  &  0.56 & 14.91  & 0.54  &   $-$1.12  &  0.40  &  $-$0.20  &    0.17 &   $-$0.14 &   0.01   \\
 52019.578  &  16.00  &  1.81 & 16.07  & 0.97  &  $-$23.82  &  0.45  &  $-$4.06  &    1.09 &   $-$1.83 &   0.02   \\ 
 52026.586  &  14.48  &  1.12 & 15.03  & 1.05  &   $-$3.80  &  0.40  &  $-$0.26  &    0.26 &   $-$0.55 &   0.05   \\
 52031.573  &  14.43  &  0.28 & 14.64  & 0.95  &   $-$9.18  &  0.68  &  $-$1.45  &    0.30 &   $-$0.64 &   0.04   \\
 52319.745  &  12.63  &  0.36 & 12.66  & 1.40  &    0.26    &  0.31  &   1.22	 &    0.12 &   $-$0.26 &   0.04   \\
52322.742$^\dag$ & 27.48 & 2.56 & 19.18 & 3.41 &   $-$8.56  &  0.40  &  $-$1.23  &    0.75 &   $-$1.02 &   0.06   \\
 52372.606  &  14.70  &     - &  9.92  & 2.40  &  $-$29.09  &  0.95  &  $-$5.60  &    0.55 &   $-$2.33 &   0.18   \\
 52385.529  &  16.44  &  0.87 & 16.54  & 0.62  &   $-$3.82  &  0.32  &  $-$0.60  &    0.31 &   $-$0.49 &   0.02   \\
 52395.617  &	4.33  &  1.22 &  9.46  & 2.23  &  $-$20.92  &  1.11  &  $-$5.22  &    0.71 &   $-$1.97 &   0.11   \\
 52396.614  &	9.93  &     - & 16.20  & 3.49  &  $-$11.51  &  1.29  &  $-$0.50  &    0.50 &   $-$1.78 &   0.23   \\
 52710.601  &  21.73  &  0.30 & 20.00  & 1.93  &   $-$0.93  &  0.22  &   0.57	 &    0.16 &   $-$0.31 &   0.03   \\
 52717.620  &  14.13  &  1.51 & 15.79  & 1.78  &  $-$13.51  &  0.45  &  $-$2.06  &    0.50 &   $-$1.16 &   0.07   \\
 52723.588  &  15.96  &  2.33 & 15.80  & 0.51  &   $-$2.09  &  0.31  &   0.62	 &    0.47 &   $-$0.35 &   0.01   \\
 53097.541  &  15.57  &  0.87 & 19.17  & 1.55  &   $-$6.14  &  0.34  &  $-$0.18  &    0.25 &   $-$0.72 &   0.02   \\
 53104.652  &  11.28  &  1.18 & 11.38  & 2.22  &   $-$0.93  &  0.21  &   0.77	 &    0.15 &   $-$0.20 &   0.10   \\
 53122.685  &  15.93  &  0.48 & 11.70  & 3.40  &  $-$22.39  &  0.50  &  $-$3.59  &    1.23 &   $-$2.00 &   0.27   \\
 53137.672  &  22.23  &  0.66 & 15.42  & 0.86  &   $-$4.21  &  0.21  &   0.40	 &    0.29 &   $-$0.23 &   0.11   \\
 53501.652  &  29.06  &  0.87 & 16.43  & 1.11  &   $-$8.22  &  0.34  &  $-$1.23  &    0.67 &   $-$0.59 &   0.06   \\
 53524.597  &	6.63  &  1.35 &  7.95  & 2.57  &   $-$5.55  &  0.27  &  $-$0.18  &    0.41 &   $-$0.51 &   0.02   \\
 53552.546  &  14.31  &  1.09 & 14.36  & 0.37  &   $-$1.06  &  0.25  &  $-$0.04  &    0.23 &   $-$0.22 &   0.04   \\
 53602.488  &  29.51  &  1.95 & 20.32  & 4.58  &   $-$8.66  &  0.34  &  $-$0.77  &    0.60 &   $-$0.68 &   0.05   \\
\hline 
\end{tabular}
\\
\noindent $\dag$ He\,I 5876 \AA\, detected
\end{table*}

\subsection{Projected rotational velocity } 
\label{sec:vseni}
We evaluated the projected rotational velocity, $v\sin{i}$, of T\,Cha  by measuring 
the full width at half maximum (FWHM) of the CCF peak for the spectra with higher signal-to-noise ratio. 
We first established a relation between the FWHM and $v\sin{i}$, as described in \citet{Covino1997}, 
using the template spectrum of HD\,152391 broadened artificially by various $v\sin{i}$ values.
The $v\sin{i}$ value determined in this way was 37$\pm$2\,km\,s$^{-1}$.            
The same measurement was repeated on an average spectrum obtained by combining a total of 
34 spectra with S/N$\gtrsim 50$ around 6000\,\AA. 
The $v\sin{i}$ determined in the average spectrum is consistent with the previous value.
However, when extending the measurements to all spectra, we identified a few with 
narrower spectral lines, yielding a $v\sin{i}$ of nearly 30\,km\,s$^{-1}$. 
Hence, we checked the $v\sin{i}$ determinations by directly fitting each spectrum with 
an artificially broadened template, which confirmed that some spectra had 
smaller line widths. 

\begin{figure} 
\centering
\resizebox{\hsize}{!}{\rotatebox[]{-90}{\includegraphics[width=7cm, height=8cm,trim = 10mm 20mm 0mm 0mm, clip]{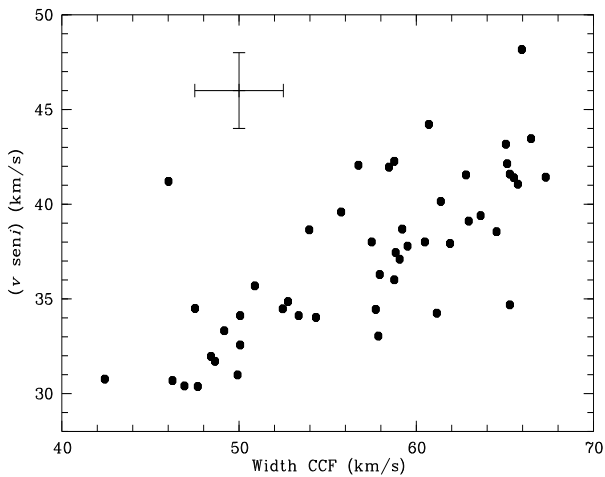}}}
\vspace{-2cm}
\caption{Projected rotational velocity from CCF Gaussian-fit calibration versus 
FWHM of the CCF peak. Only the data of the highest S/N are considered here.
} 
\label{Fig4}
\end{figure}  

In Fig.~\ref{Fig4}, we report the $v\sin{i}$ values obtained from CCF Gaussian-fit FWHM calibration versus 
the width at half maximum of the CCF peak measured directly.
In this case, it is obvious that the width of the lines does not provide a reliable measure 
of the rotational velocity. 
A variable line-width is indicative of either blended spectroscopic components
or the presence of inhomogeneities moving across the stellar disc that alter the shape of 
photospheric lines. 
However, no correlation is found between the RV (either from a Gaussian-fit or bisector method) 
and the $v\sin{i}$ measurements. 

From a periodogram analysis based on data points from the spectra of higher signal-to-noise 
ratio, 
we found a peak at about 0.25\,cycles\,day$^{-1}$, or a period close to 4\,days. 
This period does not differ dramatically from that found by \citet{MauderSosna1975} and might 
reflect a rotational modulation induced by photospheric spots or other types of inhomogeneities 
transiting over the stellar disc.

\section{The spectrum of T\,Cha}
\label{sec:Spectrum}

\begin{figure*} 
\centering
\includegraphics[width=14.5cm,angle=-90,trim = 20mm 20mm 0mm 0mm]{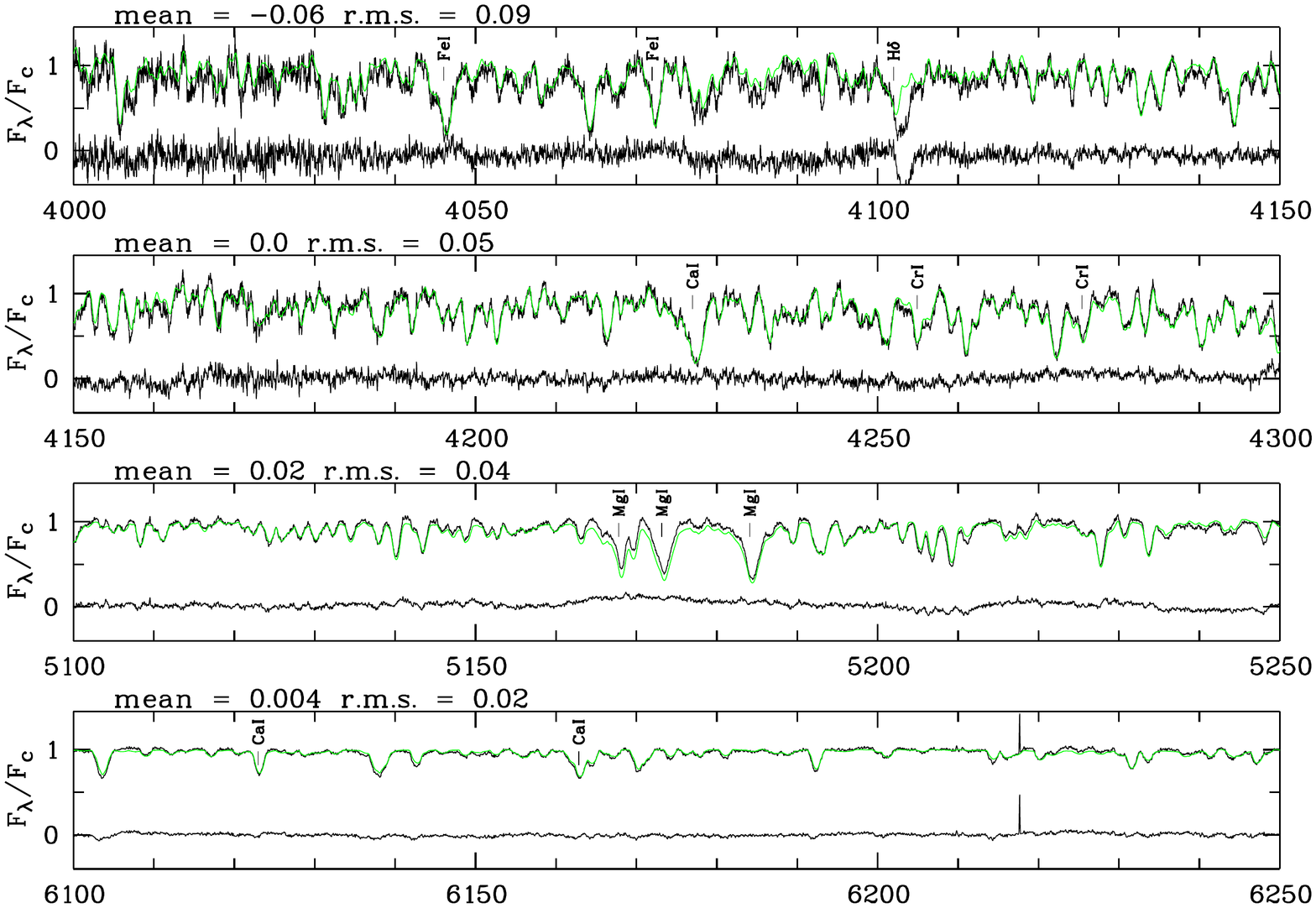}
\caption{ Spectrum of T\,Cha obtained on 14 Feb 2002. 
The green line represents the spectrum of the template (HD152391 G8.5V)
artificially broadened for rotation. 
In each panel is also shown the residual spectrum after subtraction.
Slight residuals in strong spectral features are mainly due to uncertainties 
in the normalization to the continuum. The spectrum, with a good S/N, 
shows a redshifted absorption component at H$\delta$.} 
\label{Fig5}
\end{figure*}

\subsection{Photospheric spectrum and lithium abundance}
The spectrum of T\,Cha is that of a G8\,V star with strong absorption in the 
Li{\sc i} resonance line at $\lambda$ 6708\,\AA\ and a few emission lines
typical of TTS. The spectral type does not change from one spectrum to the other,
as also demonstrated in Sect.\,\ref{sec:low-res-spec-var}. 
Figure~\ref{Fig5} shows the good match between the photospheric spectrum of T\,Cha 
and that of the template. The equivalent width (EW) of the lithium line was measured 
in each spectrum of our FEROS data-set.  The measurements were found to be internally 
consistent, within the errors, yielding a mean value of EW$_{Li}=360\pm30$\,m\AA. 
Using the calibration of \cite{Pavlenko1996} and adopting the temperature of 
5000\,K, we derived a lithium abundance compatible with the cosmic value 
($\log{N_{Li}}$ = 3.2), with an uncertainty of 0.2\,dex. 
It is important to recall that this determination is strictly dependent on 
the effective temperature. At a temperature of 5500\,K, as adopted by 
\cite{Kenyon1995} for a G8\,V star, the abundance would increase to 3.6\,dex, 
implying some lithium enrichment with respect to the interstellar value.

\subsection{Photospheric spectrum subtraction}
\label{sec:spec-sub}
One of the main goals of this work was to analyse the variability in the emission
lines and hence probe the circumstellar environment of T\,Cha.   For this
purpose, it is useful to remove the photospheric contribution. We subtracted the
the photospheric-line spectrum from the FEROS spectra of T\,Cha using 
the same template spectrum of HD152391 used in the cross-correlation analysis
presented in Sect.~\ref{sec:Analysis}.

All spectra were rebinned to the rest-wavelength frame in order to account for 
their different RV shifts, derived from the CCF analysis described in Sect.\,\ref{sec:rv}. 
The template spectrum was broadened artificially by the $v\sin{i}$ value of 
37\,km\,s$^{-1}$ and then subtracted from each of the T\,Cha spectra. 
We assumed that the photospheric spectrum of a TTS resembles that 
of a main-sequence star of the same spectral type. The photospheric subtraction
method also allows us to verify the possible presence of ``veiling''. No sign
of continuum excess emission was found in T\,Cha (see Fig.~\ref{Fig:Noveiling}), 
confirming the absence of UV-excess in the broad-band photometry  
\citep[see Sect.~\ref{sec:SED} and][]{Covino1996}.

\begin{figure*} 
\centering
\includegraphics[width=19cm, trim= 20mm 10mm 0mm 0mm, clip]{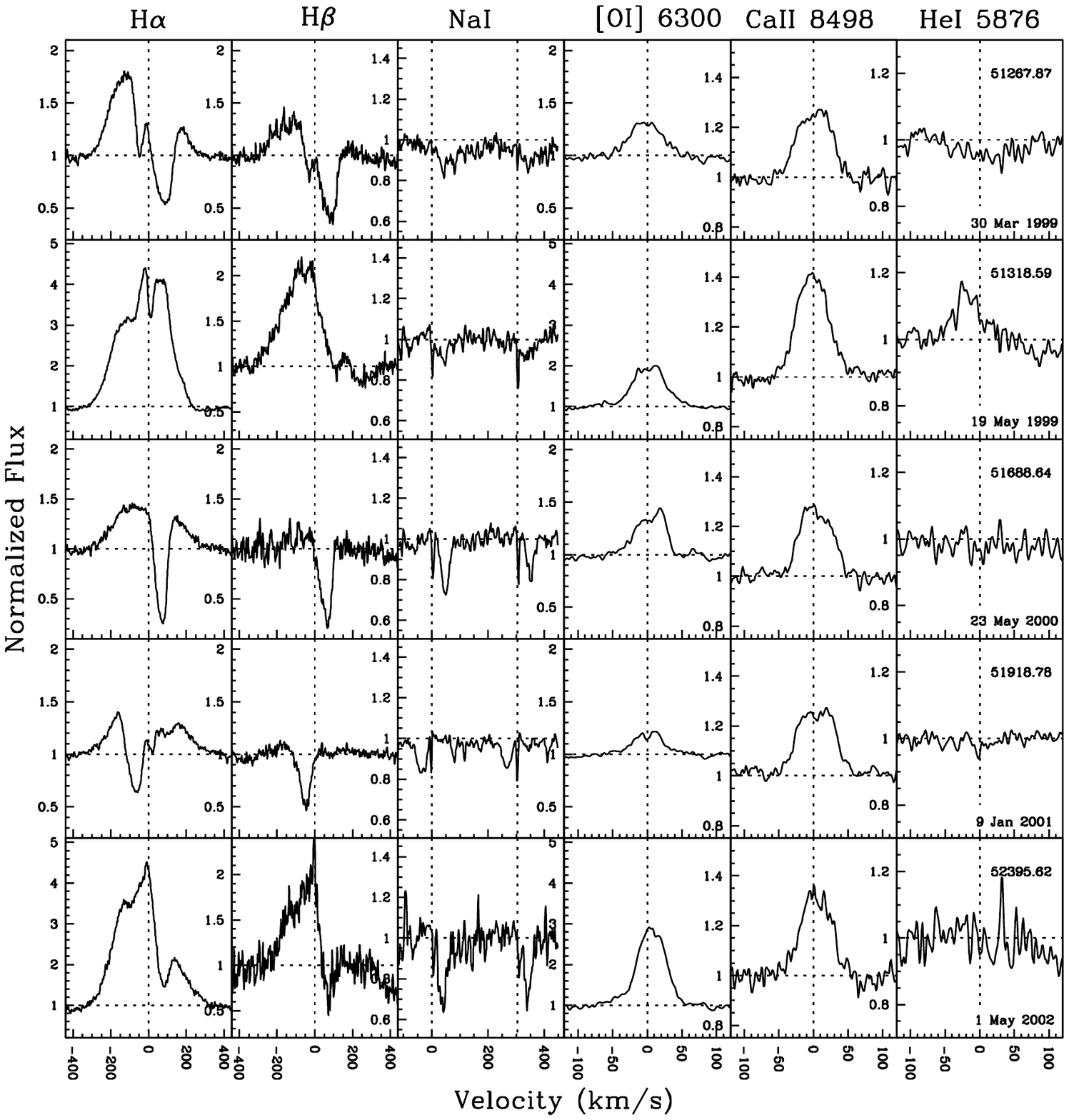}
\vspace{-5cm}
\caption{Examples of residual line profiles of T\,Cha in velocity scale. 
The vertical dotted line marks the rest position in the star frame. 
} 
\label{Fig6}
\end{figure*}

Examples of the residual profiles of the main emission lines at different dates are shown in Fig.~\ref{Fig6}. 
 Some lines show absorption components shifted toward shorter and/or longer wavelengths.
The H$\alpha$ line is the most variable in both intensity and shape. 
The residuals of the Na{\sc i}\,D lines show either redshifted or blue-shifted absorption 
components (that are easily distinguishable from the sharp, undisplaced, interstellar component), 
apparently linked to the presence of a correspondent feature in H$\alpha$ and H$\beta$. 
The sharp interstellar absorption components of the D lines from the associated cloud 
are also distinguishable, when not suppressed by the sky emission. 

\subsection{Emission line spectrum}
\label{sec:emission}
\citet{Gregorio1992}, \citet{Alcala1993} and \citet{Covino1996} already pointed out some 
outstanding properties of the variability in the emission lines, but here we exploit the 
wide spectral coverage of FEROS to analyse more lines simultaneously and identify 
the line profile changes at higher resolution than in previous works.\\

\subsubsection{The H$\alpha$ line}
\label{sec:Ha}
The most prominent emission line in the spectrum of T\,Cha is H$\alpha$, which also exhibits 
impressive variability.  
The observed equivalent width of the line measured in the FEROS spectra\footnote{reported 
in Table\,\ref{tab:high-res-spec}, only available in electronic form}, ranges from about 
0.3\,\AA\ (on 14 Feb 2002) to about $-30$\,\AA\ 
(on 8 Apr 2002)\footnote{conventionally, a negative EW indicates line in emission.}.
Besides the variations in strength, from pure emission to absorption, the line profile is also 
highly variable in its structure. 
 However, we do not observe any correlation between the profile of the line and its EW. 
The line extends approximately from $-300$ to $+300$\,km\,s$^{-1}$. However, much of the emission 
is concentrated in an interval of almost 400\,km\,s$^{-1}$ in width centred on zero velocity. 
In most of the spectra, the H$\alpha$ line exhibits both blue-shifted and redshifted absorption 
components superimposed on a nearly centred symmetric broad emission. 
The blue absorption is usually weaker than the red one, although in three spectra 
of January 2001 it appears stronger (e.g., date \,51918 in Fig.~\ref {Fig6}). 
During five different epochs, the blue absorption reached below the continuum producing a P\,Cygni profile. 
In most cases, when the emission component was not so strong, the red absorption was below the continuum, 
producing a typical inverse-P\,Cygni profile.

The red absorption is almost always present but with different velocity displacements relative to the 
rest wavelength, whereas the blue absorption sometimes becomes barely visible or disappears completely. 
In contrast, the emission on the blue side is generally more intense than the red one.
This suggests that the red wing of the line is affected by significant absorption that causes 
the emission to appear weaker than in the blue wing. 
No clear regularity appears to exist in the line changes, 
but some profiles reappear after some time.\\

The lack of both detectable veiling and UV excess indicates that the accretion rate is rather low. 
Assuming that the H$\alpha$ emission is entirely due to accretion, we estimated the mass 
accretion rate, $\dot{\rm{M}}_{accr}$, by adopting the relationship found by \citet{Natta2004} 
between $\dot{\rm{M}}_{accr}$ and the width of the H$\alpha$ line at 10\% intensity.
We emphasise that this quantity does not represent an accurate mass-accretion estimator 
for two main reasons: {\em i)} it is not easy to measure, mainly due to the difficulty 
in defining the continuum level, and {\em ii)} because the relationship, derived for 
the substellar mass regime, shows a larger dispersion at higher masses.
We determined the width at 10\% intensity of the H$\alpha$ emission only in spectra 
for which we could perform a Gaussian decomposition of the line profile into an emission component 
with overlapping absorptions. From those, we derived a mean value of 
$\dot{\rm{M}}_{accr} = 4\times 10^{-9}$\,M$_{\odot}$\,yr$^{-1}$.

\subsubsection{The H$\beta$ line and higher Balmer lines}
\label{sec:Hb}
The strength of H$\beta$ is also variable and its EW correlates well with that of H$\alpha$, 
as shown in Fig.~\ref{EqWid}.
The range of variation is between about 1.2\,\AA\ (on 14 Feb 2002) and about $-6$\,\AA\ (on 8 Apr 2002).
In most cases, the H$\beta$ line presents typical double-peak emission with higher emission 
in the blue peak than in the red one. Apart from when emission is stronger, the H$\beta$ profile 
resembles the H$\alpha$ one.
However, the inverse P~Cygni profile appears more frequently than in H$\alpha$.

Even if H$\alpha$ and H$\beta$ have different optical thicknesses and formation regions, their EWs 
are strongly correlated with each other (Fig.~\ref{EqWid} lower panel).

Because of the lower efficiency of FEROS at shorter wavelengths and the high variability of T\,Cha 
in the $U$ and $B$ bands, little can be said about the behaviour of the higher members of the Balmer 
series, since H$\gamma$ and H$\delta$ are discernible only in a few spectra.
Almost no emission is apparent in the residual spectrum, except in one or two cases, but the still 
low S/N prevents us from drawing any robust conclusion.
The most intense absorptions, either blue- or redshifted, depending on epoch, can also be identified
in the higher Balmer lines (cf. H$\delta$ in Fig.~\ref{Fig5}).

\subsubsection{Forbidden Oxygen lines}
\label{sec:OI}
 
The forbidden [O{\sc i}] 6300 and 6363\,\AA\ emission lines are observed 
in the spectrum of T\,Cha. 
The contribution of the sky emission lines to the FEROS spectra was subtracted 
after scaling the sky spectrum by a factor accounting for the different efficiencies 
of the object and sky fibers, determined from the ratio of the flat-field intensities 
in the corresponding fibers.  Figure~\ref{Fig6} shows the sky-subtracted line at 6300~\AA.
The width of $\sim$100\,km\,s$^{-1}$ places the line just on the edge 
between low- and high-velocity components for T Tauri stars \citep{Hartigan1995}. 

As in the case of the Balmer lines, the forbidden oxygen lines are of variable intensity, 
but their profiles remain about the same. The most frequently observed 
profile is a symmetrical and well-centred emission, but sometimes the line 
is slightly redshifted by about 10-20\,km\,s$^{-1}$ (see Fig.~\ref{Fig6}). 
The symmetrical profile suggests that both the approaching and receding parts 
of the flow contribute to the observed line, i.e., the star-disc system is 
viewed at relatively high inclination.  
Redshifted [O{\sc i}] 6300\,\AA\ profiles have also been observed in BP\,Tau 
and RW\,Aur \citep{Hartigan1995}, but are rather unusual for T Tauri stars. 
Some authors explain the redshifted component in terms of asymmetric 
outflows \citep[e.g.,][]{Hirth94, Hartigan1995}. 

\begin{figure} 
\centering
\includegraphics[width=10cm]{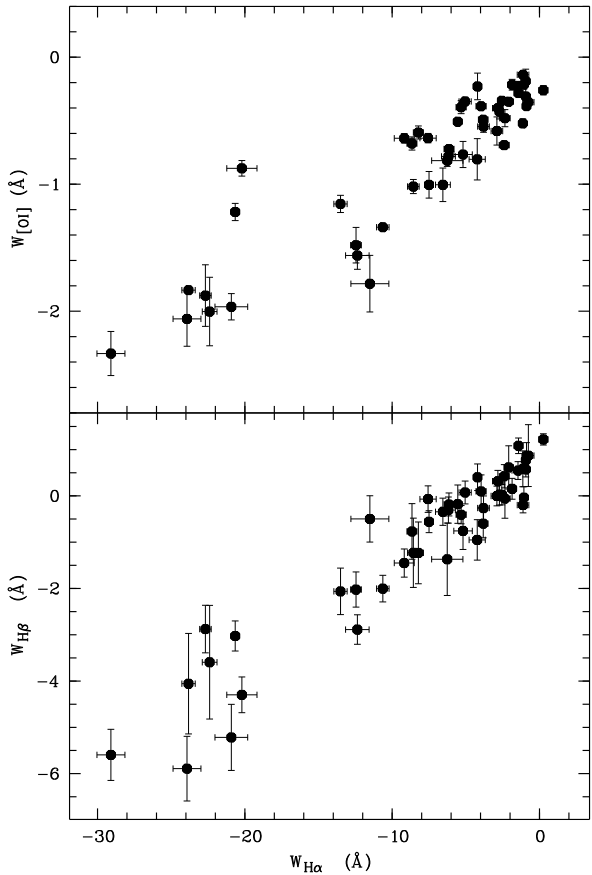}
\caption{ 
Correlations between intensities of the most prominent emission lines 
measured in the FEROS spectra.
} 
\label{EqWid}
\end{figure}

Remarkably, the equivalent width of the [O{\sc i}] 6300\,\AA\ is correlated 
with that of the H$\alpha$ emission (see Fig.~\ref{EqWid} upper panel). 
As discussed in the following sections, this correlation indicates 
that the variations must be related to changes in the intensity of the 
underlying stellar continuum due to variable circumstellar extinction.

To verify the hypothesis of variable obscuration, 
we used the only simultaneous spectroscopy and photometry data available 
to us \citep{Covino1996} to evaluate the luminosity in the [O{\sc i}] line, 

 \[ L_{\rm [O\,I]} = W_{\rm [O\,I]} \cdot f_{R} \cdot 4 \pi d^2\] 
 \[~~~~~ = W_{\rm [O\,I]} \cdot 10^{-0.4 \cdot R} \cdot 2.69 \times 10^{33} 
 {\rm erg~s^{-1}}, \]

\noindent
 where $W_{\rm [O\,I]}$ is the measured equivalent width of [O{\sc i}], 
 $d$ is the distance of 100\,pc, and $f_{R}$ is the observed flux of the star 
 in the R-band.  
 The latter is given by  $f_{R} = F^{R}_{0} \cdot 10^{-0.4 \cdot R}$, where 
 $F^{R}_{0}=2.26 \times 10^{-9} {\rm erg~s^{-1} cm^{-2} \AA^{-1}}$ is 
 the R-band flux of a zero-magnitude star \citep{Bessel1979}.
 The measured EWs and relative luminosities are reported in 
 Table\,\ref{tab:lum-OI}.
 
 \begin{table}
\caption{Equivalent widths of [O{\sc i}]\,6300\AA\ measured 
in CASPEC spectra \citep{Covino1996} and line luminosity 
derived using simultaneous R-band photometry.
}  \centering 
  \begin{tabular}{ c c c c }
  \hline \hline
     JD     &   R   & W$_[O{\sc I}]$  & L$_{[O{\sc I}]}$ \\
 $-$2440000 &  mag  &	  (\AA)~~     & (10$^{28}$erg\,s$^{-1}$)\\
  \hline
 9384.62   &  11.47 &  -0.78$\pm$0.06 & 5.4$\pm$0.4 \\        
 9384.70   &  11.70 &  -0.91$\pm$0.08 & 5.1$\pm$0.4 \\        
 9384.87   &  11.88 &  -1.24$\pm$0.16 & 5.9$\pm$0.8 \\        
 9385.62   &  10.56 &  -0.32$\pm$0.03 & 5.1$\pm$0.5 \\        
 9385.76   &  10.41 &  -0.25$\pm$0.03 & 4.6$\pm$0.5 \\        
 9385.87   &  10.39 &  -0.24$\pm$0.03 & 4.5$\pm$0.6 \\        
\hline 
\end{tabular}
\label{tab:lum-OI}
\end{table}

 The luminosity (or the flux) in the [O{\sc i}] line remains relatively constant 
 with time, as in the case of the UXOR star RR\,Tauri \citep{Rodgers2002}.
 Therefore, the strong variability in the emission-line intensity is caused by
 changes in the stellar continuum flux level rather than the intrinsic 
 variability of the line-emitting region. 
 This explains the correlation between H$\alpha$ and [O{\sc i}] discussed 
 in Sect.~\ref{sec:OI}, justifying the fact that the intensities of 
 the two lines, originating in distinct regions, respond simultaneously to 
 the brightness variations of the star.
 As the analysis of flux-calibrated low-resolution spectra presented 
in Sect.~\ref{sec:low-res-spec-var} shows, the intensities of the 
H$\alpha$ and [O{\sc i}] lines are also found to be related directly to the 
amount of circumstellar extinction affecting the photospheric continuum.

\subsubsection{Other emission lines}
\label{sec:Other}

Franchini et al. (1992) detected a weak variable emission component 
in the core of the Na{\sc i}\,D lines.
In our spectral subtraction analysis, we found no sign of this emission, but instead, 
were able to identify strong redshifted absorption in many residual spectra. 
that is visible only when the redshifted 
absorption in H$\alpha$ is also present (although the reverse is untrue), independently 
of the equivalent width of the H$\alpha$ line. 
On three dates, namely 9 Jan 2001, 21 Apr 2002, and 9 Apr 2005, when the blue-shifted 
absorption component in H$\alpha$ was at least as strong as the red one, the feature 
appeared blue-shifted, but there was no sign of absorption on 14 Jan 2001, when 
H$\alpha$ showed a clear P\,Cygni profile. 

The [N{\sc ii}] 6583\,\AA\ forbidden emission line is occasionally observed in the 
low-resolution spectra when the star reaches its faintest state (see 
Sect.~\ref{sec:low-res-spec-var}).  

The He\,I emission line at 5876\,\AA\ was seen on only two dates (19 May 1999 and 17 Feb 2002), 
while missing in other spectra close to those dates. 
This line is probably caused by episodic, short-lived flare activity, common 
in young solar-type stars. As shown in Fig.~\ref{Fig6}, the line appears blue-shifted 
by about $-30$\,km\,s$^{-1}$.

Finally, the emission cores of the two lines 8498 and 8662\,\AA\ of the Ca{\sc ii} IR triplet 
are clearly visible. 
The other line, at 8541\,\AA, falls in a wavelength gap in-between the echelle orders of FEROS.
The equivalent widths of the two measurable lines remain almost constant, with small variations 
presumably related to chromospheric activity. No correlation is found with other 
emission lines. The profiles appear always symmetrical and well centred on the star velocity, 
indicating  presumably a chromospheric origin, typical of young stars. 

As for higher Balmer lines, the low S/N prevented us from analysing 
the Ca{\sc ii} H and K lines, but in some spectra of sufficiently high S/N, 
a strong emission core, indicative of intense chromospheric activity, is observed.

\begin{figure*} 
\begin{center}
\begin{minipage}{.50\linewidth}
\centering
\resizebox{\hsize}{!}{\includegraphics[trim = 10mm 20mm 0mm 0mm, clip,angle=-90,origin=c]{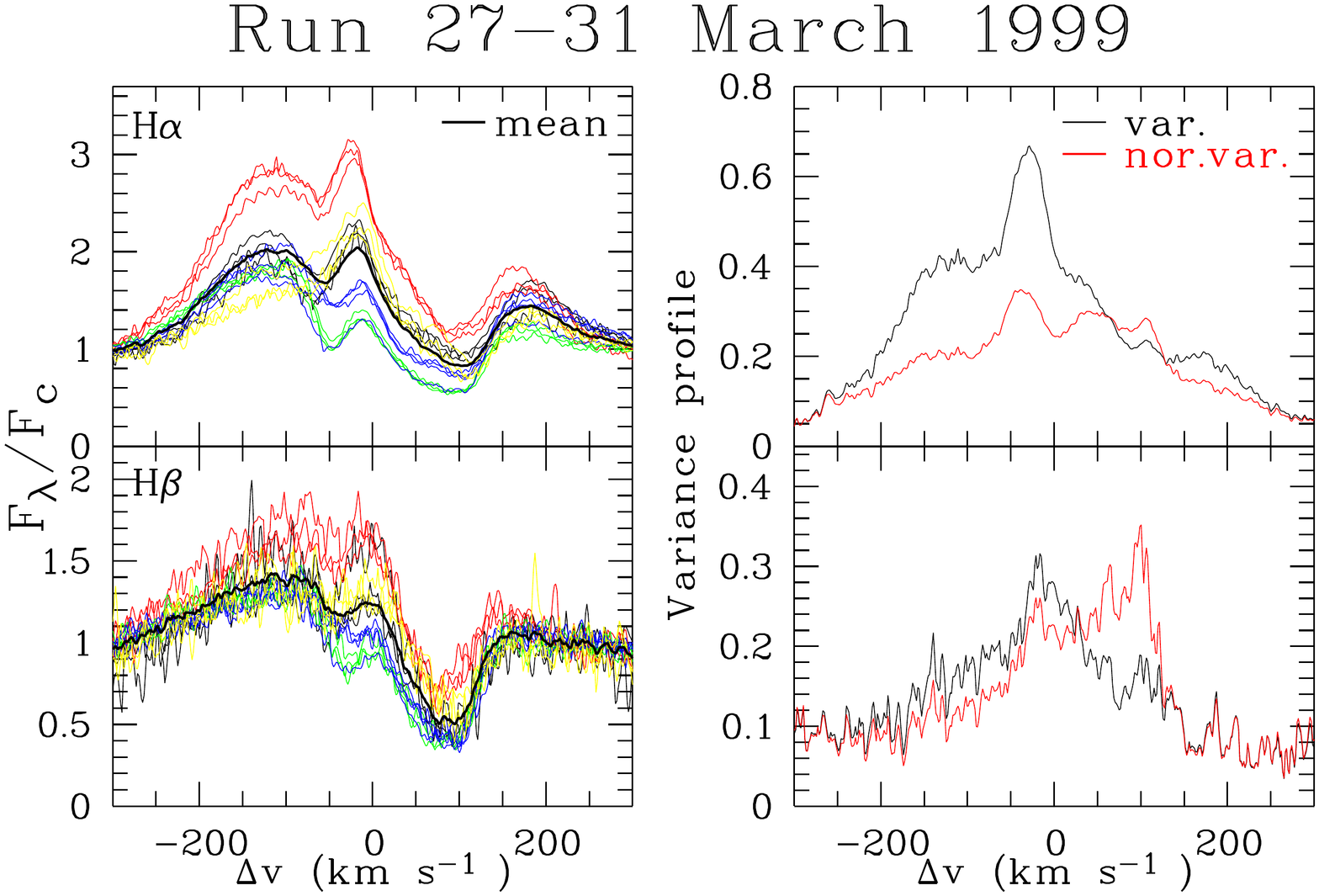}}
\end{minipage}
\hspace{5mm}
\begin{minipage}{.45\linewidth}
\centering
\resizebox{\hsize}{!}{\includegraphics[]{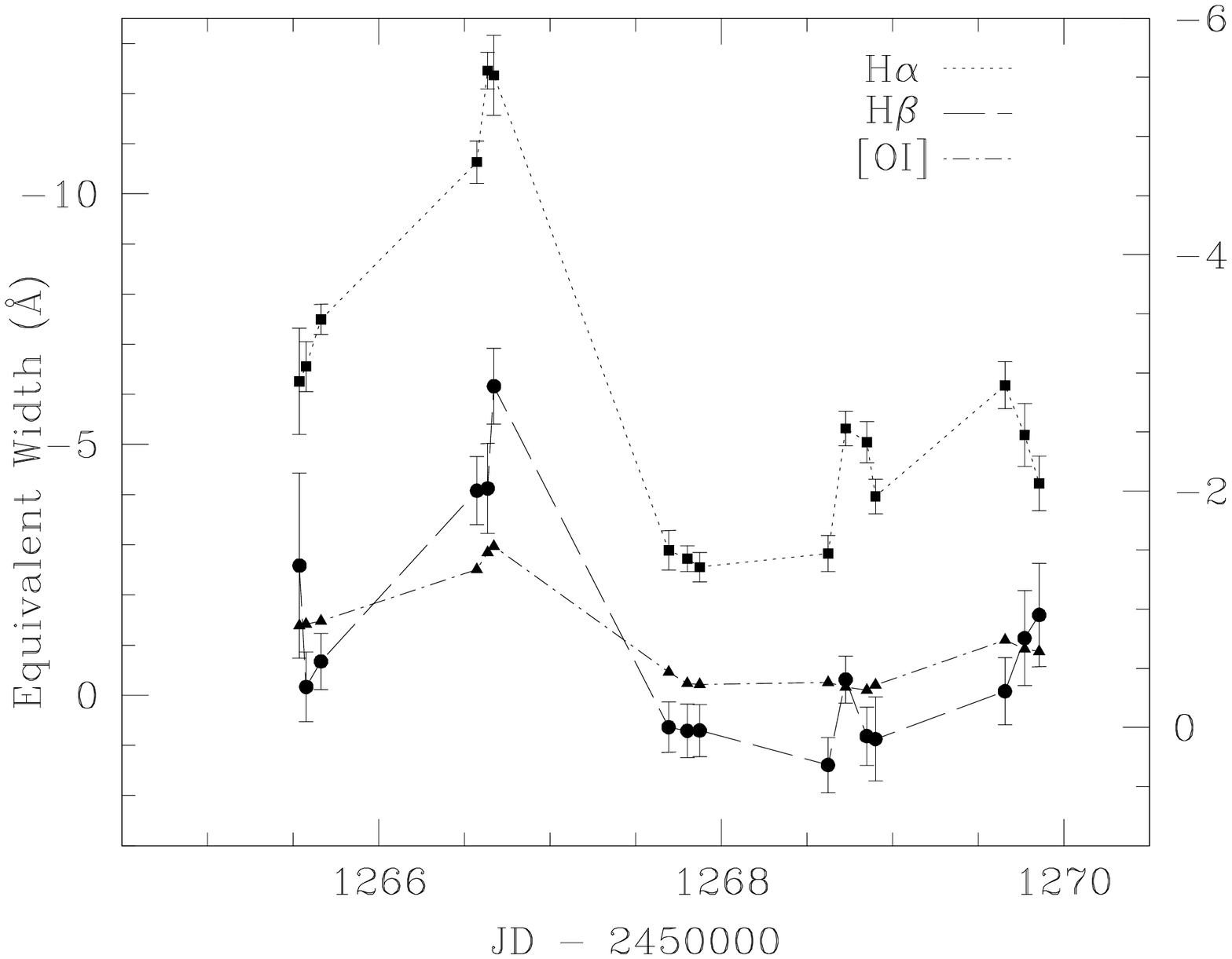}}
\end{minipage}\\
\vspace{-3.5cm}
\begin{minipage}{.50\linewidth}
\centering
\resizebox{\hsize}{!}{\includegraphics[trim = 10mm 20mm 0mm 0mm, clip,angle=-90,origin=c]{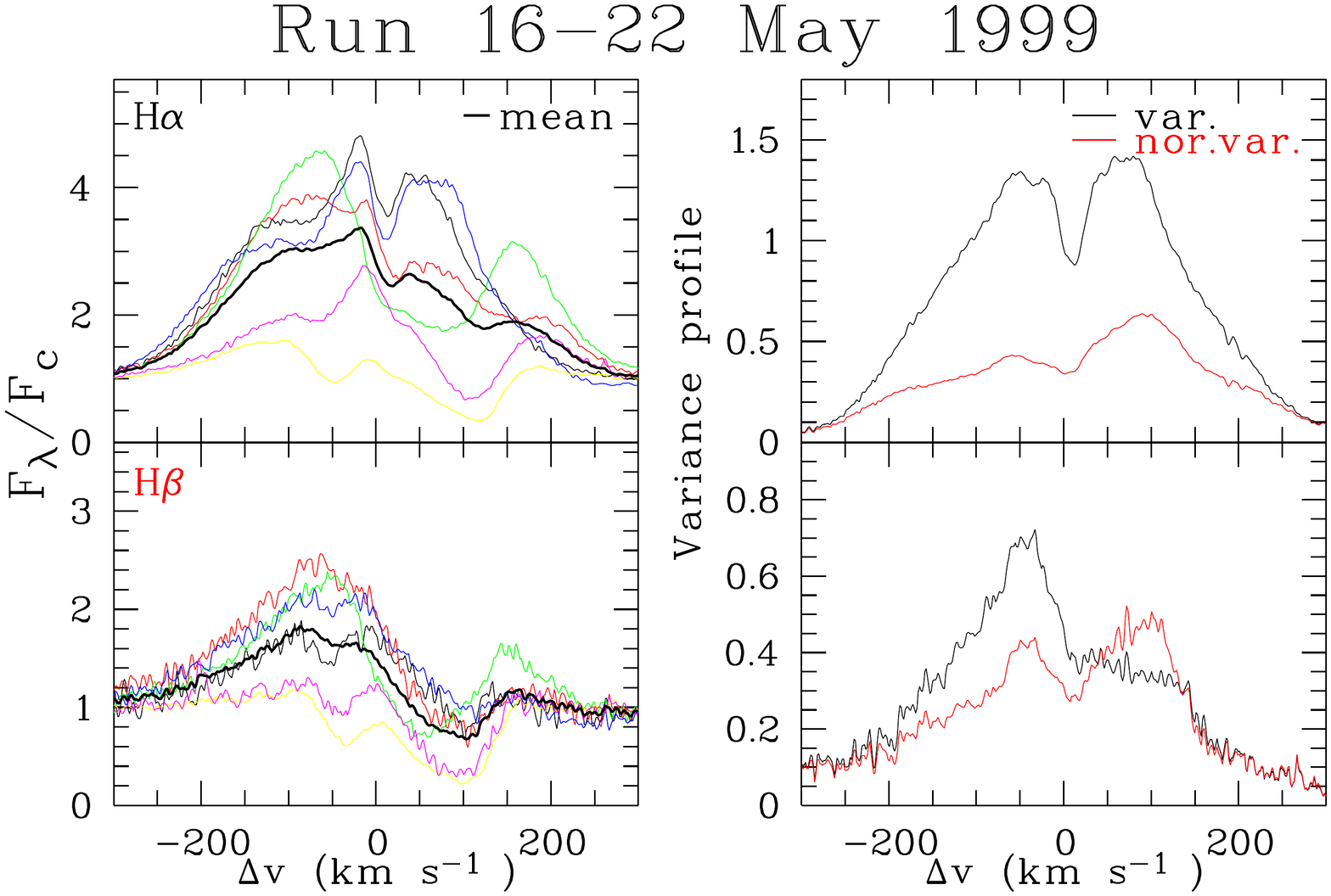}}
\end{minipage}
\hspace{5mm}
\begin{minipage}{.45\linewidth}
\centering
\resizebox{\hsize}{!}{\includegraphics[]{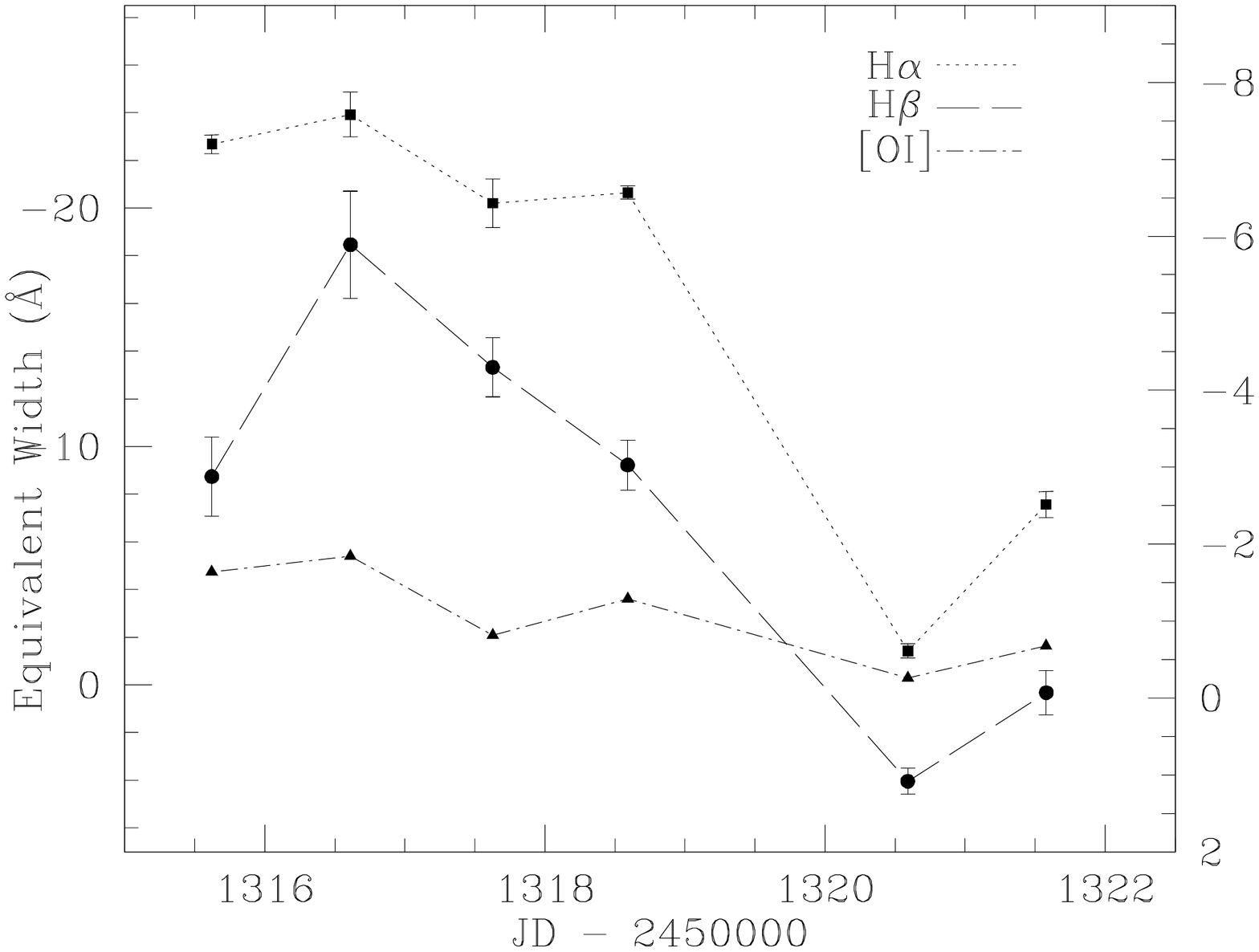}}
\end{minipage}\\
\vspace{-3.5cm}
\begin{minipage}{.50\linewidth}
\centering
\resizebox{\hsize}{!}{\includegraphics[trim = 10mm 20mm 0mm 0mm, clip, angle=-90,origin=c]{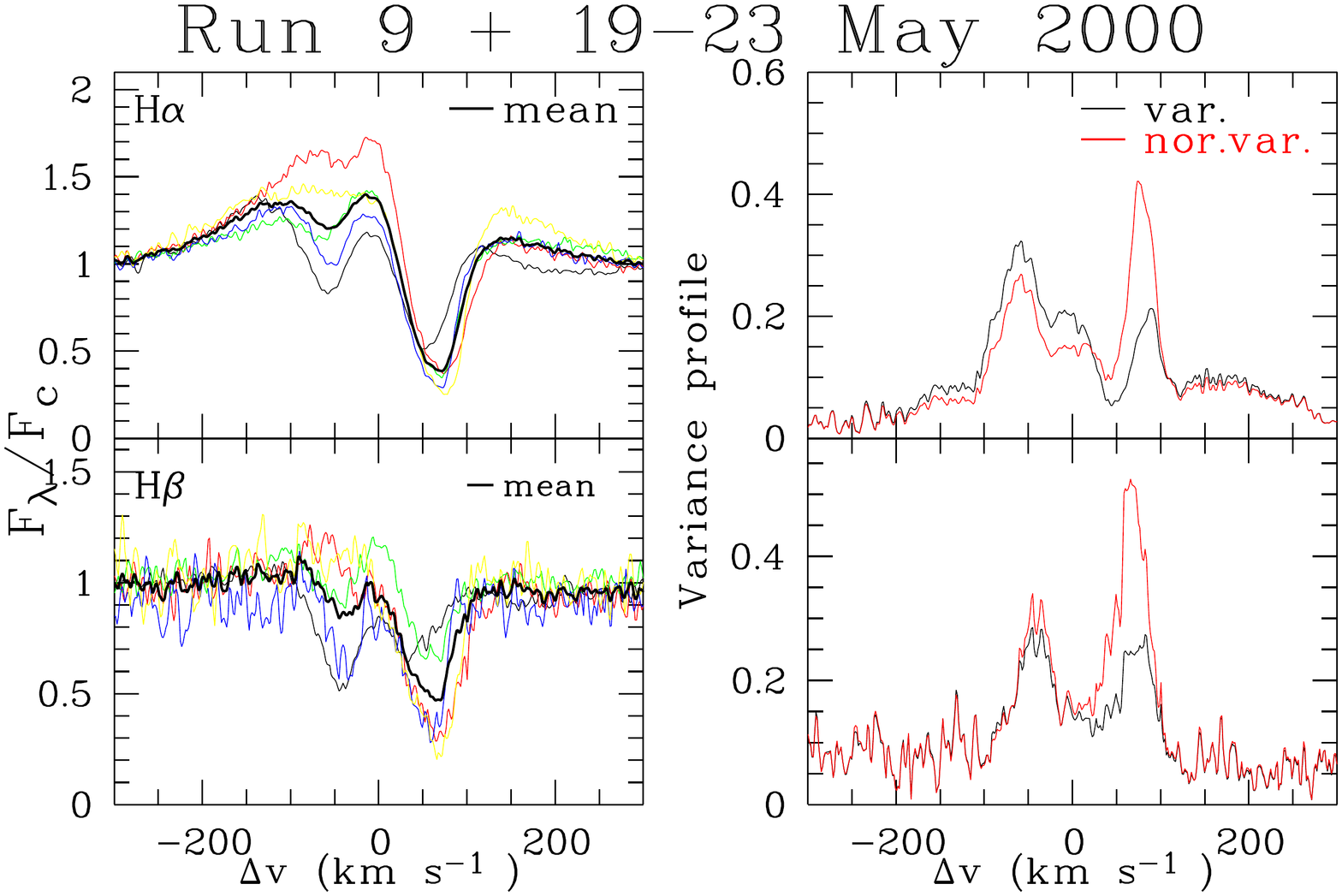}}
\end{minipage}
\hspace{5mm}
\begin{minipage}{.45\linewidth}
\centering
\resizebox{\hsize}{!}{\includegraphics[]{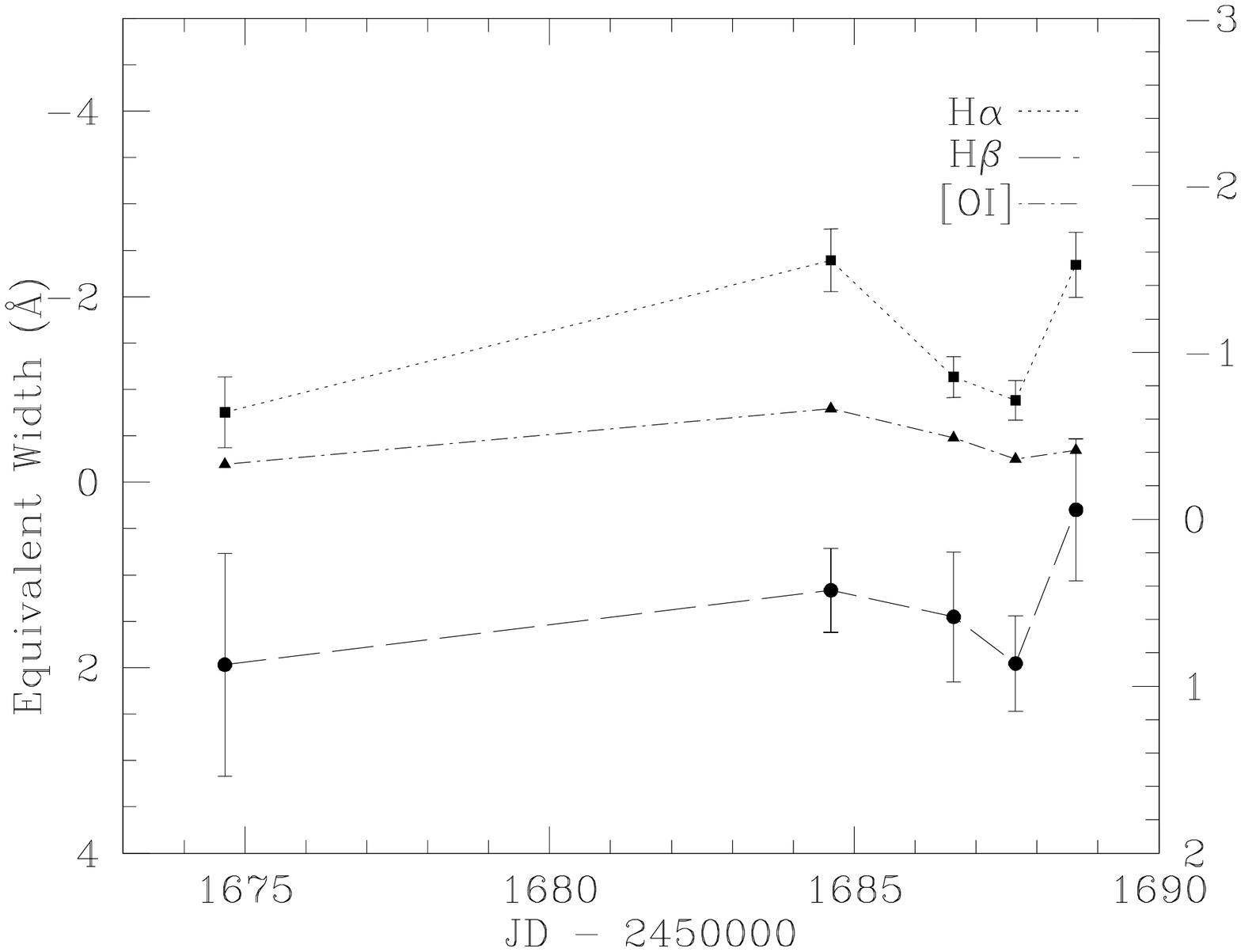}}
\end{minipage}\\
\vspace{-1.5cm}
\begin{minipage}[t]{.50\linewidth}
\caption{Line profiles ({\it left panels}) and corresponding variance and normalized variance profiles 
({\it right panels}) for H$\alpha$ and H$\beta$ in three different runs. }
\label{Fig7a}
\end{minipage}
\hspace{5mm}
\begin{minipage}[t]{.45\linewidth}
\caption{Equivalent widths of the main emission lines during the runs displayed 
in Fig.~\ref{Fig7a}. The left scale refers to H$\alpha$, while the right one 
is for H$\beta$ and [O{\sc i}] lines.}
\label{Fig7b}
\end{minipage}
\end{center}
\end{figure*}

\begin{figure} 
\begin{center}
\vspace{-1.5cm}
\resizebox{\hsize}{!}{{\includegraphics[trim = 0mm 0mm 0mm 0mm, clip]{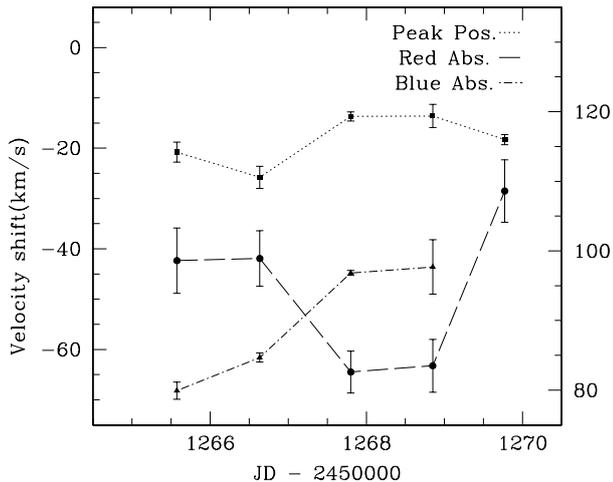}}}
\vspace{-3cm}
\caption{Variations in position of the central peak emission and the two 
absorptions of H$\alpha$ line during the period from 27 to 30 March 1999. 
The scale on the right refers to the red absorption, while the one on left 
refers to the central emission and the blue-shifted absorption.}
\label{FigAbs}
\end{center}
\end{figure}

\section{Line profile variability on short timescale}
\label{sec:Runs}

The large variety of H$\alpha$ profiles shown by T\,Cha appear to occur rather 
erratically, although this may be caused at least partly by the quite uneven 
temporal sampling of our spectra. 

We used three different series of spectra acquired with a time sampling 
shorter than one day to investigate variations on short timescales.
We analysed separately each of the following three runs: 
1) from 27 March 1999 to 1 April 1999 (total of 16 spectra in 5 nights);
2) from 16 to 22 May 1999 (total of 6 spectra in 6 nights); and 
3) from 9 to 23 May 2000 (total of 5 spectra in 5 nights). 
Figure~\ref{Fig7a} shows in each panel the overlap of H$\alpha$ (top-left box) 
and H$\beta$ (bottom-left box) line profiles for each of the three runs.  
The average normalized profile $\langle I_{n}(\lambda)\rangle$ is also superposed,
and is represented by a thicker line. The profiles, and their behaviour, appear to 
differ in each run, being more erratic in the second period than in the other two. 
The H$\alpha$ and H$\beta$ line intensities and profiles change on a daily timescale, 
with only minor changes occurring during the night. 
We report in Fig.~\ref{FigAbs}, the variability in the positions of the central peak 
and the two absorption components in H$\alpha$ during the first run.

To quantify the variability in each line, we examined the variance, 
$\sigma$, and the normalized variance, $\sigma_{n}$, in the different runs, 
defined respectively as:

\begin{equation}
\sigma^{2} = \Sigma_{i=1}^{N} [I_{n,i}(\lambda) - \langle I_{n}(\lambda)\rangle]^{2}/(N - 1) \\
\end{equation}
and
\begin{equation}
\sigma_{n}^{2}(\lambda) = \sigma^{2}(\lambda)/\langle I_{n}(\lambda)\rangle \\
\end{equation}

\noindent
where $I_{n,i}(\lambda)$ is the $i$-th spectrum. 
These quantities are plotted in the right panels of Fig.\,\ref{Fig7a}. 
 If variability were merely caused by the brightening or fading of the stellar continuum flux, 
the shape of normalized variance profiles would be the same as the average line profile 
\citep{Johns1995}. 
This is generally not the case, suggesting that some other changes in the H$\alpha$ 
forming region also occur.

Figure~\ref{Fig7b} shows the trends of the equivalent widths of the main emission lines with time.  
H$\alpha$ and H$\beta$ mainly vary in similar ways. 
On two dates, their behaviour is different, i.e., the last day of the first run, when the 
H$\alpha$ intensity increased as the H$\beta$ intensity decreased, 
and the third day of the second run, when an increase in H$\alpha$ intensity did not correspond 
to an increase in that of H$\beta$.
The [O{\sc i}] line trend is similar to that of H$\alpha$ except for during the fourth night 
of the first run and the last night of the third run, when both Balmer lines strengthened 
in emission, while the [O{\sc i}] lines did not change. 
The similarities in the variability of these lines indicate that the processes producing the 
observed changes must affect contemporaneously circumstellar zones located at different distances 
from the star, i.e., on length scales of between a few tenths of AU for H$\alpha$ and a few AUs 
for [O{\sc i}].

\subsection{Correlation Matrices} \label{sec:CCM}

\begin{figure*} 
\begin{center}
\begin{minipage}{.45\linewidth}
\centering
\resizebox{\hsize}{!}{\includegraphics[trim = 25mm 45mm 35mm 0mm, clip,angle=0,origin=c]{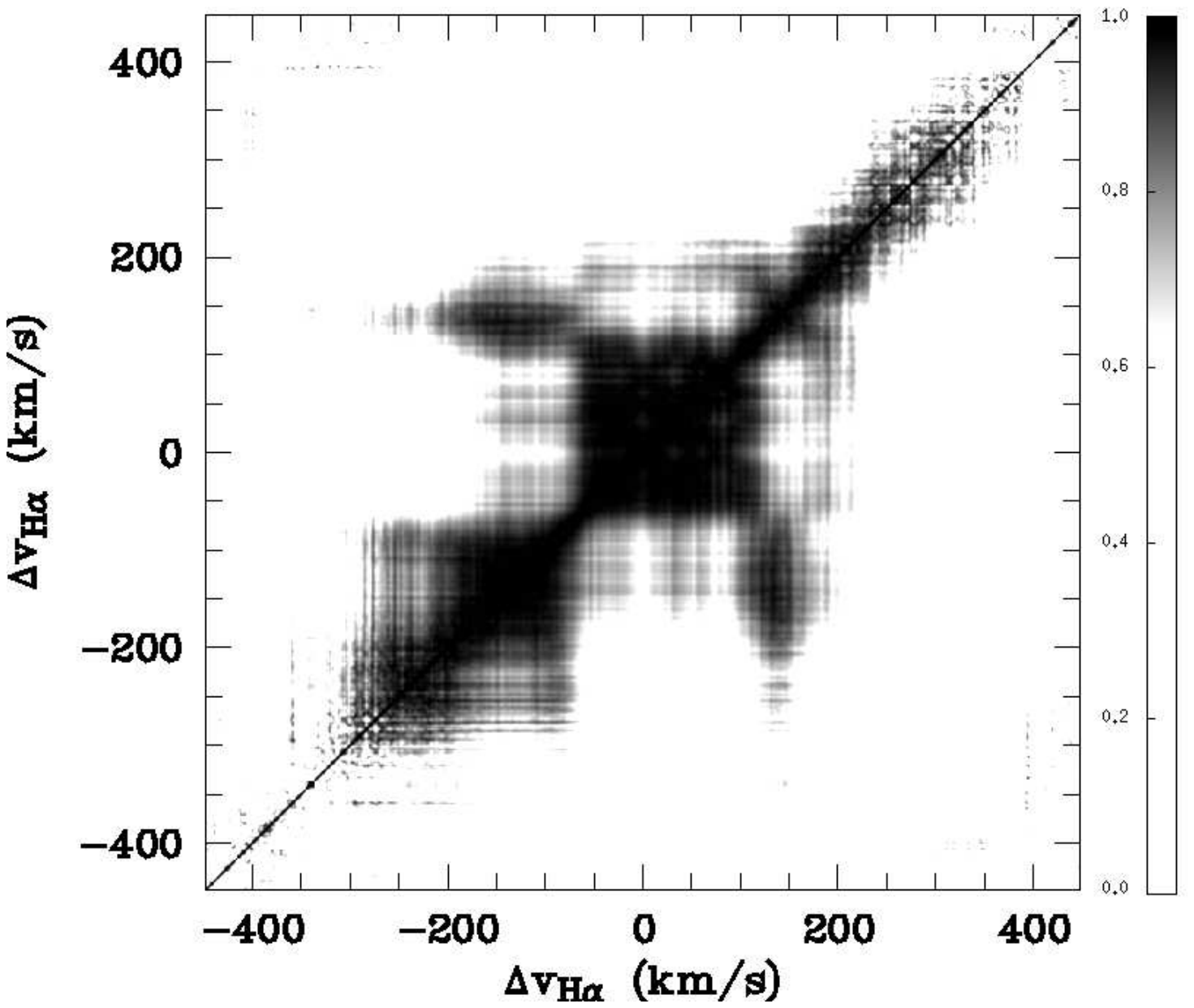}}
\end{minipage}
\hspace{0mm}
\begin{minipage}{.45\linewidth}
\centering
\resizebox{\hsize}{!}{\includegraphics[trim = 25mm 45mm 35mm 0mm, clip,angle=0,origin=c]{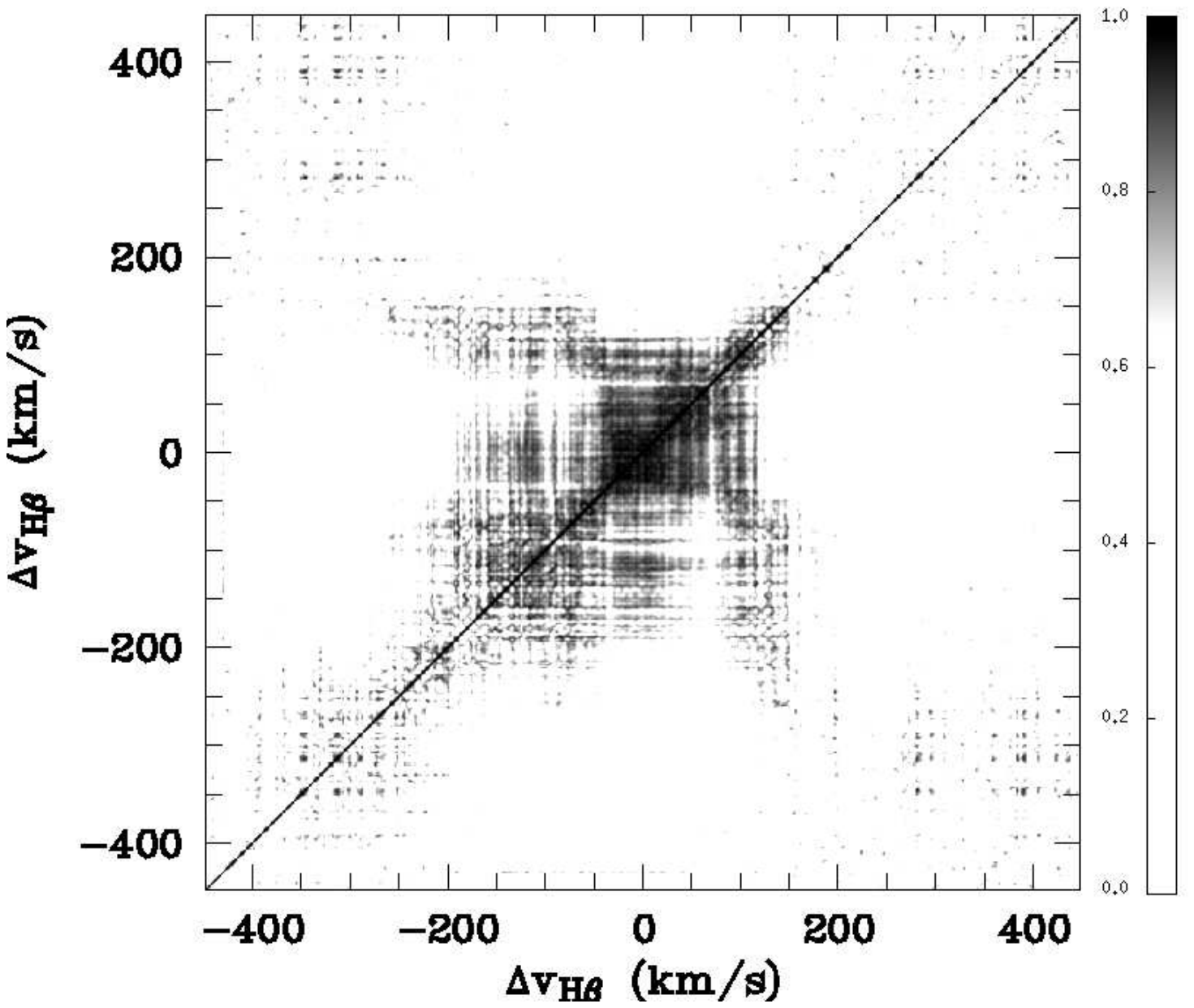}}
\end{minipage}\\
\vspace{-3.5cm}
\begin{minipage}{.45\linewidth}
\centering
\resizebox{\hsize}{!}{\includegraphics[trim = 25mm 45mm 35mm 15mm, clip,angle=0,origin=c]{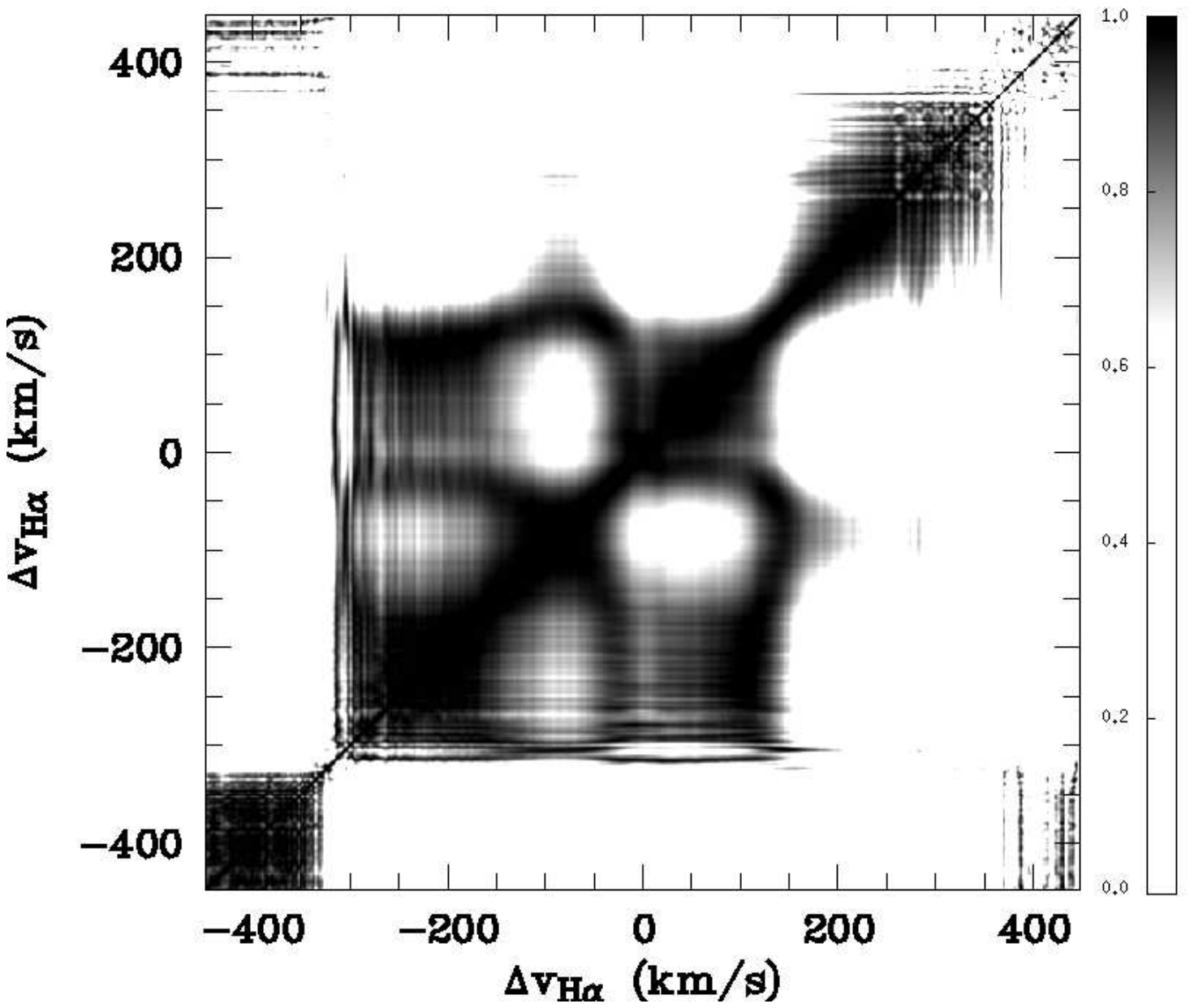}}
\end{minipage}
\hspace{0mm}
\begin{minipage}{.45\linewidth}
\centering
\resizebox{\hsize}{!}{\includegraphics[trim = 25mm 45mm 35mm 15mm, clip,angle=0,origin=c]{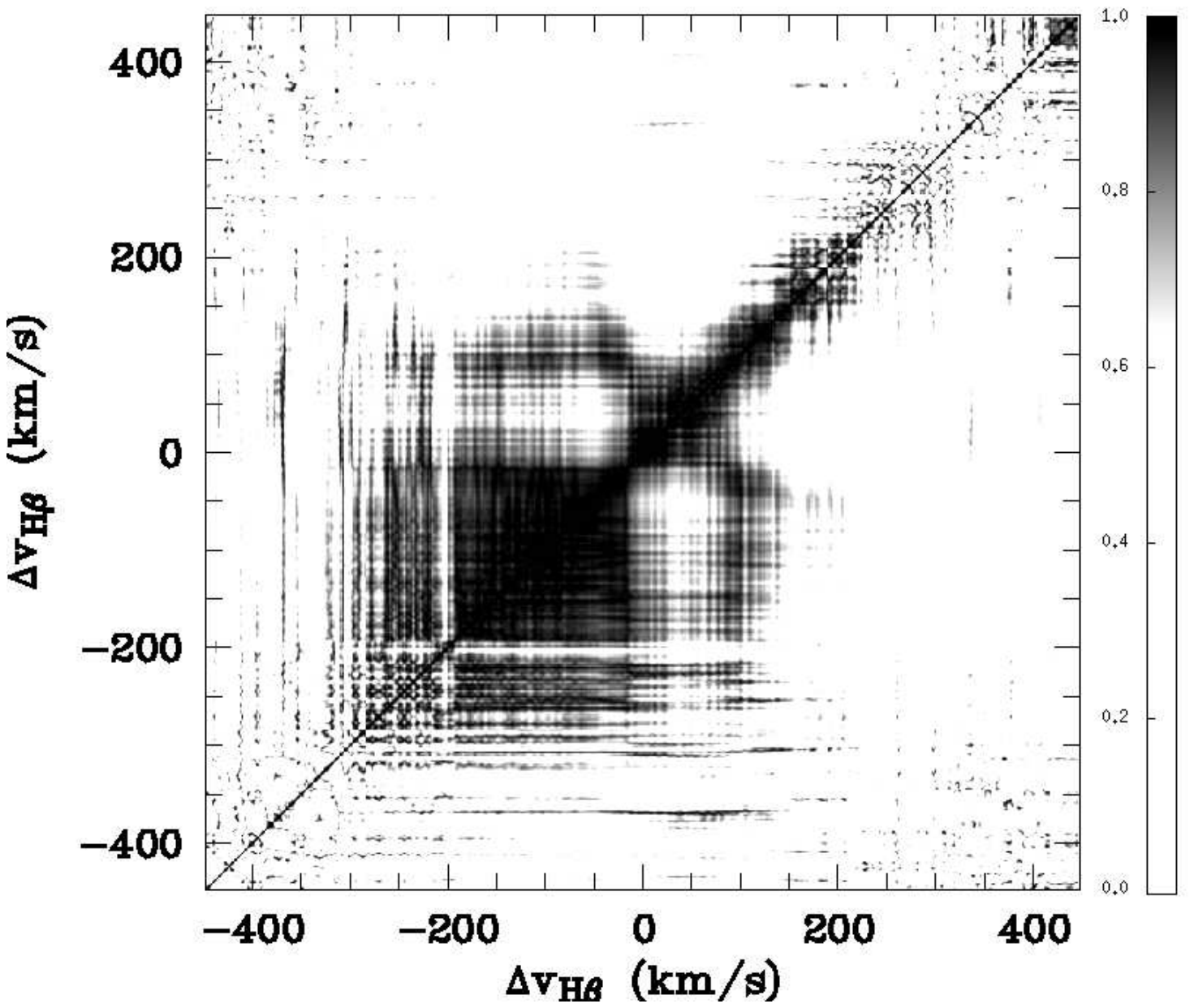}}
\end{minipage}\\
\vspace{-3.5cm}
\begin{minipage}{.45\linewidth}
\centering
\resizebox{\hsize}{!}{\includegraphics[trim = 25mm 45mm 35mm 15mm, clip, angle=0,origin=c]{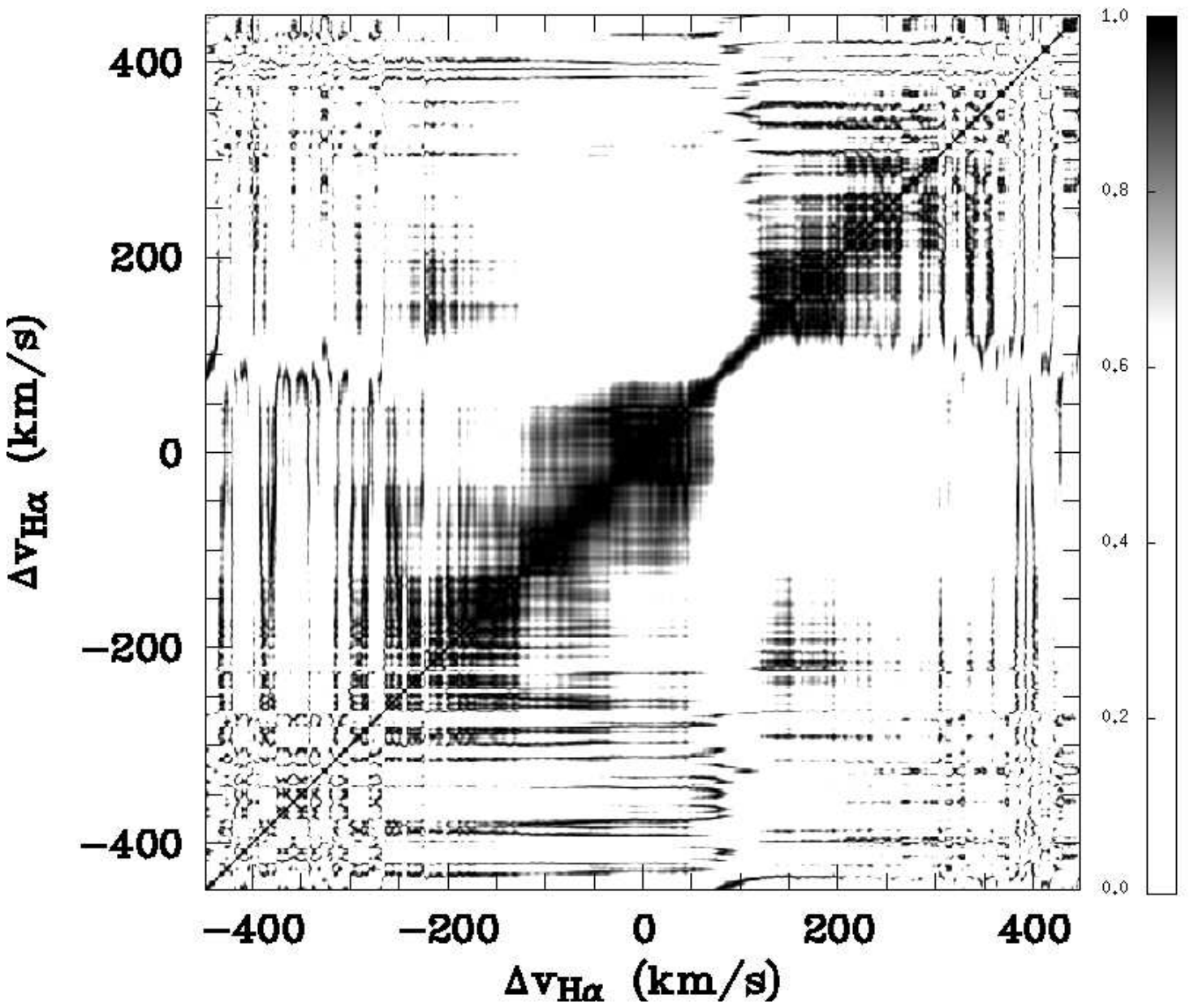}}
\end{minipage}
\hspace{0mm}
\begin{minipage}{.45\linewidth}
\centering
\resizebox{\hsize}{!}{\includegraphics[trim = 25mm 45mm 35mm 15mm, clip, angle=0,origin=c]{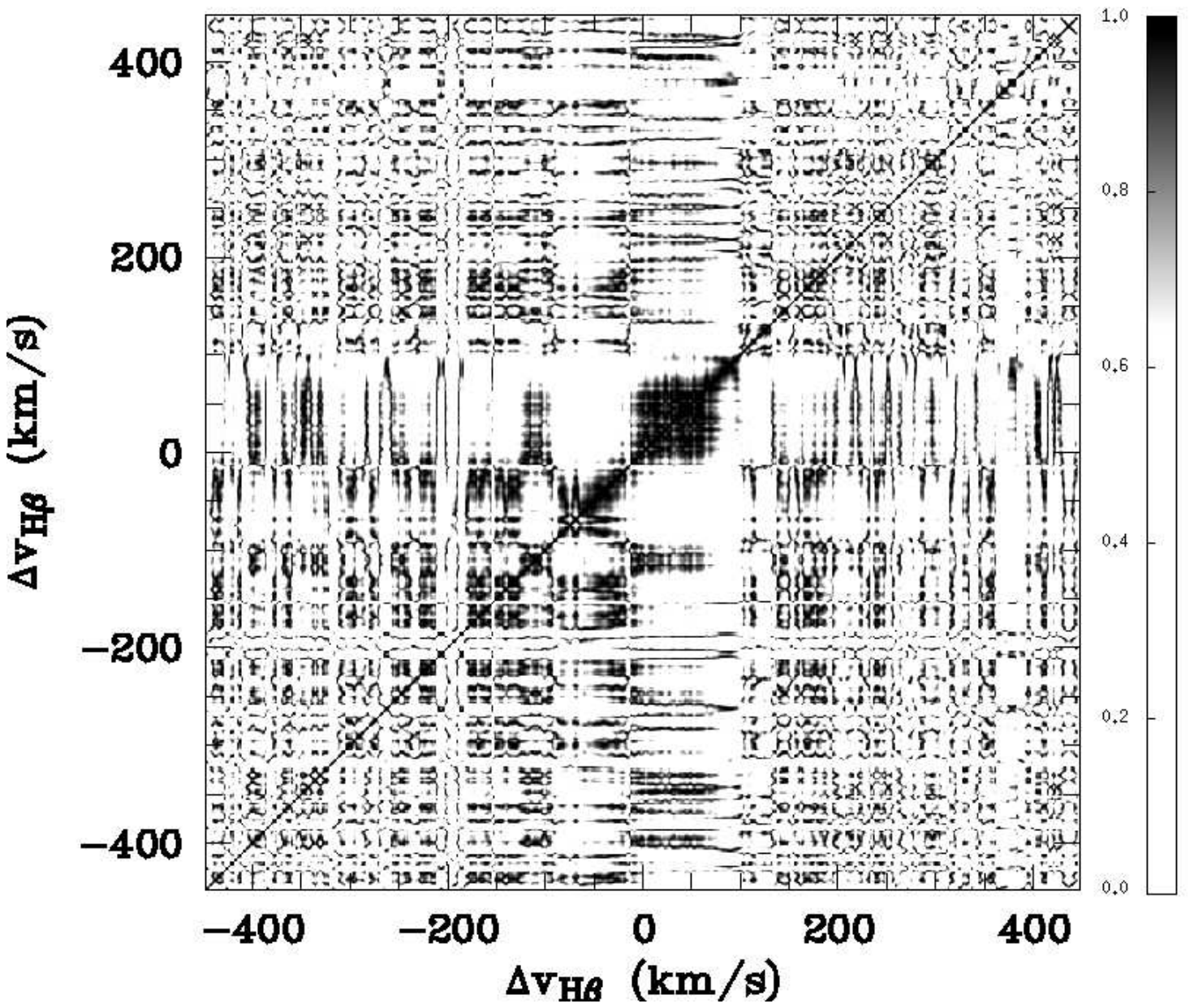}}
\end{minipage}\\
\vspace{-3.5cm}
\begin{minipage}[t]{.45\linewidth}
\end{minipage}
\hspace{0mm}
\begin{minipage}[t]{.45\linewidth}
\end{minipage}
\end{center}
\caption{Two-dimensional maps of the correlation matrices of H$\alpha$ (left panels)
 and H$\beta$ (right panels) line profiles with themselves, for the three runs of 
 Figs.~\ref{Fig7a} and \ref{Fig7b}. 
Different squarish regions are distinguishable in the first two runs, indicating 
a coherence in the variations of H$\alpha$ intensity in the velocity interval from 
$-300$ to $120$\,km\,s$^{-1}$.
} 
\label{fig:corr_mtrx}
\end{figure*}

To test whether the variations across the emission line profiles have a common origin, 
we computed their correlation matrices (CMs). 
CMs indicate the linear correlation coefficient between the variation in each velocity bin of 
the spectral line profile and variations in all other bins for the same or for two different lines.
A strong correlation between different velocity bins is indicative of a common origin, or 
emitting region.
If the variability in the emission line is linked to variability in the continuum, then the line 
profile is highly correlated over a wide range of velocity, producing a typical squarish shape 
in the CM \citep{Johns1995,Alencar2002}.

We computed the CMs of H$\alpha$ and H$\beta$ with respect to themselves and to each other. 
 Two-dimensional plots of these matrices for H$\alpha$ and H$\beta$ in three different runs 
are shown in Fig.~\ref{fig:corr_mtrx}. 
No anticorrelation is evident in the matrices. 
In each run, there are regions that correlate well with themselves, but do not correlate with 
the remainder of the line. No significant correlation is seen between the red 
($\ge200$\,km\,s$^{-1}$) and the blue ($\le-200$\,km\,s$^{-1}$) wings of the profile.  
However, in the second run, the H$\alpha$ line exhibits a correlation over a wide velocity range, 
its squarish form extending from about $-$300 to nearly 150\,km\,s$^{-1}$. The correlation
is disrupted by the blue-shifted absorption, from $-$150 to $-$50\,km\,s$^{-1}$, and the 
redshifted component, up to about $-$100\,km\,s$^{-1}$, which do not correlate with the remainder 
of the line profile but with the redshifted emission peak, around 130\,km\,s$^{-1}$. 
This behaviour is not observed in the other epochs.
Similarly, in the first run, the broad red absorption from $-$20 to 150\,km\,s$^{-1}$ breaks the 
correlation with the blue emission wing, which still correlates weakly with the red emission wing. 
Interestingly, the square regions in the CMs correspond to some of the highest peaks in the 
variance profiles, i.e., the blue emission wing, the central peak and the red absorption, 
respectively. 

The H$\beta$ CMs are more affected by noise due to lower S/N at those wavelengths, but again 
traces of a squarish form can be seen in particular in data from the second run.
We also computed the CM between H$\alpha$ and H$\beta$, but, although the profiles of the two 
lines appear to be similar, the corresponding correlation matrices are affected too significantly 
by noise and no useful information can be extracted from them. 

In conclusion, coherent changes in the velocity interval between $-$300 and 120\,km\,s$^{-1}$ 
appear consistent with variations dominated by variable circumstellar obscuration of 
the stellar photosphere, giving rise, due to a contrast effect, to changes in the 
relative intensity of emission lines. 
In the case of magnetospheric accretion, parts of the H$\alpha$ profile, i.e., the redshifted 
component, are expected to originate in regions close to the star and eventually to be affected 
by direct obscuration. 
These may be identified kinematically with the intervals where the correlation breaks down 
whereas the differences from one epoch to the other possibly reflect changes in the geometry
of the emitting region.

\begin{table}
\caption{Equivalent widths of H$\alpha$ and [O{\sc i}]\,6300\AA\ 
     measured on the low-resolution spectra (see the atlas in electronic form)
     and corresponding total extinction, A$_V$, obtained as described in 
     Sect.~\ref{sec:SED}. {\bf Only in electronic form. }
     }\label{tab:low-res-spec} 
 \centering
\begin{tabular}{ c  c  r  r  r  r  c }
\hline \hline
    Date &   JD      & W$_{H\alpha}$ & $\sigma_{H\alpha}$ & W$_{[O{\sc i}]}$ &
$\sigma_{[O{\sc i}]}$ & A$_V$ \\
         &$-$2440000 &       (\AA)   & (\AA)              &  (\AA)           &
(\AA)                &   (mag) \\
\hline     
07 May 92 & 8749.668 &  $-$4.58 & 0.30 & -0.33 & 0.05 & 2.2  \\
08 May 92 & 8750.606 &  $-$3.63 & 0.40 & -0.35 & 0.05 & 2.4  \\
10 May 92 & 8752.628 &     0.40 & 0.10 & -0.13 & 0.03 & 1.8  \\
11 May 92 & 8753.634 &  $-$4.87 & 0.70 & -0.22 & 0.05 & 2.0  \\
12 May 92 & 8754.600 &  $-$6.42 & 0.80 & -0.87 & 0.10 & 2.8  \\
27 Mar 93 & 9073.698 &  $-$1.09 & 0.20 & -0.23 & 0.05 & 1.7  \\
27 Mar 93 & 9073.847 &  $-$0.51 & 0.20 & -0.19 & 0.03 & 1.4  \\
28 Mar 93 & 9074.689 &     1.80 & 0.20 &   -   &  -   & 1.5  \\
28 Mar 93 & 9074.843 &     2.08 & 0.20 & -0.06 & 0.05 & 1.5  \\
29 Mar 93 & 9075.680 &     0.28 & 0.20 & -0.25 & 0.05 & 1.7  \\
29 Mar 93 & 9075.873 &  $-$0.14 & 0.10 & -0.11 & 0.03 & 1.7  \\
30 Mar 93 & 9076.758 & $-$23.70 & 2.00 & -4.16 & 0.50 & 3.3  \\
30 Mar 93 & 9076.840 & $-$34.00 & 4.00 & -6.29 & 0.55 & 3.6  \\
31 Mar 93 & 9077.693 &  $-$9.20 & 1.00 & -1.27 & 0.15 & 2.6  \\
01 Apr 93 & 9078.700 &  $-$1.80 & 0.10 & -0.26 & 0.05 & 1.9  \\
01 Apr 93 & 9078.879 &  $-$0.32 & 0.10 & -0.14 & 0.05 & 1.3  \\
02 Apr 93 & 9079.687 &     1.90 & 0.10 & -0.06 & 0.05 & 1.5  \\
02 Apr 93 & 9079.879 &     1.64 & 0.10 &   -   &  -   & 1.2  \\
30 May 93 & 9137.531 &     1.40 & 0.10 &   -   &  -   & 1.2  \\
31 May 93 & 9138.515 &     0.26 & 0.10 & -0.26 & 0.05 & 2.0  \\
31 May 93 & 9138.738 &  $-$0.25 & 0.10 & -0.35 & 0.05 & 2.4  \\
01 Jun 93 & 9139.545 & $-$44.60 & 5.00 & -9.20 & 0.55 & 4.6  \\
03 Jun 93 & 9142.495 &     0.16 & 0.10 & -0.36 & 0.05 & 1.5  \\
04 Jun 93 & 9143.494 &  $-$0.23 & 0.10 & -0.48 & 0.05 & 2.3  \\
05 Jun 93 & 9143.736 &     0.66 & 0.10 & -0.23 & 0.05 & 2.2  \\
05 Jun 93 & 9144.496 &  $-$6.20 & 0.50 & -0.57 & 0.05 & 2.8  \\
06 Jun 93 & 9144.718 &  $-$5.81 & 0.60 & -0.68 & 0.05 & 3.2  \\
28 Jun 93 & 9167.463 &  $-$2.40 & 0.30 & -0.35 & 0.05 & 1.8  \\
29 Jun 93 & 9168.478 &  $-$19.3 & 2.00 & -4.18 & 0.50 & 3.1  \\
02 Jul 93 & 9170.519 &  $-$7.21 & 0.60 & -0.93 & 0.10 & 2.1  \\
03 Jul 93 & 9172.467 &  $-$57.0 & 5.50 & -9.00 & 0.60 & \ddag \\
24 Mar 94 & 9435.745 &     1.73 & 0.10 & -0.09 & 0.05 & 1.8  \\
24 Mar 94 & 9435.887 &     2.20 & 0.20 & -0.10 & 0.03 & 2.1  \\
25 Mar 94 & 9436.698 &     0.55 & 0.10 & -0.23 & 0.05 & 1.5  \\
25 Mar 94 & 9436.900 &  $-$1.03 & 0.10 & -0.36 & 0.05 & 1.6  \\
26 Mar 94 & 9437.532 &     0.96 & 0.10 & -0.24 & 0.05 & 1.7  \\
26 Mar 94 & 9437.672 &     0.21 & 0.10 & -0.36 & 0.05 & 1.5  \\
27 Mar 94 & 9438.519 &  $-$1.10 & 0.10 & -0.30 & 0.05 & 1.8  \\
27 Mar 94 & 9438.662 &  $-$0.95 & 0.10 & -0.27 & 0.12 & 1.6  \\
28 Mar 94 & 9439.520 &     1.61 & 0.20 &   -   &   -  & 1.3  \\
28 Mar 94 & 9439.661 &     0.95 & 0.10 &   -   &   -  & 1.7  \\
29 Mar 94 & 9440.518 &     0.25 & 0.10 & -0.03 & 0.05 & 1.2  \\
29 Mar 94 & 9440.661 &     1.05 & 0.20 & -0.03 & 0.05 & 1.6  \\
09 Apr 95 & 9816.807 &  $-$0.19 & 0.10 & -0.04 & 0.05 & 1.6  \\
10 Apr 95 & 9817.789 & $-$38.70 & 3.50 & -4.31 & 0.45 & 3.4  \\
11 Apr 95 & 9818.805 &     0.70 & 0.10 & -0.05 & 0.05 & 1.5  \\
12 Apr 95 & 9819.788 &     1.05 & 0.10 & -0.14 & 0.05 & 1.8  \\                                                                                    
\hline
\end{tabular}
\\
\noindent $\ddag$ A$_V$ not determined due to very faint continuum \\
\end{table}

\begin{figure*}[ht]  
\resizebox{\hsize}{!}{\rotatebox{-90}{\includegraphics[trim = 15mm 30mm 0mm 0mm]{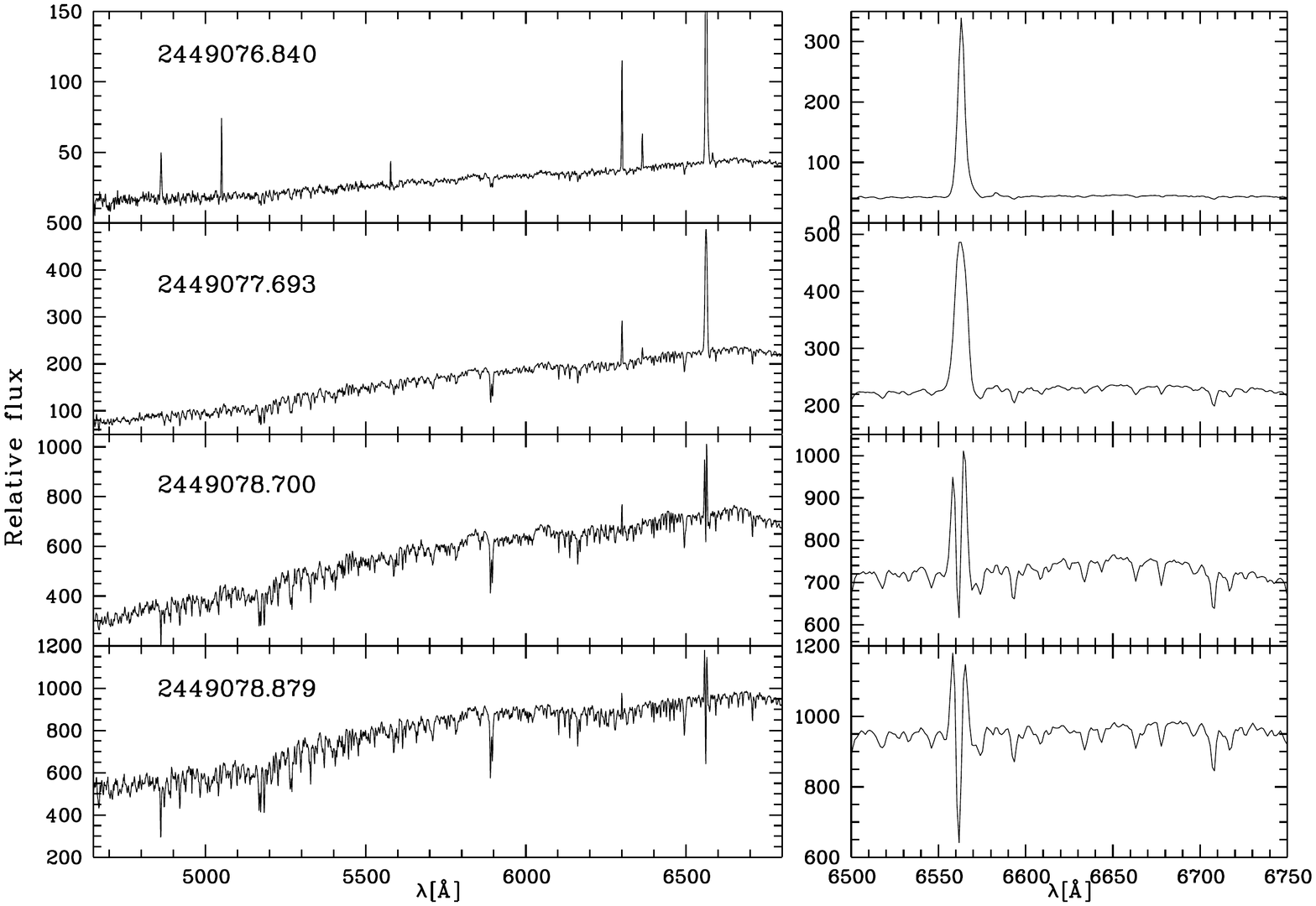}}}
 \vspace{-1cm}
  \caption{ Low-resolution spectra of T\,Cha. The panels on the left side display the whole range, 
 those on the right show the interval containing H$\alpha$ and the Lithium line. The complete 
 atlas of low-resolution spectra is only available electronically.}
 \label{fig:lowres-ex}
\end{figure*}

\section{Variability in the low-resolution spectra}
\label{sec:low-res-spec-var}
 
Strong variability was also detected in our low-resolution spectroscopy. 
We refer in particular to Table~\ref{tab:low-res-spec}\footnote{available
only in electronic form}, where we report the equivalent widths of 
the H$\alpha$ and [O{\sc i}]$\lambda$6300\,\AA\ lines, and to 
Fig.~\ref{fig:lowres-ex}, where a sample of low-resolution 
spectra is shown\footnote{the full atlas of low-resolution spectra 
is reported in Fig.~{fig:lowres-atlas}, only in electronic form}. 

We note that during the May-June 1993 run, the star exhibited 
spectacular changes between one night and the next. For instance, H$\alpha$ 
was observed to vary between weak absorption (EW$\approx 0$) on May 31, 
to a very strong emission line (EW$\approx-50$\AA) on June 1. 
The latter spectrum was also characterised by strong [O{\sc i}] lines, 
and the [N{\sc ii}] line was also present, while no emission lines were 
detected in data acquired the two nights before. 

\begin{figure}[ht]  
 \hspace{0.3cm}
 \resizebox{\hsize}{!}{\rotatebox{-90}{\includegraphics[trim = 15mm 30mm 0mm 0mm]{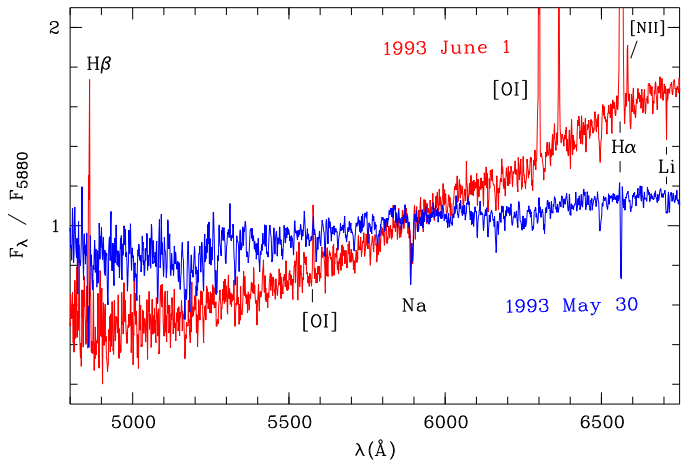}}}
 \vspace{-1cm}
 \caption{ Low-resolution spectra of T\,Cha near minimum (June 1, in red) 
 and maximum brightness (May 30, in blue). 
 The spectra are calibrated in relative flux and normalised to the flux at $\lambda=5880$\,\AA.
 Emission features, only visible in the spectrum corresponding to the fainter level, are marked.
 }
 \label{fig:lr-spectra}
\end{figure}

The spectacular variability of these lines is illustrated in 
Fig.~\ref{fig:lr-spectra}, where two low-resolution spectra, corresponding to 
close to the maximum and minimum brightness of the star, just two nights apart, 
are compared after normalising each of them to the flux at 
$\lambda=5880$\,\AA. We recall that these spectra are calibrated in 
relative flux (see Sect.~\ref{sec:low-res-obs}). 
The variation in the continuum slope between the two dates is also remarkable, 
in addition to the emission features appearing prominent when the 
photospheric continuum looks fainter and heavily reddened. 

\begin{figure}[ht] 
\includegraphics[width=9cm]{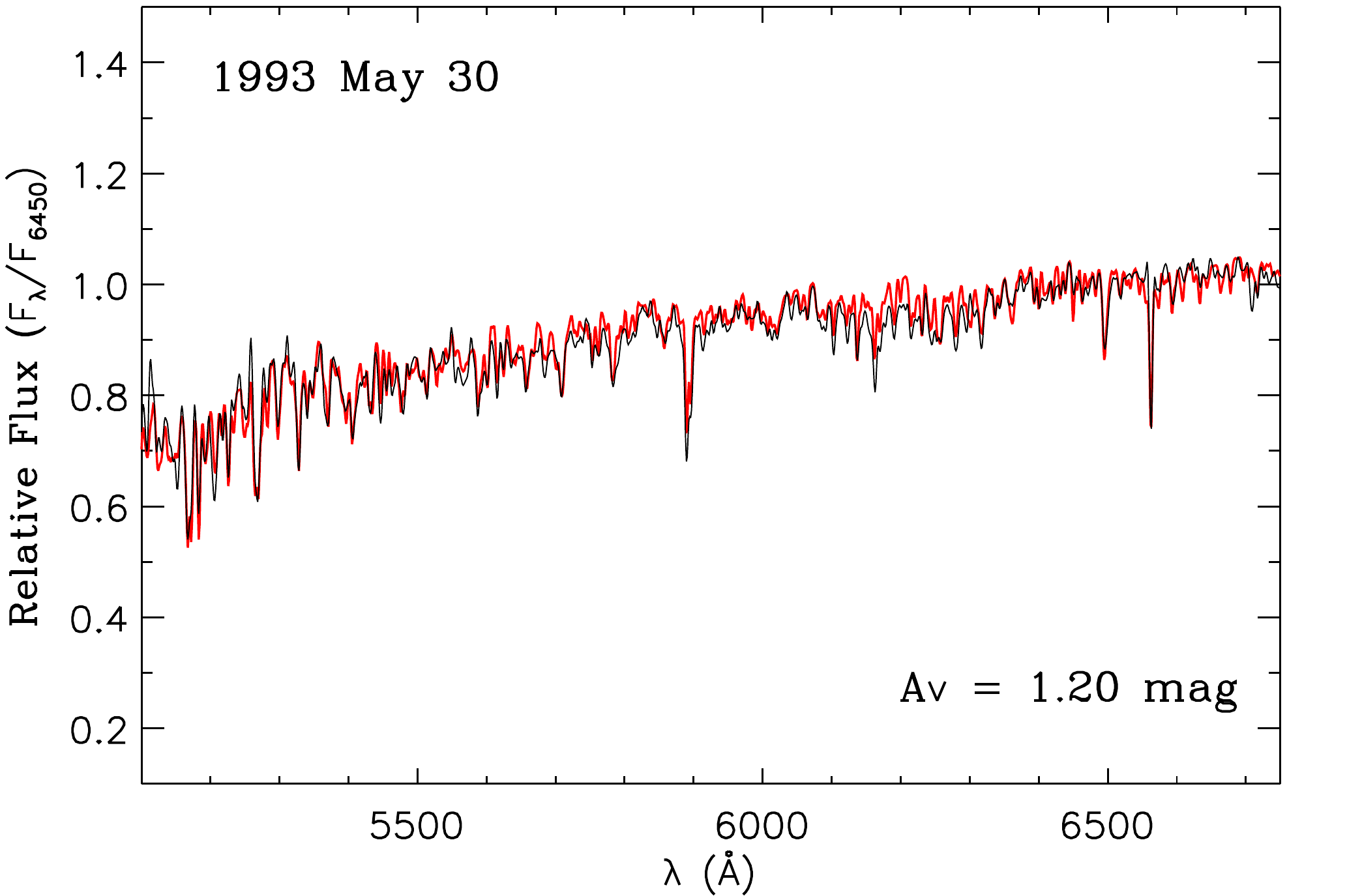}
\includegraphics[width=9cm]{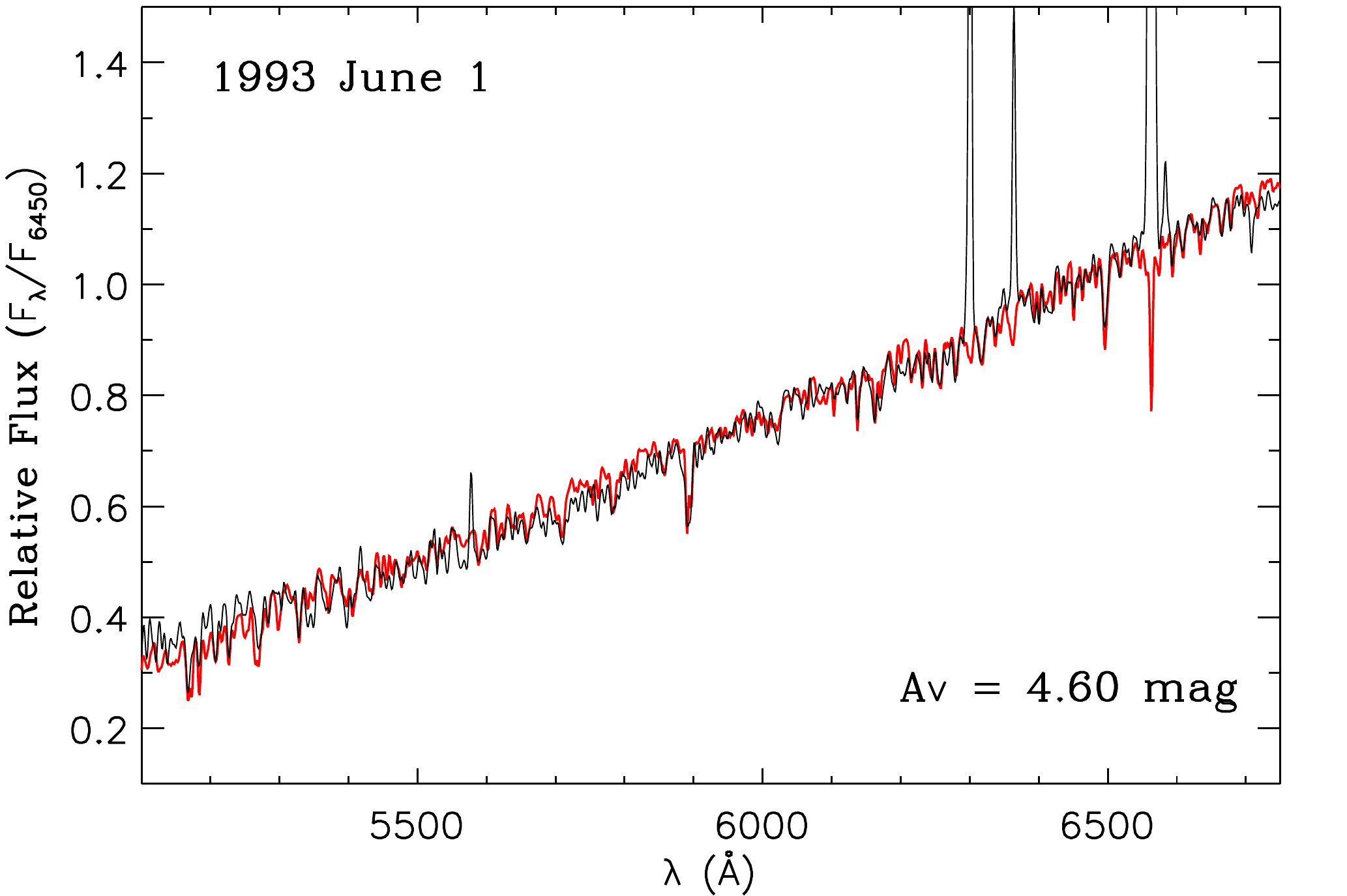}
 \caption{Best-fit of spectral type templates to the spectra shown
 in Figure~\ref{fig:lr-spectra}. The spectra of T\,Cha are shown
 as thin (black) lines, while the templates are shown as thick (red) lines.
 The corresponding values of total visual extinction, as derived from 
 the two-parameter fit adopting a normal interstellar extinction law 
 with $R_V=3.1$, 
 are also indicated for both the faint and bright stage, respectively.}
 \label{fit-spec-type}
\end{figure}

\begin{figure}[ht] 
\resizebox{\hsize}{!}{\includegraphics{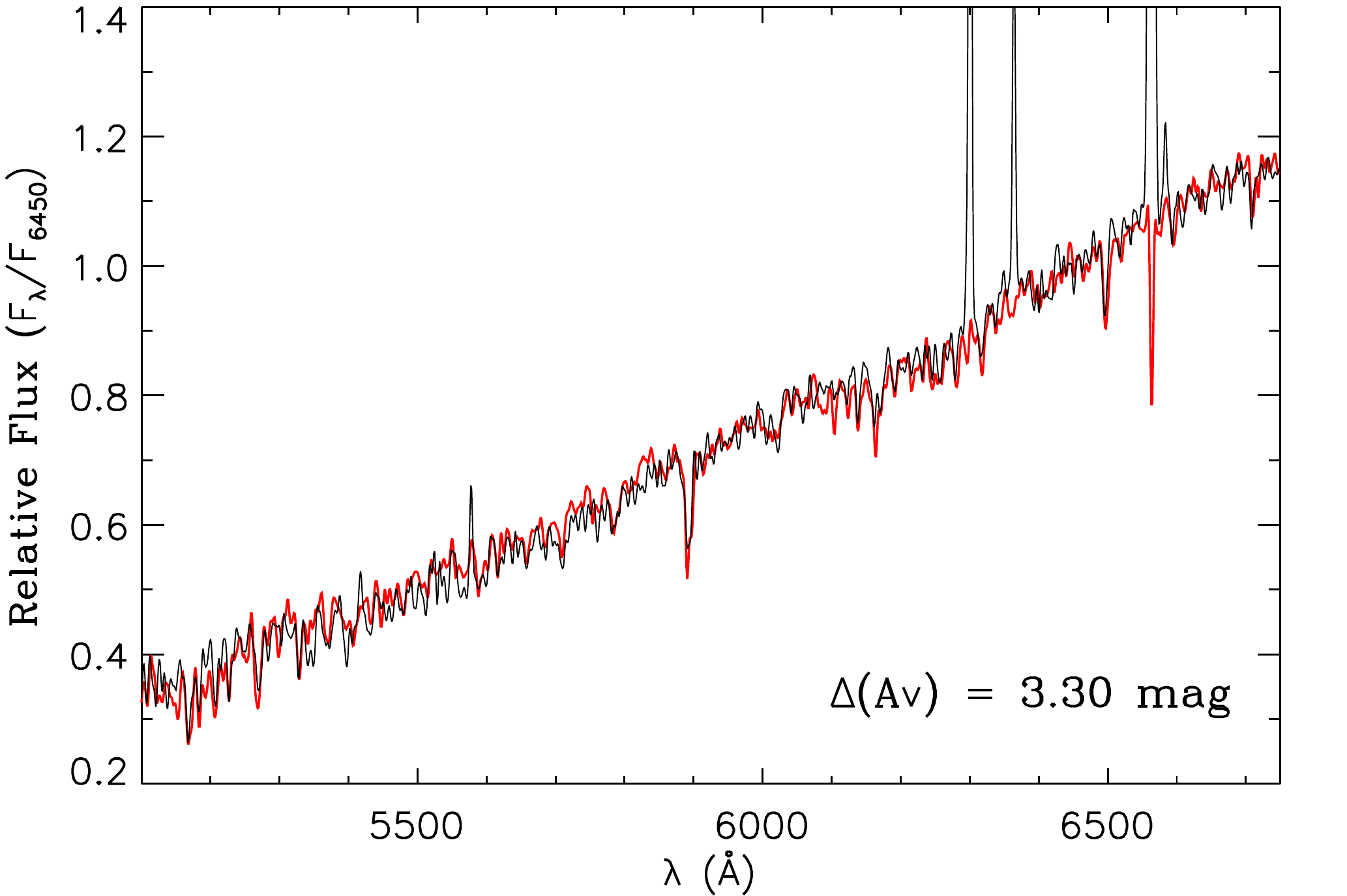}}
\vskip -0.74cm
\resizebox{\hsize}{!}{\rotatebox[]{-90}{\includegraphics[width=8.cm,height=8.2cm,trim = 10mm 20mm 0mm 10mm, clip]{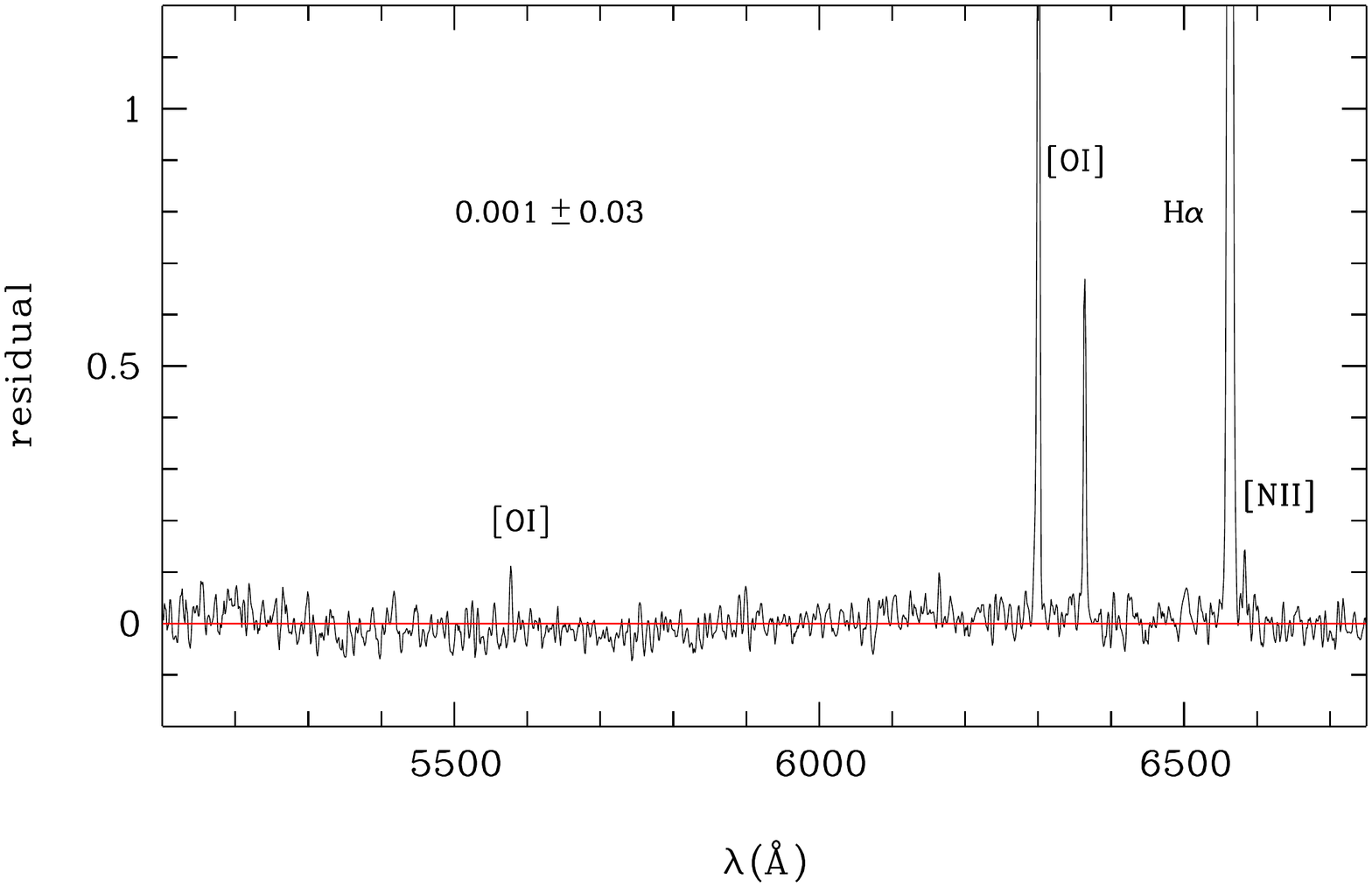}}}
\vspace{-1.5cm}
\caption{The top panel shows the spectrum at the faint state as a thin (black) line, 
 and the spectrum at the bright stage, used as a template, 
 is shown as a thick (red) line.
 The lower panel shows the residual. The average residual and rms are 
 indicated. The emission lines, not included in the deriving the average 
 value of the residual, are marked. 
 }
 \label{fit-self}
\end{figure}

\begin{figure*}[h!]  
\centering
\vspace{-2cm}
\resizebox{\hsize}{!}{\rotatebox[]{-90}{\includegraphics[width=7cm]{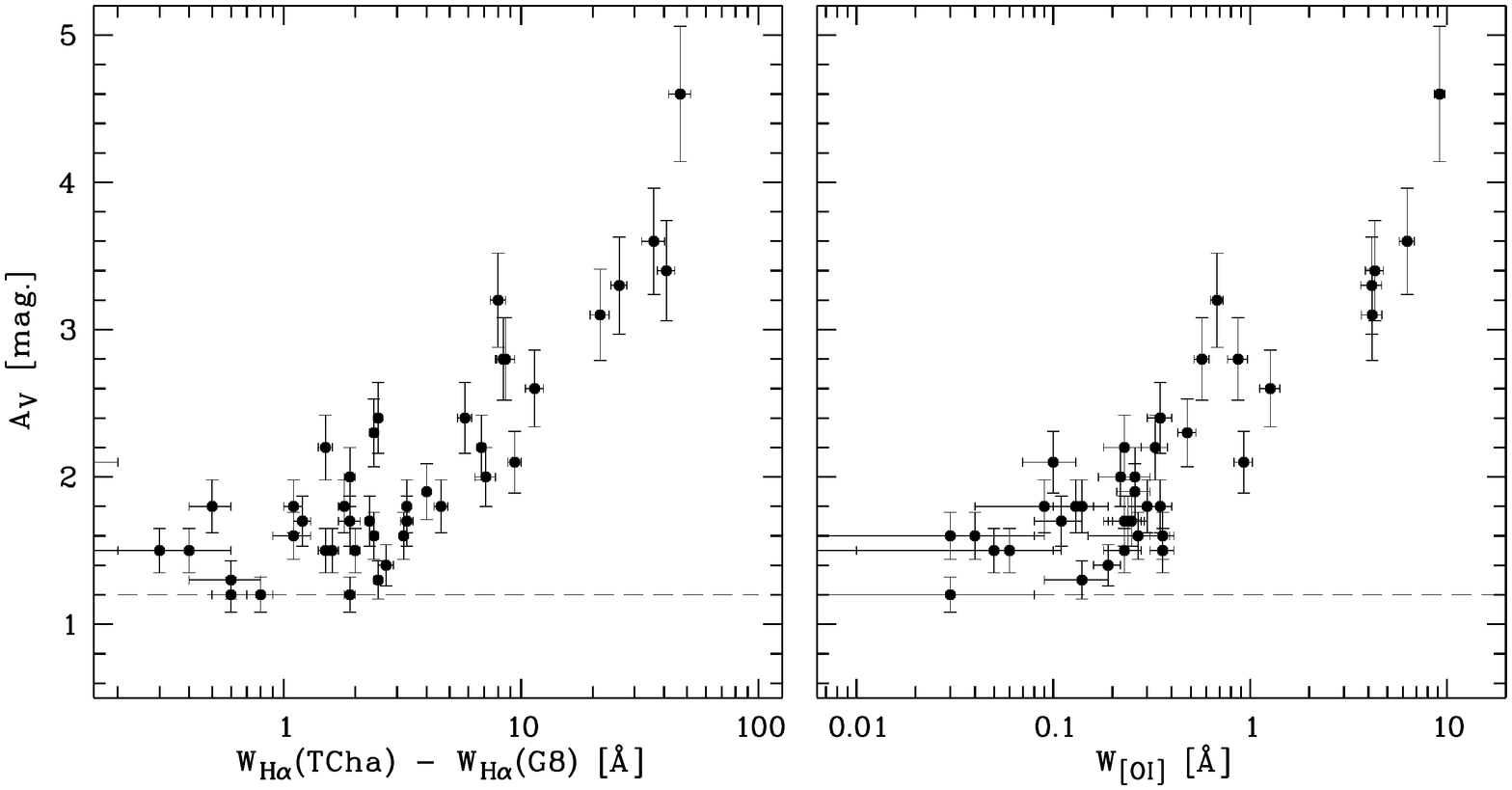}}}
\vspace{-2cm}
\caption{ 
Visual extinction derived from low-resolution flux-calibrated spectra versus the 
equivalent width of H$\alpha$ (left panel) and [O{\sc i}]\,6300\AA\ (right panel) 
emission. For H$\alpha$, the abscissa refers to the net emission after subtracting 
a constant value corresponding to the photospheric absorption for a G\,8 star. 
The horizontal dashed line indicates the minimum value of A$_V$ obtained for T\,Cha 
at maximum brightness. 
} 
\label{Fig:Av}
\end{figure*}
 
 However, no change in the spectral type of the star is observed. 
 To verify this, we used low-resolution spectra calibrated in relative flux, 
 and applied the methods described in \citet{Alcala2006} and \citet{Gandolfi2008} 
 to determine simultaneously the spectral type and the visual extinction 
 at the phases of minimum and maximum brightness, respectively. 
 The result, shown in Fig.~\ref{fit-spec-type}, is that the spectral type did not vary 
 to within about half a sub-class, while the visual extinction changed 
 from 1.2\,mag in the bright phase to about 4.6\,mag in the faint state, respectively. 
 This corresponds to an extinction increase of at least 3.4\,mag between the bright and 
 faint states.  By instead using the spectrum corresponding to the 
 bright level as a template, an extinction of 3.3 mag is needed to reproduce 
 the faint spectrum, and, apart from emission lines, 
 the residual is indeed quite low (cf. Fig.~\ref{fit-self}).
   
\begin{figure}[h!]  
\centering
\resizebox{\hsize}{!}{\rotatebox[]{-90}{\includegraphics[width=3.5cm]{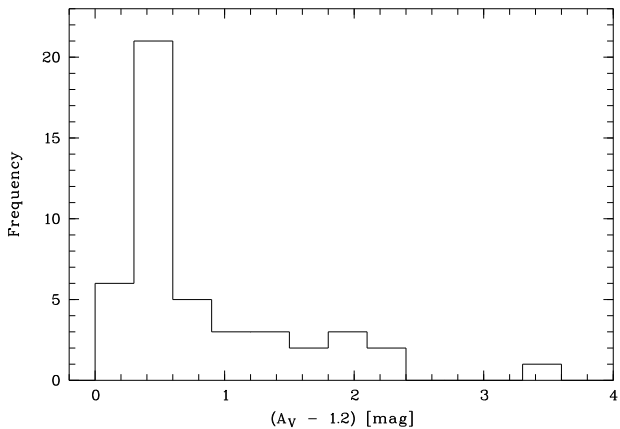}}}
\vspace{-2cm}
\caption{Histogram of differential visual extinction. 
The most frequent value is around 0.5\,mag, while the maximum observed is 3.3\,mag.
} 
\label{hist_Av}
\end{figure}

We applied the same procedure to each of the low-resolution spectra and 
determined the corresponding value of A$_V$ (see Table~\ref{tab:low-res-spec}),
with estimated errors of about 10\%. 
This takes account of the uncertainty in the fit, as well as the relative 
flux calibration, but not systematic effects caused by deviation of the 
circumstellar extinction from the normal interstellar law, as shown in 
Sect.\,\ref{sec:SED}. 
The two panels of Fig.~\ref{Fig:Av} show that a clear trend exists between the 
amount of visual extinction and the intensity of the H$\alpha$ and 
[O{\sc i}]\,6300\AA\ emission lines.  
This indicates that the observed variations do not reflect intrinsic changes
in the stellar photosphere, but arise presumably from variable circumstellar 
extinction, as can be inferred in Fig.~\ref{fig:lr-spectra} from the different 
continuum slopes of two spectra taken near maximum (on 30 May 1993) and minimum 
(on 1 Jun 1993) brightness. 
 The histogram in Fig.~\ref{hist_Av} indicates that the highest extinction 
events are relatively rare, while events with differential extinction below 
1\,mag are more frequent.

\section{The spectral energy distribution}
\label{sec:SED}
 
\begin{figure}[t]  
\centering
\resizebox{\hsize}{!}{\rotatebox[]{-90}{\includegraphics[trim = 10mm 20mm 0mm 0mm, clip]{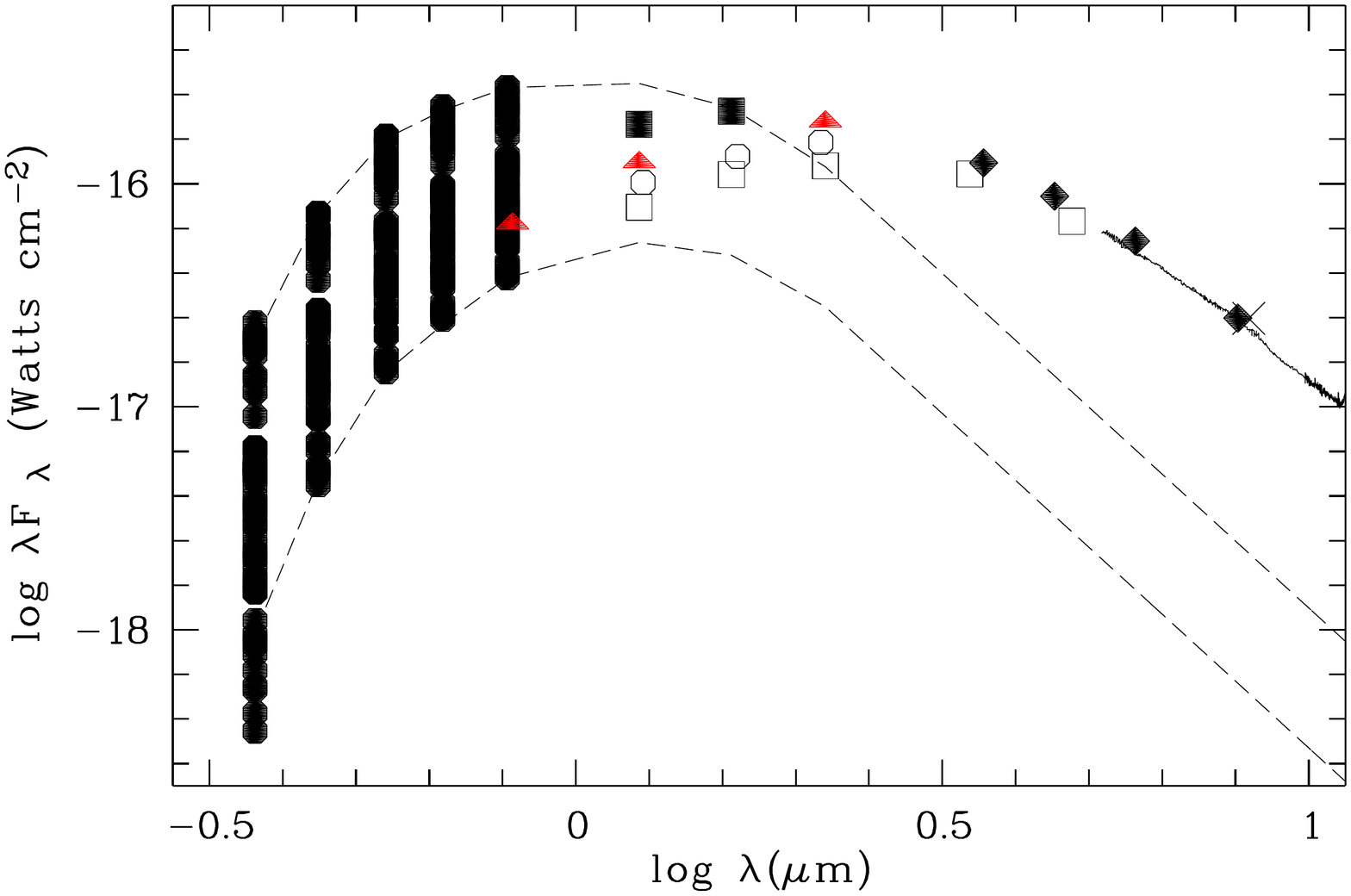}}}
\vspace{-2cm}
\caption{Observed SED of T\,Cha for $\lambda < 1\mu$m. 
The $U$,$B$,$V$,$R$,$I$ data (filled circles) are from \citet{Alcala1993,Covino1996}. 
The IR fluxes come from different papers:  \citet{Alcala1993} (open squares), 
2MASS fluxes (open circles), DENIS survey (filled triangles), 
\citet{Morel1978} (filled squares), 
Spitzer IRAC bands (filled diamonds), MSX6C (crosses).
IRS spectra are also superposed. 
The upper and lower dashed lines represent the SEDs of a G8\,V standard, reddened 
to the corresponding ($B-V$) values that match the brightest and faintest 
$I$-band observed fluxes, respectively. 
} 
\label{FigSED-data}
\end{figure}

We constructed the observed SED of T\,Cha using all optical and 
near-IR photometry available to us \citep{Alcala1993, Covino1996}, 
as well as data from public catalogues. In Fig.~\ref{FigSED-data}, 
the SED is shown for wavelengths shorter than $10\,\mu$m. 
We remark that photometric errors are smaller than the symbol size, 
and the considerable scatter of the optical ($UBVRI$ bands) 
data points reflects the significant variability of the star. 
Although of smaller amplitude, variability in the near-IR ($JHK$ bands) 
can also be appreciated. 
Therefore, relying on data that have not been acquired simultaneously 
may be a limitation in the SED analysis.

\begin{figure}[t]  
\centering
\resizebox{\hsize}{!}{\rotatebox[]{-90}{\includegraphics[trim = 10mm 20mm 0mm 0mm]{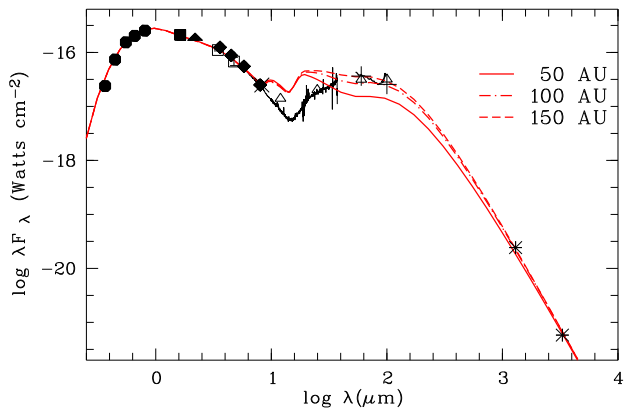}}}
\vspace{-2cm}
\caption{ Observed data-points used for SED fitting.
Optical: symbols as in Fig.\,\ref{FigSED-data};  IRAS: open triangles. 
Also superposed are IRS spectra (thin line). 
Three disc models, whose parameters are reported in Table~\ref{tab:param}, are represented
by the solid, dashed and dotted lines, for different disc radii.} 
\label{FigSED-fit}
\end{figure}

At the brightest level, the observed $J$ flux was lower than the expected 
stellar photospheric flux. However, we note that T\,Cha has not been 
monitored as extensively in the near-IR as in the optical. 
Therefore, the available data probably did not detect the brightest $J$ flux. 
For that reason, the $J$ flux was excluded from our SED-fitting.
Excess emission is instead observed for wavelengths 
longer than the $H$ band. The agreement of the Spitzer spectroscopy with 
the photometric data in the mid- and far-IR (cf. Fig.\,\ref{FigSED-fit}) 
indicates that there is no significant variability at longer wavelengths. 
This is expected in the case that the strong brightness and colour 
variations affecting the star are due to variable extinction of the 
stellar photosphere from inhomogeneous, circumstellar material.

By assuming variable extinction, we used relationships derived from $UBVRI$ 
photometry (Covino et al. 1996) to probe the dust column affecting the brightness 
of T\,Cha. 
In Table~\ref{tab:extinc}, we report the circumstellar extinction law for T\,Cha, 
expressed by the total to selective extinction ratios, $R_\lambda = A_\lambda /E_{B-V}$, 
derived from the differential brightness variations measured in the $UBVRI$ bands 
(Covino et al. 1996).
The resulting value of $R_V$ deviates from the normal value ($R_V=3.1$) for the diffuse 
interstellar medium, which indicates different physical properties of the circumstellar
grains. In particular, a higher value of $R_V$ means a flatter extinction curve 
(i.e., grayer extinction), and reveals the presence of larger grains, thus implying 
probable grain growth and depletion of small grains \citep{Draine2008}.   

\begin{table}[pt]
\caption{ 
Total to selective extinction ratios, $R_\lambda = A_\lambda / E_{B-V}$, for T\,Cha,
derived from $UBVRI$ photometry (Covino et al. 1996).
}
 \centering 
\begin{tabular}{c c c c c }
\hline \hline
 $R_U$   &  $R_B$  &  $R_V$   &  $R_R$   &  $R_I$  \\
\hline
 $8.4\pm0.4$ & $6.5\pm0.3$ & $5.5\pm0.2$ & $4.9\pm0.2$ & $4.5\pm0.2$ \\
\hline 
\end{tabular}
\label{tab:extinc}
\end{table}

Therefore, in the following SED analysis, we considered more suitable the flux 
measurements corresponding to the brightest level of the star, shown in 
Fig.~\ref{FigSED-fit}. 

\begin{figure*}  
\centering
\includegraphics[width=16cm]{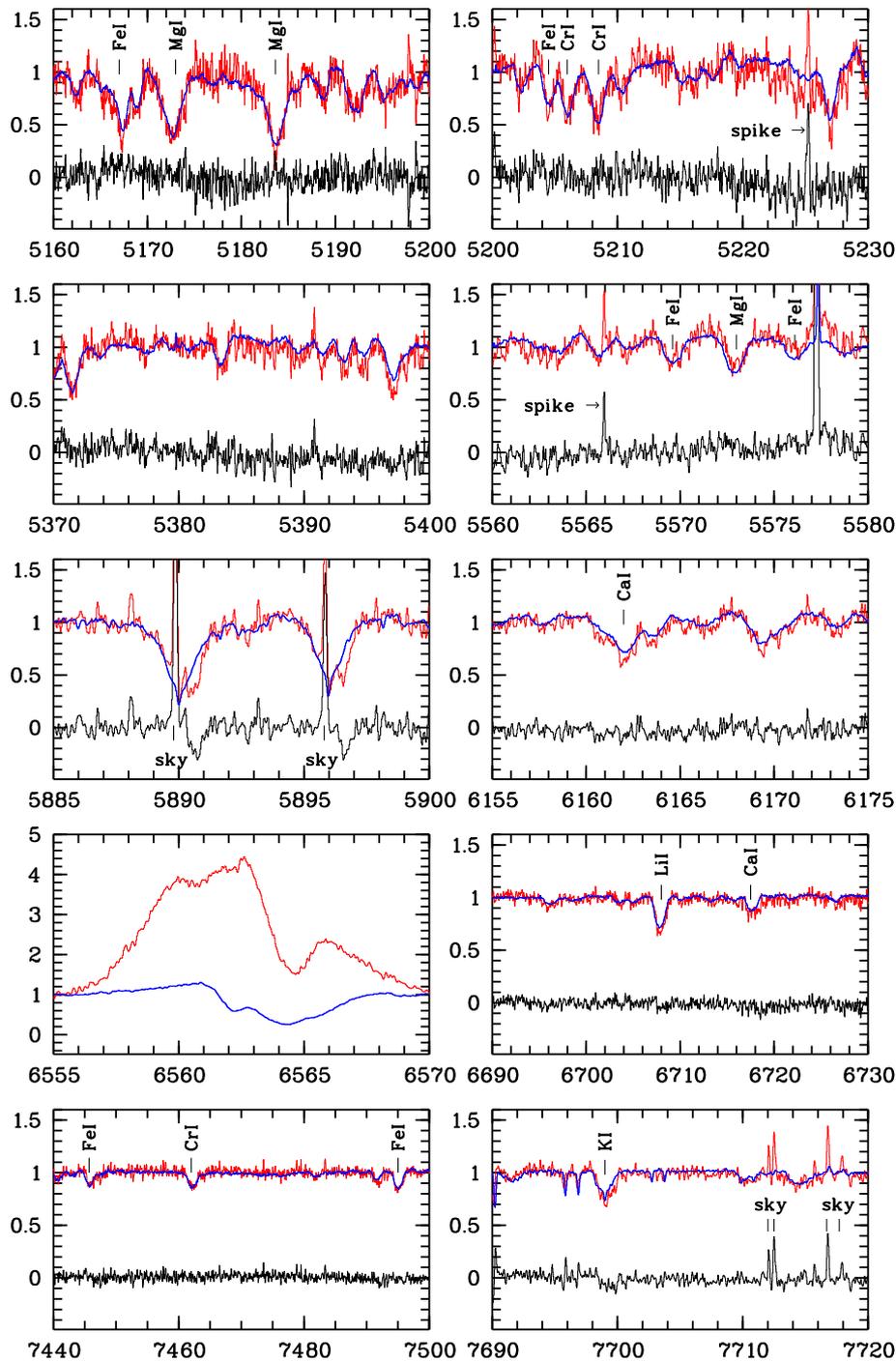}
\caption{ Comparison between T\,Cha spectra observed in two different epochs:
{\bf in blue } the spectrum observed on 14 Feb 2002, in red the one observed on 
26 Apr 2004, in black the difference between the two. 
Various spectral lines at different wavelengths are marked. 
} 
\label{Fig:Noveiling}
\end{figure*}

\begin{table}
\caption{In the upper table, we report the estimated stellar properties, and 
 in the lower table, the stellar and disc parameters resulting from 
 SED-fitting using \cite{Dullemond2001} model. 
}
\begin{tabular}{ l l }
\hline \hline
Stellar Parameter & value \\

\hline
 ${\rm L}$  (${\rm L}_{\odot}$)  &   2.6$^\dag$      \\
 ${\rm T}_{\rm eff}$ (K)  		&  5520$^\ddag$      \\
 ${\rm R}$ (${\rm R}_{\odot}$)	        &   1.8$^*$      \\
 ${\rm M}$ (${\rm M}_{\odot}$)	        &   1.3$^{**}$      \\
 ${\rm Age}$  (Myr)                     &   10$^{**}$       \\
\hline
\hline
Model Disc Best-Fit Parameters          & 	     \\
\hline
 ${\rm L}_{\star}$  (${\rm L}_{\odot}$) &   2.5      \\
 ${\rm T}_{\star}$ (K)  	        &  5400      \\
 ${\rm M}_{\star}$ (${\rm M}_{\odot}$)  &   1.5      \\
 ${\rm M}_{disc}$ (${\rm M}_{\odot}$)   &  5$\times 10^{-3}$ \\
 ${\rm R}_{out}$ (AU)                   &   150 	    \\
 ${\rm R}_{in}$ (AU)  		        &   0.13	    \\
Temperature of rims                     &  1300 	   \\
Inclination $i$	                        &  75$^{\rm o}$    \\
\hline 

\end{tabular}
\label{tab:param}

{\footnotesize 
$\dag$ from SED integration and adopting $d$=100\,pc and A$_V$=1.3; 
$\ddag$ from the relations by \citet{Kenyon1995} for a G8\,V star;
$*$  from the Stefan-Boltzman law;
$**$ from comparison with \citet{Palla1999} PMS tracks. 
}
\end{table}

Disc models such as those by \citet{Robitaille2006,Robitaille2007} do not 
take account of the presence of a partially evacuated inner hole or gap, 
so may be inadequate in reproducing the SED of transitional objects 
such as T\,Cha. \citet{Brown2007} found that the mid-IR spectrum 
of T\,Cha can be reproduced by a disc truncated at 0.08\,AU with 
a gap between 0.2 and 15\,AU, but they did not consider the strong 
variability of the object nor use the available millimetre data.

We explored the possibility of modelling the SED, simultaneously from 
optical to millimetre wavelengths, using the CGPLUS prescription 
by \cite{Dullemond2001}. Starting from the stellar parameters in 
Table~\ref{tab:param}, we constructed a grid of SEDs. 
The results of the SED fitting are reported in Table~\ref{tab:param} 
and the best-fit models are overplotted on the observed SED in 
Fig.~\ref{FigSED-fit}. 
The dust opacities calculated by \citet{Laor1993}, modified to match the 
\citet{Beckwith1990} opacity law at wavelengths longer than 100\,$\mu$m 
were used. We note that a blackbody with ${\rm T}_{\rm eff}=$5400\,K better 
reproduces the optical data. 
Previous estimates of the disc mass based on single measurements at millimetre 
wavelengths are in the range of 3$\times 10^{-3}$--1$\times 10^{-1}{\rm M}_{\odot}$ 
\citep{Henning1993, Lommen2007}, but our estimate reproduces well the observed data 
for wavelengths longer than 60\,$\mu$m. The disc radius is more poorly constrained, 
but values between 100 and 150\,AU provide quite reasonable results. 
The latter is consistent with the 3.3\,mm observations by \cite{Lommen2007} 
that did not resolve the disc around T\,Cha. 
We note that the disc radius adopted by \citet{Brown2007} 
is a factor of two higher than the value derived here. 

Although the model with a disc radius of 150\,AU reproduces well 
the observed SED at optical, far-IR, and millimeter wavelengths, 
all the models predict too high flux between 10 and 30\,$\mu$m 
relative to the observed SED. 
Thus, despite our attempt to minimize the effects of colour changes by adopting 
the brightest flux level in each band, we were unable to fit the complete SED 
of T\,Cha with a simple reprocessing-disc model. This failure is ascribed to 
excess emission in the near-IR. Therefore, the most plausible disc model 
for reproducing the SED of T\,Cha is a disc with a gap, as proposed 
by \citet{Brown2007}.

\section{Discussion and conclusions}
\label{sec:conclusion}
 
Before discussing our interpretation of T\,Cha, we briefly summarise the most 
significant properties of the star derived from analysis of the spectroscopic 
observations presented here, complemented with published photometry and other 
data from public archives: 
 \\
a) the spectral type G8 and a lithium abundance indicative of a very young age 
are confirmed
  \\
b) RV variations of almost 10\,km\,s$^{-1}$ on relatively short timescale are 
   detected, although without revealing any periodicity that could be releted
   to orbital motion. 
  Besides that, the $v\sin{i}$ determinations obtained from FEROS observations 
  are consistent with the value of 37$\pm$2\,km\,s$^{-1}$, except for a few 
  spectra in which the photospheric lines are significantly narrower, 
  yielding a $v\sin{i}$ of 30$\pm$2\,km\,s$^{-1}$. 
  \\
c) strong variability in the main emission features (e.g., H$\alpha$, H$\beta$, and 
   [O{\sc i}] lines) on timescales comparable to that of photometric variability,  
 and a close correlation between the H$\alpha$ and [O{\sc i}] line intensities are found.    
  \\
d) variations in the photospheric continuum slope are observed in connection with 
   intensity changes in the main emission features (i.e., the line intensities strengthen 
   when the star reddens and fades). 
  \\
e) clearly evident relationships between brightness and colours based on optical 
   photometry \citep{Alcala1993,Covino1996} have allowed us to derive a circumstellar 
   extinction law characterised by $R_V=5.5$ and infer the presence of large 
   circumstellar dust grains.
\\
f)  the observed SED, showing a dip at mid-IR wavelengths, is typical 
  of a transitional disc. 
  SED modelling has confirmed the presence of a disc with a gap and a high 
  inclination angle, as first indicated by \citet{Brown2007}.

Several of the properties outlined above provide clues (e.g., items {\em c}, {\em d}, and 
{\em e}) that a mechanism of variable circumstellar extinction is most probably operating 
in T\,Cha, as in UX\,Ori-like stars \citep{Natta2000, Rodgers2002}. 
The simultaneous brightening of H$\alpha$ emission and oxygen forbidden lines as 
reddening increases, as well as the existence of a tight correlation between brightness 
and colour variations, can be most plausibly explained in terms of variable extinction 
produced by obscuring clumpy material moving across the line of sight to the star. 
First of all, whereas the Hydrogen Balmer lines originate in regions closer to the star, 
the oxygen forbidden lines form instead farther away from the star and are hence not 
expected to be sensitive to variations occurring in the inner parts of the disc. 
Moreover, the dominant timescales and ranges of the observed variability phenomena 
indicate that the occulting clumps must be small compared to the size of the disc 
and more likely concentrated within a distance of a few tenths AU from the star.
These clumps occasionally obscure the star favoured by the high inclination of the disc, 
while not affecting either the outer, extended, low-density wind region traced by 
the [O{\sc i}], or the bulk of the circumstellar hydrogen emitting zone. 
The consequence is a strong contrast enhancement of all emission lines relative 
to the photospheric continuum when the star is highly attenuated, while 
the lines weaken or even disappear as the photospheric contribution turns brighter. 
Furthermore, these dusty clumps might also be associated with patchy gas structures 
orbiting close to the star ($\lesssim 0.2$\,AU) that cause daily changes in the 
H$\alpha$ line profile, similar to those we observe (see Sect.~\ref{sec:Ha}), 
due to variations in the absorbing column of material shielding the star. 
 If the H$\alpha$ variability were merely a consequence of the brightening and fading 
of the stellar continuum flux, the shape of the line profile should not change with time. 
This is generally not the case, suggesting that some other modifications 
in the H$\alpha$ forming region must also intervene. 
In particular, we expect that the parts of the H$\alpha$ profile originating 
closer to the star (e.g., the redshifted absorption components, in the case of 
magnetospheric accretion) also show the effects of occultation by orbiting material 
as well as eventual modulation induced by rotation. This is consistent with
the nightly line profile changes seen in the FEROS high-resolution spectra and 
the continuum flux (i.e. extinction) variations sharing a common timescale. 
However, while we can estimate that the time required for an occulting clump to appear 
in front of the star likely ranges between about one-third and two-thirds of a day, 
it is difficult, from the present data, to quantify in a reliable way the fraction 
of H$\alpha$ emission that may originate close to the star, as we are unable 
to determine information about the detailed geometry of the occultation.

But what is the source for the bulk H$\alpha$ emission in T\,Cha?
It is generally accepted that the strong H$\alpha$ emission line in 
T\,Tauri stars is due to mass accretion, which leads to UV excess 
emission and photospheric continuum veiling \citep{Hartigan1995}. 
Transitional objects, however, have more evolved discs in which accretion 
should significantly decrease with respect to the earlier CTTS phase.  
In T\,Cha, no veiling is apparently detected, not even in the stages of 
strong H$\alpha$ emission (see Figs.~\ref{Fig5}, \ref{fit-self}, 
and \ref{Fig:Noveiling}) suggesting very low or no mass accretion. 
However, the non-detection of veiling in T\,Cha may also caused by
the relatively high continuum flux emitted by its G8 photosphere. 
Hence, we cannot exclude the possibility that erratic accretion episodes 
are caused by instabilities in the inner disc, in which infalling gas 
produces the variable redshifted absorption features that we observe 
in the H$\alpha$ line profile.
However, the [O{\sc i}] $\lambda$6300\,\AA\ emission, which corresponds 
to a luminosity of $\log(L_{\rm [O\,I]}/L_{\odot})\approx -4.8$, implies
the presence of strong winds, although the width of the line 
($\approx$100\,km\,s$^{-1}$, see Fig.~\ref{Fig6}) is at the lower limit 
of a typical high-velocity component \citep{Hartigan1995}.  
It is thus probable that the bulk of the H$\alpha$ emission observed in T\,Cha 
is also produced in a wind, as reported for some Herbig Ae stars \citep{Kraus2008}. 
Moreover, the H$\alpha$ versus [O{\sc i}] correlation shown in 
Fig.~\ref{EqWid} is more easily explained in terms of variable 
circumstellar extinction, rather than variable accretion.

By both assuming variable circumstellar extinction, and adopting the relation 
$A_V = 1.086\,C_{ext}(V) N_H$ and the extinction cross-section per H nucleon 
for $R_V=5.5$, $C_{ext}(V)=6.715\times 10^{-22}$cm$^2/H$, given by 
\citet{Draine2003}, we estimate the column density for an obscuring clump 
to be $N_H \approx 1.4\times 10^{21} A_V$\,cm$^{-2}$. 
By also assuming the maximum observed differential extinction of 3.3~mag, we obtain 
a hydrogen column density $N_H = 4.6\times 10^{21}$ cm$^{-2}$. 
We can estimate the mass of the obscuring clumps, $M_{cl}$, by assuming 
that they cover completely the photospheric disc of the star. 
Using the stellar radius reported in Table~\ref{tab:param}, the result is 
$M_{cl}\approx 4\times10^{20}$\,g, i.e., of the order of 
$2\times10^{-13}$\,M$_{\odot}$. 
Similarly, assuming the most frequent differential circumstellar extinction 
from the histogram in Fig.~\ref{hist_Av}, the result would be 
$M_{cl}\approx 3\times 10^{-14}$\,M$_{\odot}$. In contrast, the mass 
of asteroid 253\,Mathilde is $5\times 10^{-14}$\,M$_{\odot}$ \citep{Yeomans98}.

 Finally, could the RV variations in T\,Cha becaused by a low-mass companion?
 Some PMS stars with transitional discs, such as CoKu\,Tau/4, have been 
 discovered to be binaries \citep[e.g.,][]{Ireland2008}. The inner 
 edge of circumbinary discs is believed to be set by dynamical truncation. 
 In the case of CoKu\,Tau/4, where the binary separation is about 8\,AU, 
 the truncation radius would be of the order of 13-16\,AU, depending on 
 the eccentricity of the binary \citep{Ireland2008}. 
 As concluded from our analysis of the CCF bisector, the RV variation of T\,Cha 
 is about 10\,km\,s$^{-1}$. A very low-mass object, say a 0.05\,M$_{\odot}$ 
 brown dwarf on a circular orbit with a radius of about 0.1\,AU may cause
 this variation, producing a disc truncation radius of about 0.2\,AU, 
 consistent with the inner disc edge proposed by \citet{Brown2007}. 
 However, no obvious periodicity is detected in our RV measurements. 
 On the other hand, the gap proposed by \citet{Brown2007} cannot be explained 
 in terms of truncation by a companion at a separation of about 10-15\,AU.
 At such separation, a 0.9\,M$_{\odot}$ star would be necessary to produce 
 the observed RV variation. An object of that mass and separation 
 would have been eventually detected either by us as a spectroscopic binary, 
 or by \citet{Kohler2001} in his {\em speckle} survey. Thus, the 
 gap of the transitional disc in T\,Cha is probably not the result of dynamical 
 truncation due to a binary companion.    
 In any case, the presence of a lower-mass (stellar or substellar) companion 
 cannot be definitely excluded because a quasi-periodicity of about 3\,days 
 is clearly evident throughout the entire data-set of T\,Cha. 
On the other hand, T\,Cha is a relatively fast rotator ($v\sin{i}\approx 37$\,km\,s$^{-1}$) 
and a magnetically active ($\log L_X \approx$ 30\,erg\,s$^{-1}$) PMS star. 
Therefore, the observed RV variations might be alternatively induced by stellar spots, 
as in the case of the WTTs V410\,Tau \citep{stelzer03}.  
Otherwise, we cannot exclude the possibility that the same circumstellar 
inhomogeneities responsible for the variable extinction may also alter the symmetry 
of the photospheric lines, and cause apparent RV shifts due to a sort of Rossiter 
effect, as in eclipsing binaries, which occurs when the stellar photosphere is 
partially obscured. 
In all cases, the presence of an inner disc characterised by a clumpy structure 
and the evidence of grain growth both indicate that T\,Cha is a likely candidate 
to host planet formation.

\begin{acknowledgements}      
We thank the user support group of ESO/La Silla. We are grateful to 
Rainer Wichmann for providing us with the series of FEROS spectra of March 1999. 
This research has made use of SIMBAD and VIZIER databases, operated at CDS, 
Strasbourg, France. This publication makes use of data products from the 
Two Micron All Sky Survey, which is a joint project of the University of 
Massachusetts and the Infrared Processing and Analysis Center/California 
Institute of Technology, funded by the National Aeronautics and Space 
Administration and the National Science Foundation. Financial support 
from INAF (PRIN 2007: {\em From active accretion to debris discs})
is acknowledged.
We thank the anonymous referee for constructive criticism and comments 
that helped to improve this paper. 
\end{acknowledgements}

\Online
\onecolumn
\begin{landscape}
\begin{figure*}
\resizebox{\hsize}{!}{\rotatebox{-90}{\includegraphics[trim = 15mm 30mm 0mm 0mm]{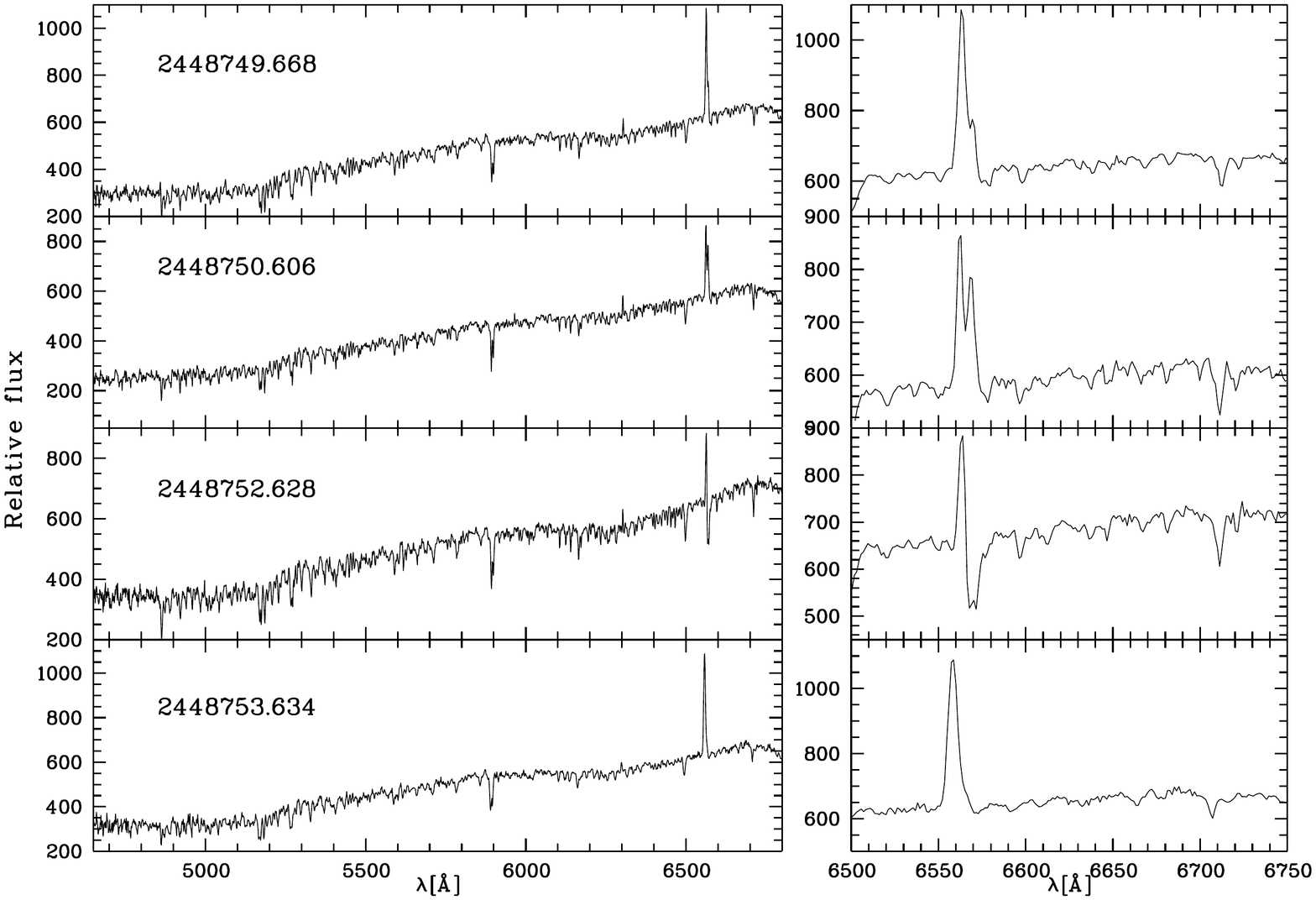} }}
 \vspace{-1cm}
  \caption{ Atlas of low-resolution spectra of T\,Cha. The panels on the left side display the whole range, 
            those on the right show the interval containing H$\alpha$ and the Lithium line. }
 \label{fig:lowres-atlas}
\end{figure*}
 
\begin{figure*}
\resizebox{\hsize}{!}{\rotatebox{-90}{\includegraphics[trim = 15mm 30mm 0mm 0mm]{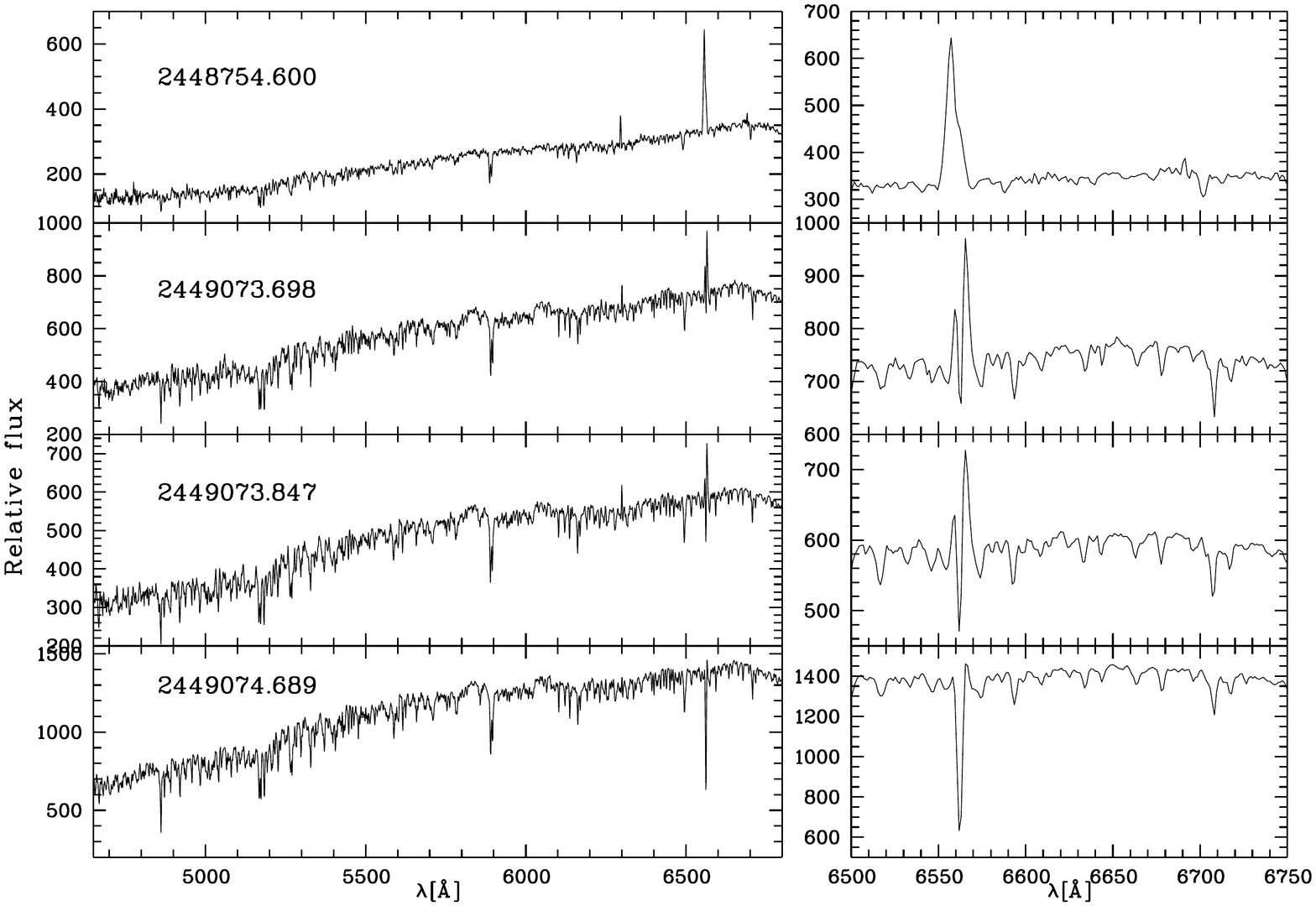} }}
 \vspace{-1cm}
  \caption{ Atlas of low-resolution spectra of T\,Cha, continue.}
 \label{fig:lowres-atlas}
\end{figure*}
\begin{figure*}
\resizebox{\hsize}{!}{\rotatebox{-90}{\includegraphics[trim = 15mm 30mm 0mm 0mm]{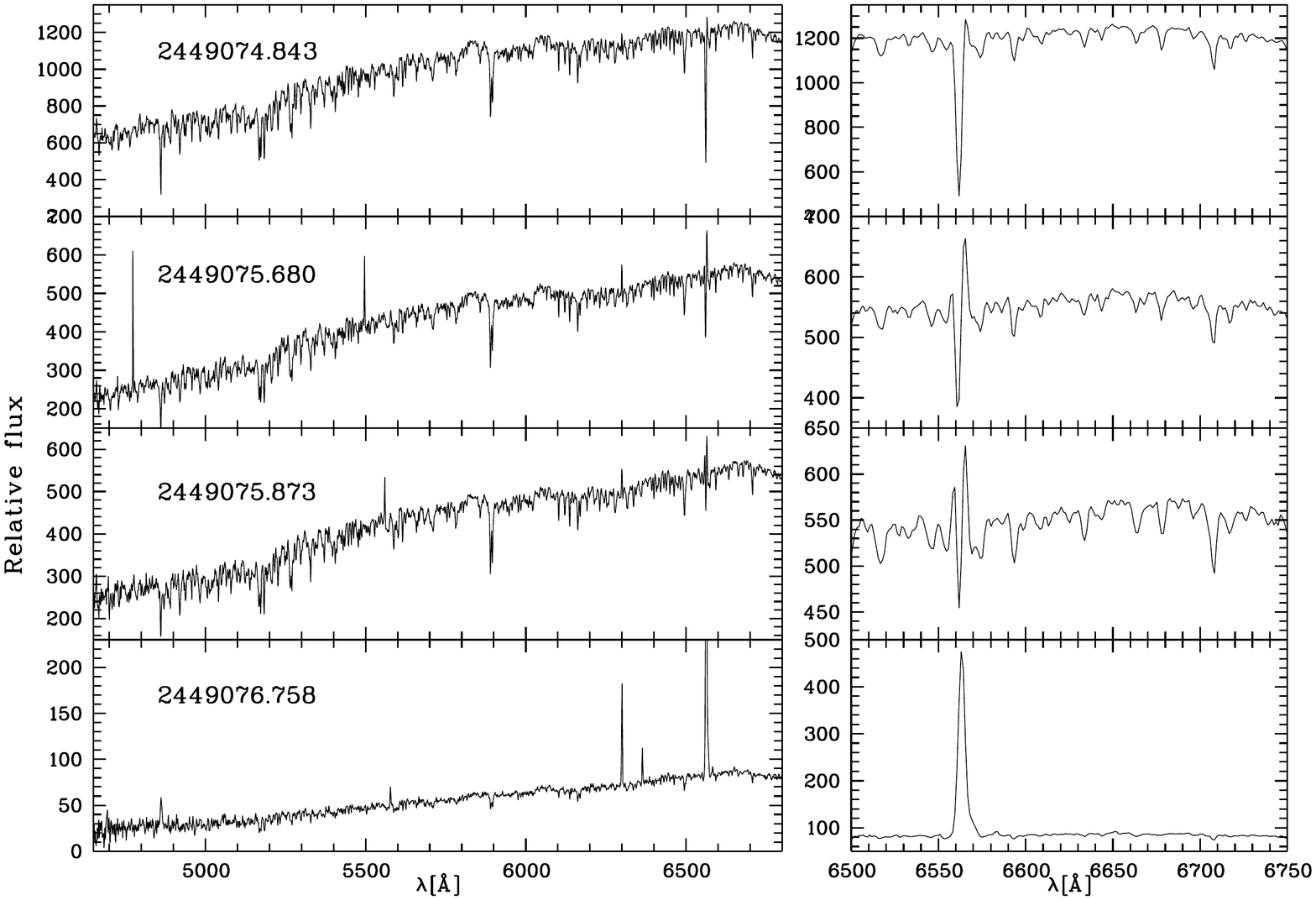} }}
 \vspace{-1cm}
  \caption{ Atlas of low-resolution spectra of T\,Cha, continue.}
 \label{fig:lowres-atlas}
\end{figure*}
\begin{figure*}
\resizebox{\hsize}{!}{\rotatebox{-90}{\includegraphics[trim = 15mm 30mm 0mm 0mm]{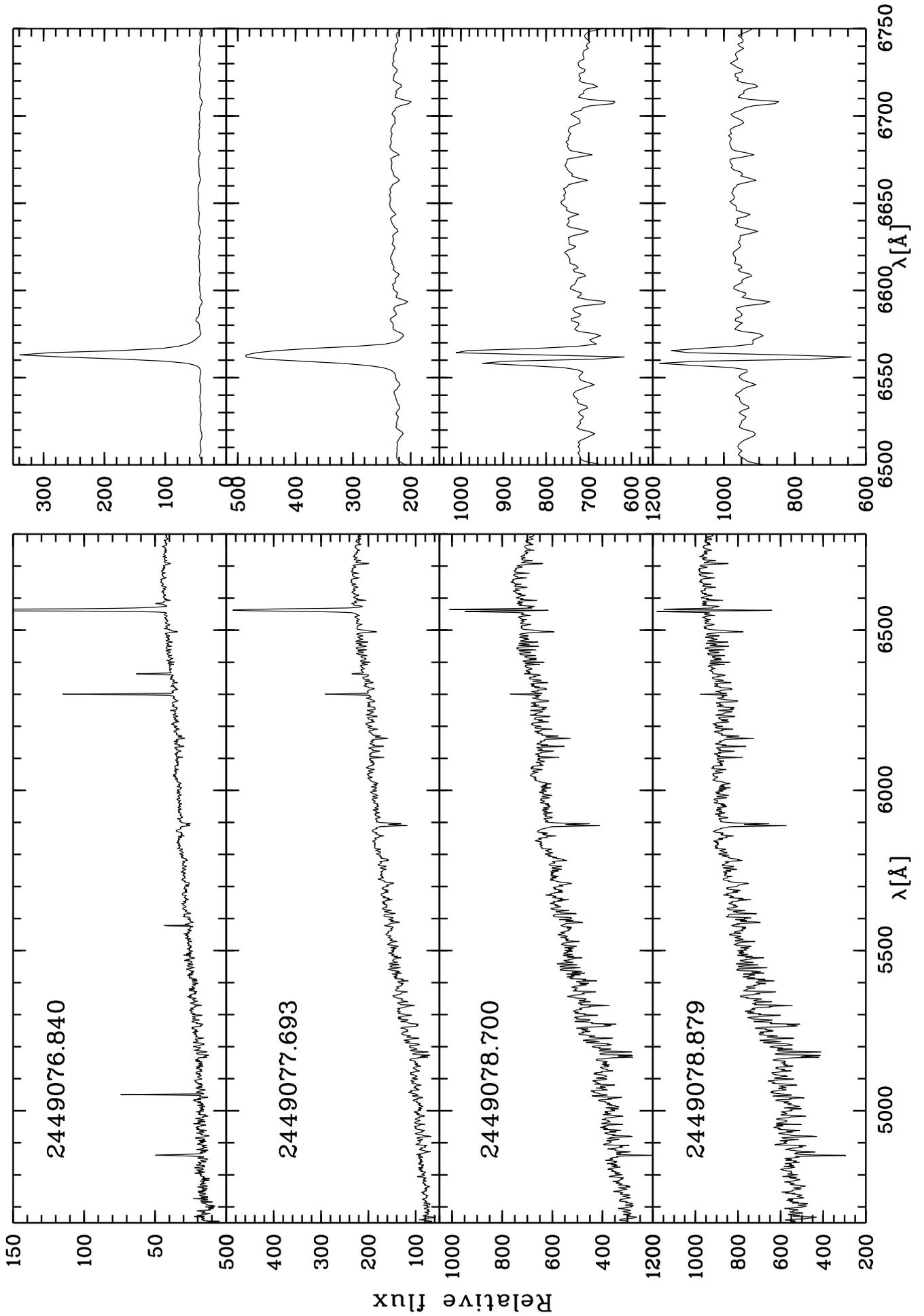} }}
 \vspace{-1cm}
  \caption{ Atlas of low-resolution spectra of T\,Cha, continue.}
 \label{fig:lowres-atlas}
\end{figure*}
\begin{figure*}
\resizebox{\hsize}{!}{\rotatebox{-90}{\includegraphics[trim = 15mm 30mm 0mm 0mm]{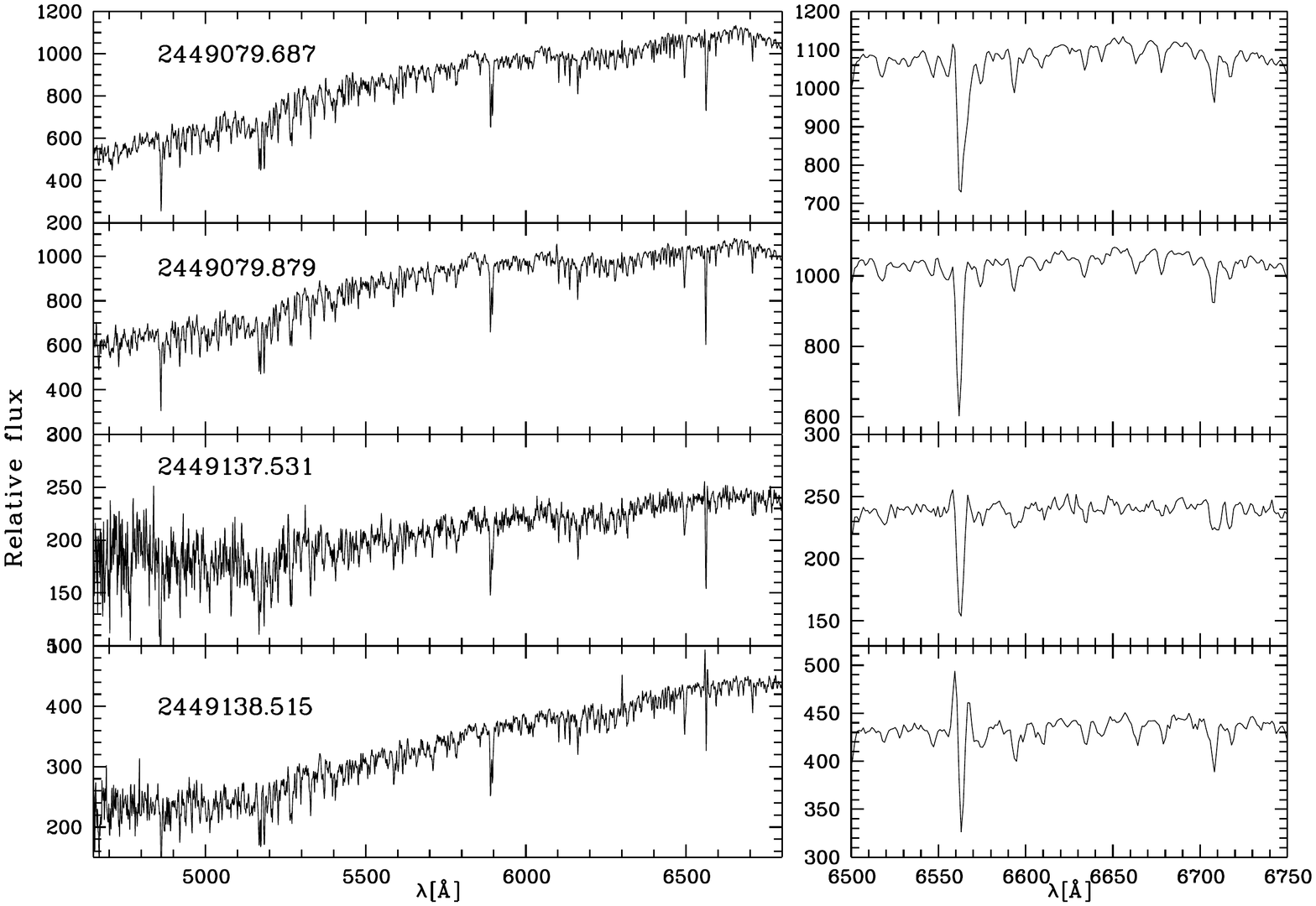} }}
 \vspace{-1cm}
  \caption{ Atlas of low-resolution spectra of T\,Cha, continue.}
 \label{fig:lowres-atlas}
\end{figure*}
\begin{figure*}
\resizebox{\hsize}{!}{\rotatebox{-90}{\includegraphics[trim = 15mm 30mm 0mm 0mm]{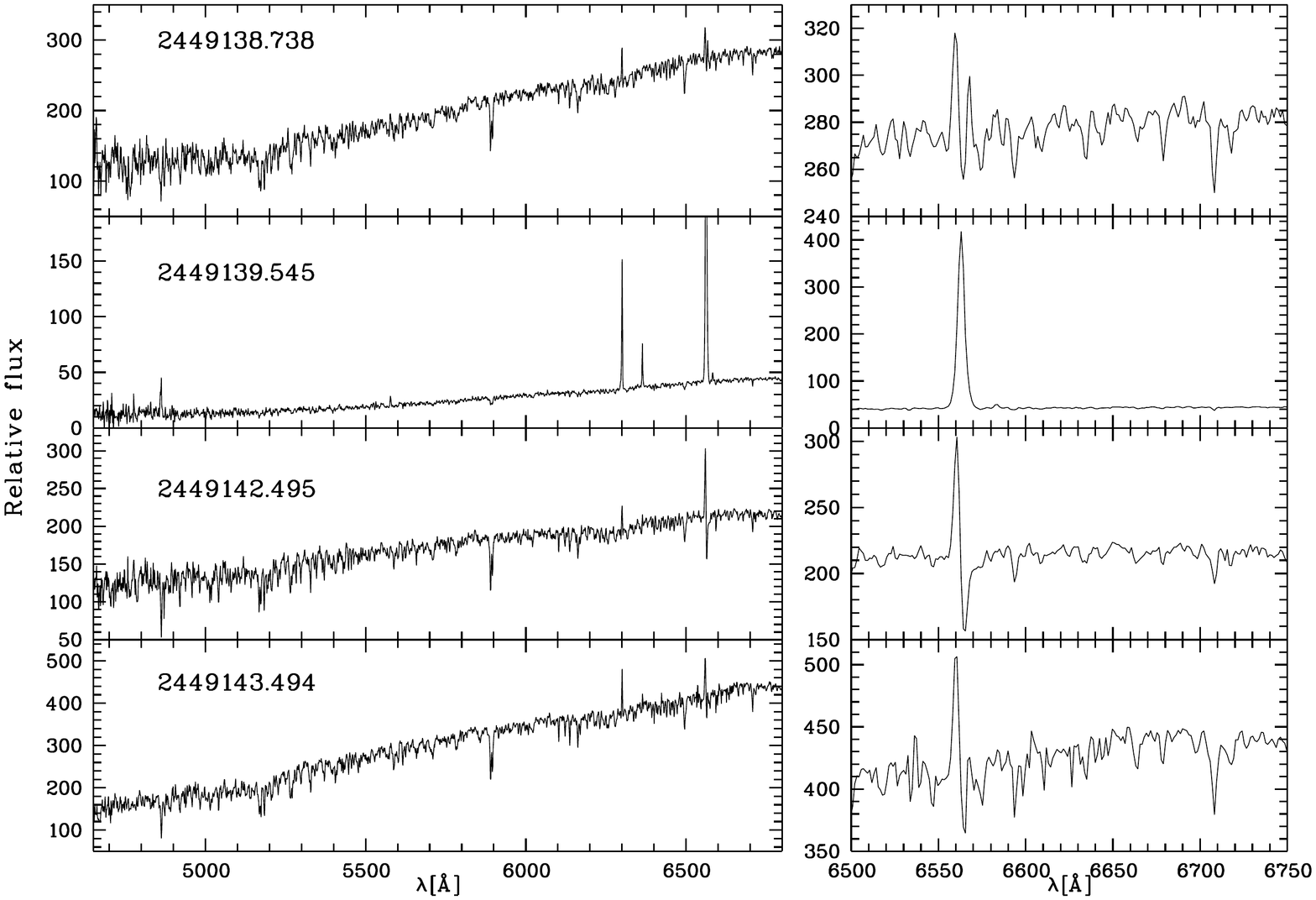} }}
 \vspace{-1cm}
  \caption{ Atlas of low-resolution spectra of T\,Cha, continue.}
 \label{fig:lowres-atlas}
\end{figure*}
\begin{figure*}
\resizebox{\hsize}{!}{\rotatebox{-90}{\includegraphics[trim = 15mm 30mm 0mm 0mm]{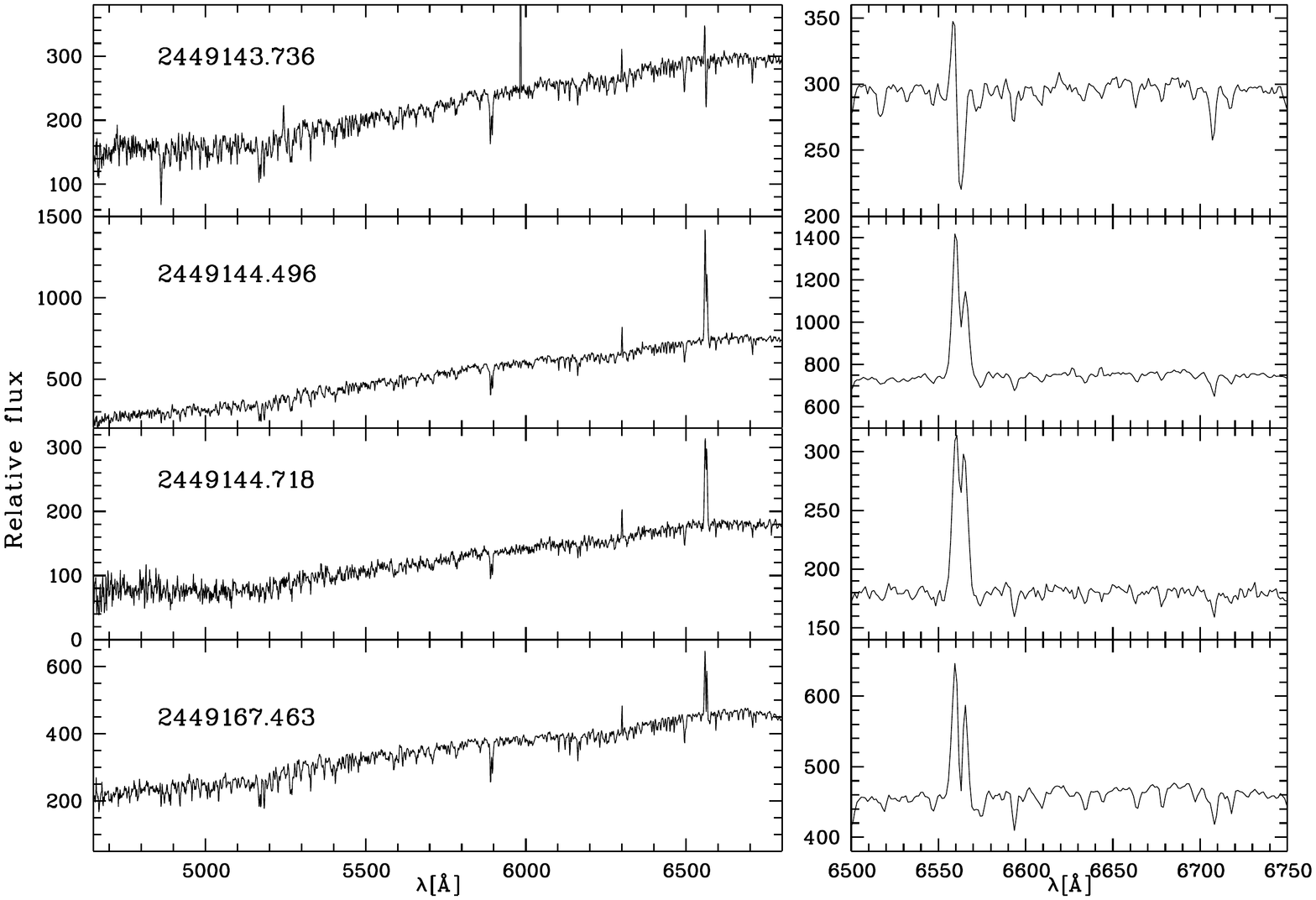} }}
 \vspace{-1cm}
  \caption{ Atlas of low-resolution spectra of T\,Cha, continue.}
 \label{fig:lowres-atlas}
\end{figure*}
\begin{figure*}
\resizebox{\hsize}{!}{\rotatebox{-90}{\includegraphics[trim = 15mm 30mm 0mm 0mm]{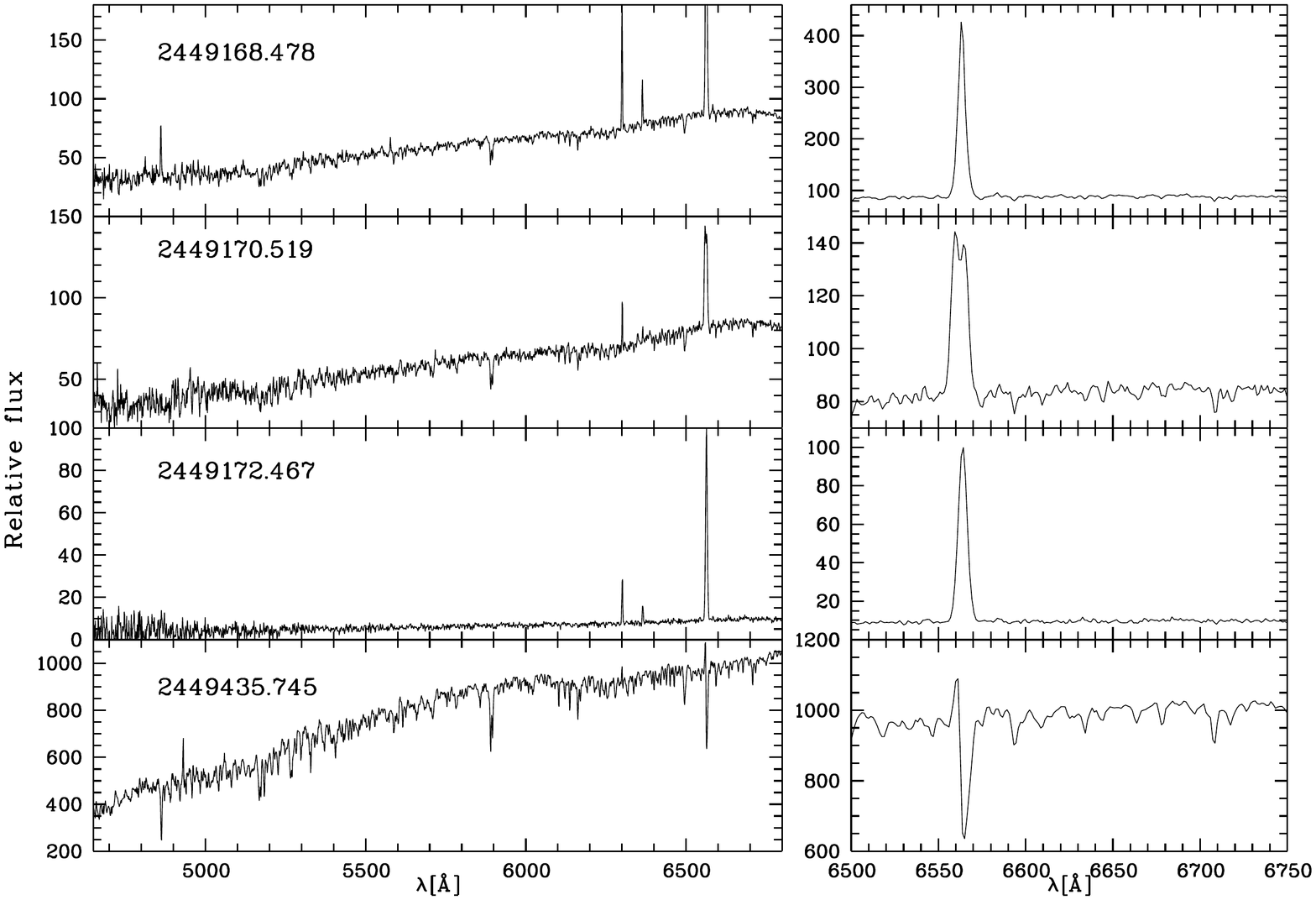} }}
 \vspace{-1cm}
  \caption{ Atlas of low-resolution spectra of T\,Cha, continue.}
 \label{fig:lowres-atlas}
\end{figure*}
\begin{figure*}
\resizebox{\hsize}{!}{\rotatebox{-90}{\includegraphics[trim = 15mm 30mm 0mm 0mm]{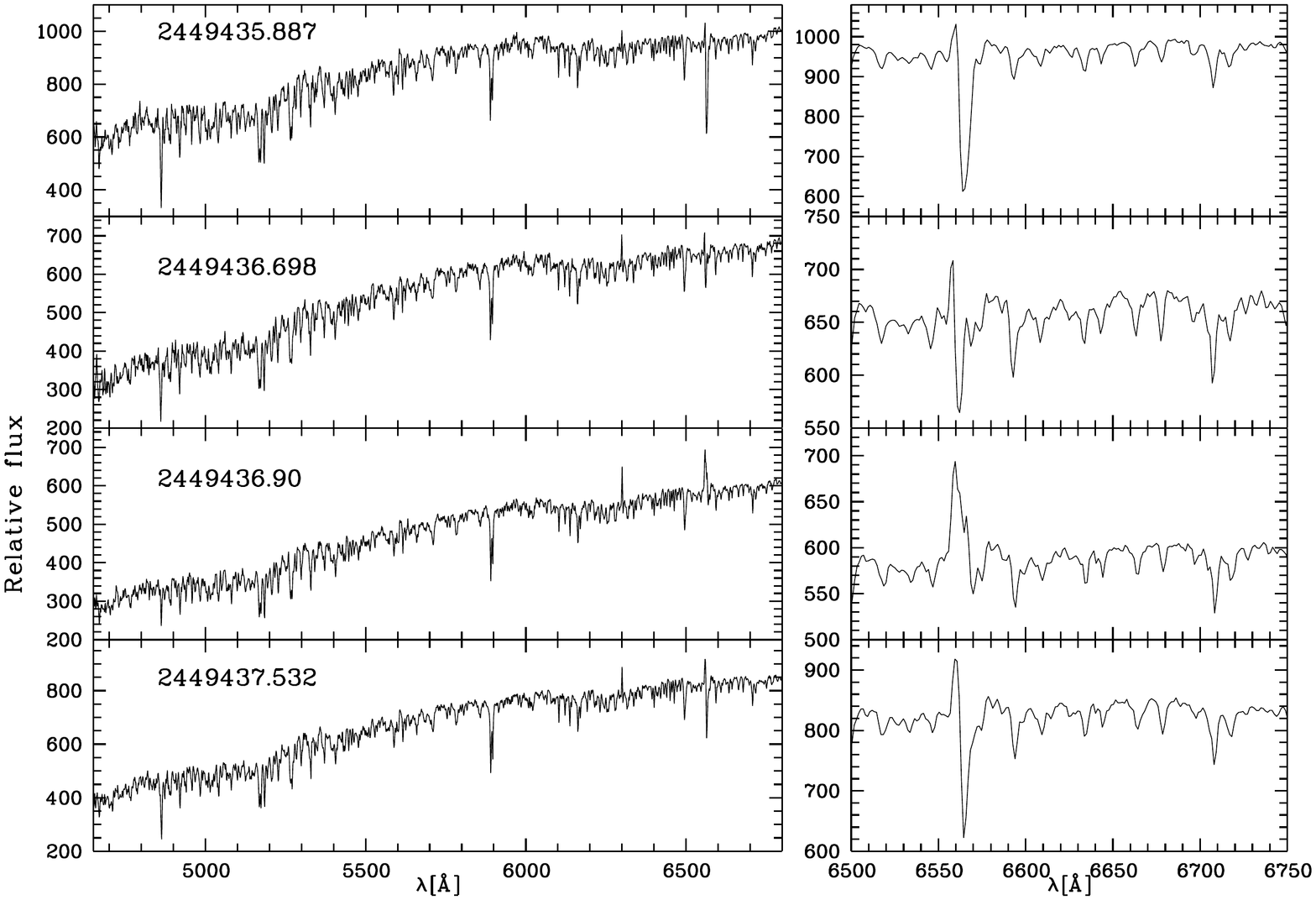} }}
 \vspace{-1cm}
  \caption{ Atlas of low-resolution spectra of T\,Cha, continue.}
 \label{fig:lowres-atlas}
\end{figure*}
\begin{figure*}
\resizebox{\hsize}{!}{\rotatebox{-90}{\includegraphics[trim = 15mm 30mm 0mm 0mm]{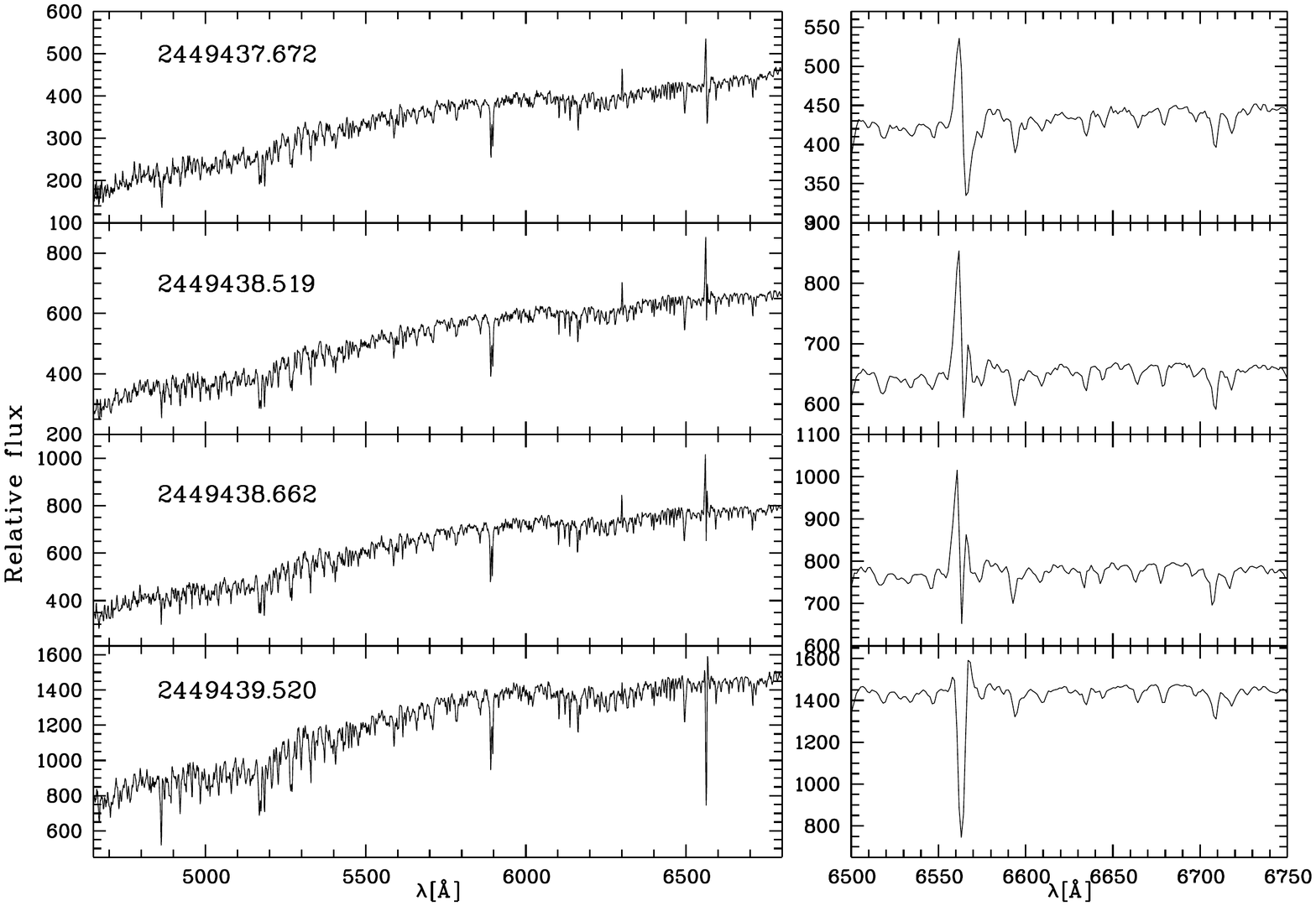} }}
 \vspace{-1cm}
  \caption{ Atlas of low-resolution spectra of T\,Cha, continue.}
 \label{fig:lowres-atlas}
\end{figure*}
\begin{figure*}
\resizebox{\hsize}{!}{\rotatebox{-90}{\includegraphics[trim = 15mm 30mm 0mm 0mm]{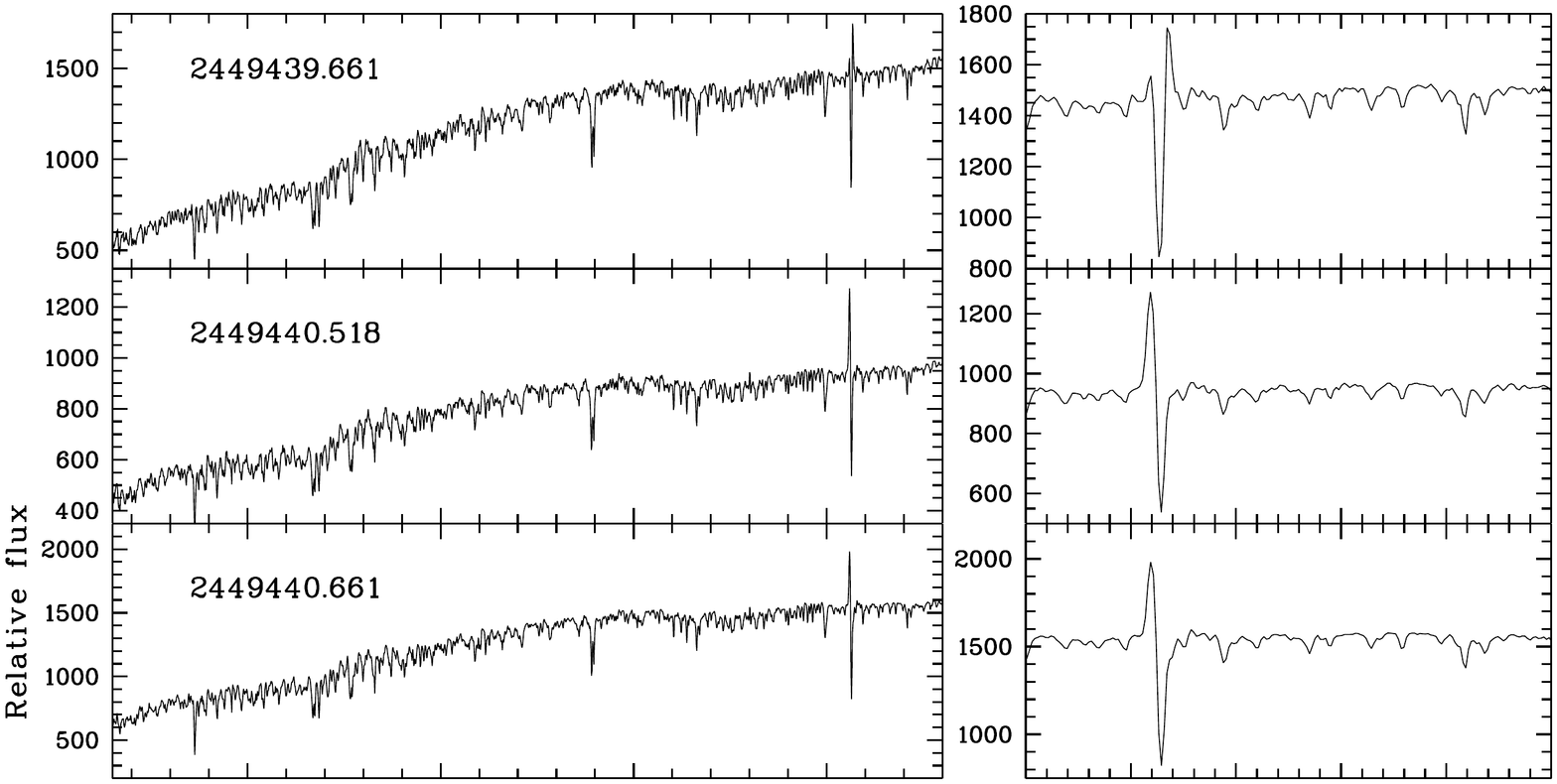} }}
 \vspace{-1cm}
  \caption{ Atlas of low-resolution spectra of T\,Cha, continue.}
 \label{fig:lowres-atlas}
\end{figure*}
\begin{figure*}
\resizebox{\hsize}{!}{\rotatebox{-90}{\includegraphics[trim = 15mm 30mm 0mm 0mm]{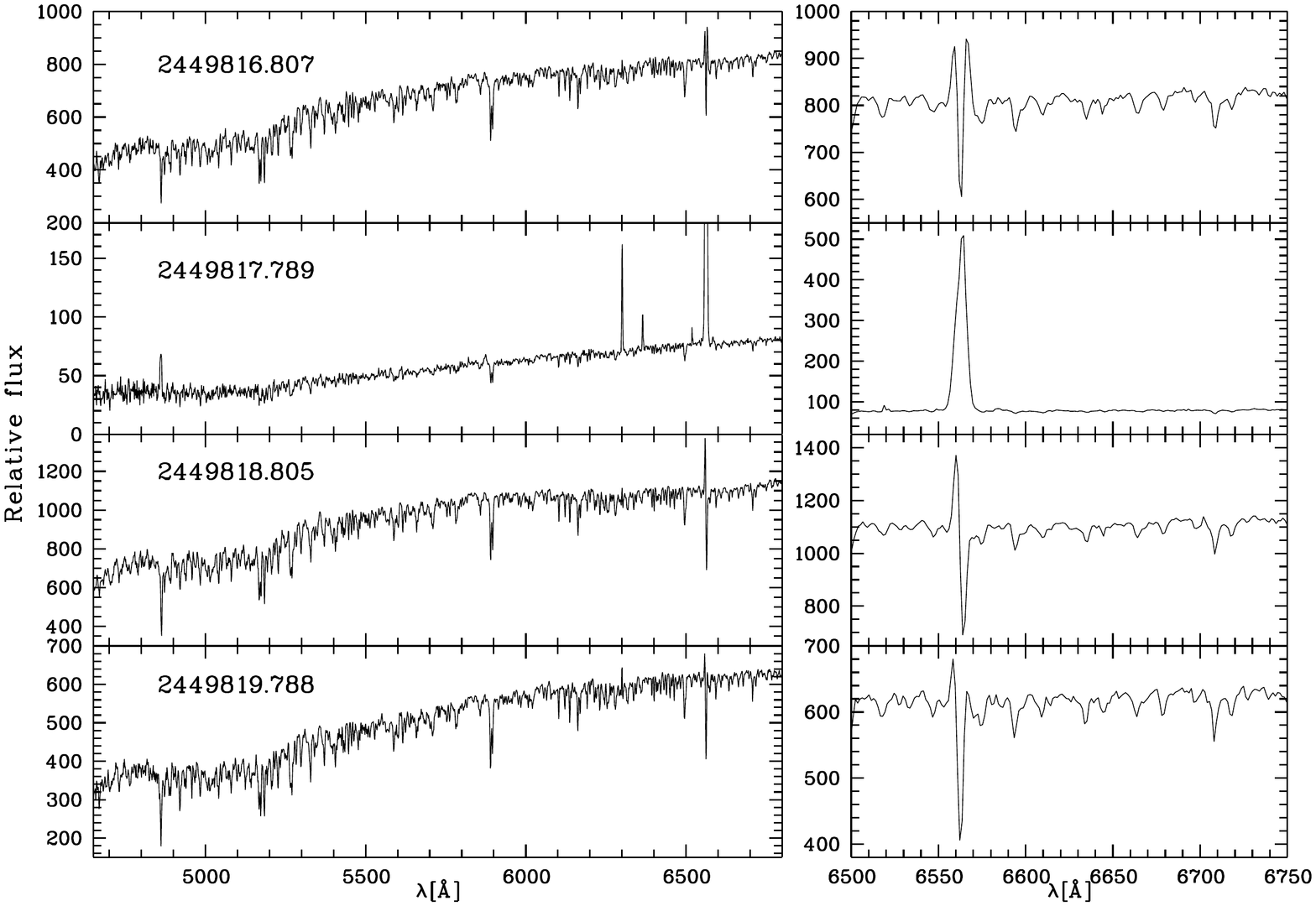} }}
 \vspace{-1cm}
  \caption{ Atlas of low-resolution spectra of T\,Cha, continue.}
 \label{fig:lowres-atlas}
\end{figure*}
\end{landscape} 
 
\end{document}